\documentclass[useAMS,usegraphicx]{mn2e}
\usepackage{amsmath,amssymb}
\usepackage{color,graphicx,colortbl,url,multirow}
\usepackage[utf8]{inputenc}
\usepackage{rotating}
\usepackage{caption}
\usepackage[longnamesfirst]{natbib}
\usepackage{subfig}
\usepackage{multirow}
\bibpunct[, ]{(}{)}{;}{a}{}{,}

\shortcites{Auger2009,Baldry2004,Bethermin2010,Blain1999,Bourne2011,Cowie1996,Cunha2010a,Damen2009,Damen2009a,Draine2007,Driver2007,Dunne2000,Dye2010,Dye2010a,Fixsen1998,Franceschini2001,Gott2001,Hill2006,Holwerda2009,Hopkins2010,James2002,Leeuw2004,Lilly1996,Madau1996,Magnelli2009,Negrello2005,Negrello2007,Popescu2002,Popescu2011,Puget1996,Santini2010,Saunders1990,Silva1998,Temi2004,Tojeiro2011,Tresse1999,White2007,Wuyts2008,Zheng2006}

\arrayrulecolor{black}

\newcommand{\degree}{\ensuremath{^\circ}}
\newcommand{\mum}{\ensuremath{\,\mu\text{m}}}
\renewcommand{\thefootnote}{\fnsymbol{footnote}}
\newcommand{\Mr}{\ensuremath{M_{r}}}
\newcommand{\hatlas}{\mbox{H-ATLAS}}

\title[\hatlas/GAMA: dust in optically selected galaxies]{\textit{Herschel}\footnotemark-ATLAS/GAMA: a census of dust in optically selected galaxies from stacking at submillimetre wavelengths}

\author[N.\ Bourne et al.]
       {N.\ Bourne,$\!^1$\footnotemark\ 
	S.\,J.\ Maddox,$\!^1$ 
	L.\ Dunne,$\!^1$  
	R.\ Auld,$\!^{2}$
	M.\ Baes,$\!^{3}$
	I.\,K.\ Baldry,$\!^{4}$  \newauthor
	D.\,G.\ Bonfield,$\!^{5}$ 
	A.\ Cooray,$\!^{6}$
	S.\,M.~Croom,$\!^{7}$ 
	A.\ Dariush,$\!^{8}$
	G.\,de~Zotti,$\!^{9,10}$  \newauthor
	S.\,P.~Driver,$\!^{11,12}$ 
	S.\ Dye,$\!^{1}$  
	S.\ Eales,$\!^{2}$
	H.\,L.~Gomez,$\!^{2}$
	J.~Gonz{\'a}lez-Nuevo,$\!^{10}$  \newauthor
	A.\,M.~Hopkins,$\!^{13}$
	E.\ Ibar,$\!^{14}$ 
	M.\,J.\ Jarvis,$\!^{5,15}$
	A.~Lapi,$\!^{10,16}$  
	B.~Madore,$\!^{17}$  \newauthor
	M.\,J.~Micha{\l}owski,$\!^{18}$ 
	M.\ Pohlen,$\!^{2}$ 
	C.\,C.~Popescu,$\!^{19}$ 
	E.\,E.\ Rigby,$\!^{1,18}$
	M.~Seibert,$\!^{17}$  \newauthor
	D.\,J.\,B.\ Smith,$\!^{1,5}$ 
	R.\,J.~Tuffs,$\!^{20}$ 
	P.~van~der~Werf,$\!^{21}$
	S.~Brough,$\!^{13}$ 
	S.\ Buttiglione,$\!^{9}$\newauthor
	A.\ Cava,$\!^{22}$ 
	D.\,L.\ Clements,$\!^{8}$
	C.\,J.~Conselice,$\!^{1}$  
	J.\ Fritz,$\!^{3}$ 
	R.\ Hopwood,$\!^{8}$ \newauthor
	R.\,J.\ Ivison,$\!^{14,18}$  
	D.\,H.~Jones,$\!^{23}$  
	L.\,S.~Kelvin,$\!^{11}$
	J.~Liske,$\!^{24}$ 
	J.~Loveday,$\!^{25}$
	P.~Norberg,$\!^{18}$ \newauthor
	A.\,S.\,G.~Robotham,$\!^{11}$
	G.\ Rodighiero,$\!^{26}$
	P.\ Temi$^{27}$
	\vspace*{1mm}\\
	$^1$School of Physics \& Astronomy, University of Nottingham,
	University Park, Nottingham NG7 2RD, UK\\
	$^{2}$School of Physics and Astronomy, Cardiff University, The Parade, Cardiff, CF24 3AA, UK\\
	$^{3}$Sterrenkundig Observatorium, Universiteit Gent, Krijgslaan 281 S9, B-9000 Gent, Belgium\\ 
	$^{4}$Astrophysics Research Institute, Liverpool John Moores University,
	Twelve Quays House, Egerton Wharf, Birkenhead,\\ CH41 1LD, UK\\
	$^{5}$Centre for Astrophysics, Science \& Technology Research Institute,
	University of Hertfordshire, Hatfield, Herts, AL10 9AB, UK \\
	$^{6}$Department of Physics and Astronomy, University of California, Irvine, CA 92697, USA\\ 
	$^{7}$Sydney Institute for Astronomy, School of Physics, University of
	Sydney, NSW 2006, Australia\\
	$^{8}$Physics Department, Imperial College London, Prince Consort Road, London SW7 2AZ, UK\\ 
	$^{9}$INAF-Osservatorio Astronomico di Padova, Vicolo Osservatorio 5, I-35122
	Padova, Italy\\ 
	$^{10}$Astrophysics Sector, SISSA, Via Bonomea 265, I-34136 Trieste, Italy\\
	$^{11}$Scottish Universities' Physics Alliance (SUPA), School of Physics and 
	Astronomy, University of St Andrews, North Haugh,\\ St Andrews, KY16 9SS, UK\\
	$^{12}$International Centre for Radio Astronomy Research (ICRAR), 
	University of Western Australia, Crawley, WA 6009, Australia\\
	$^{13}$Australian Astronomical Observatory, PO Box 296, Epping, NSW 1710, Australia\\
	$^{14}$UK Astronomy Technology Centre, Royal Observatory, Blackford Hill, Edinburgh EH9 3HJ, UK\\
	$^{15}$Physics Department, University of the Western Cape, Cape Town, 7535, South Africa\\
	$^{16}$Dip. Fisica, Univ. ``Tor Vergata'', Via Ricerca Scientifica 1, I-00133 Roma, Italy\\
	$^{17}$Carnegie Institution for Science, 813, Santa Barbara Street, 
	Pasadena, CA 91101, USA\\
	$^{18}$Institute for Astronomy, University of Edinburgh, Royal
	Observatory, Blackford Hill, Edinburgh EH9 3HJ, UK\\
	$^{19}$Jeremiah Horrocks Institute, University of Central Lancashire,
	Preston PR1 2HE, UK\\
	$^{20}$Max Planck Institute for Nuclear Physics (MPIK), Saupfercheckweg
	1, 69117 Heidelberg, Germany\\
	$^{21}$Leiden Observatory, Leiden University, P.O. Box 9513, NL - 2300 RA Leiden, The Netherlands\\
	$^{22}$Departamento de Astrof\'{\i}sica, Facultad de CC. F\'{\i}sicas,
 	Universidad Complutense de Madrid, E-28040 Madrid, Spain\\
	$^{23}$School of Physics, Monash University, Clayton, Victoria 3800,
	Australia\\
	$^{24}$European Southern Observatory, Karl-Schwarzschild-Str.~2, 85748
	Garching, Germany\\
	$^{25}$Astronomy Centre, University of Sussex, Falmer, Brighton BN1 9QH, UK\\
	$^{26}$University of Padova, Vicolo dell'Osservatorio 3, I-35122 Padova, Italy\\ 
	$^{27}$Astrophysics Branch, NASA/Ames Research Center, MS 245-6, Moffett Field, CA 94035, USA\\
	}
\begin{document}
\maketitle
\clearpage

\begin{abstract}
{
We use the {\it Herschel}-ATLAS survey to conduct the first large-scale statistical study of the submillimetre properties of optically selected galaxies. Using $\sim80,000$ $r$-band selected galaxies from $126\,\text{deg}^2$ of the GAMA survey, we stack into submillimetre imaging at 250, 350 and 500\mum\ to gain unprecedented statistics on the dust emission from galaxies at $z<0.35$. 
We find that low redshift galaxies account for 5\% of the cosmic 250\mum\ background ({4\% at 350\mum;} 3\% at 500\mum), of which approximately 60\% comes from `blue' and 20\% from `red' galaxies (rest-frame $g-r$).
We compare the dust properties of different galaxy populations by dividing the sample into bins of optical luminosity, stellar mass, colour and redshift. 
In blue galaxies we find that dust temperature and luminosity correlate strongly with stellar mass at a fixed redshift, but red galaxies do not follow these correlations and overall have lower luminosities and temperatures.  
We make reasonable assumptions to account for the contaminating flux from lensing by red sequence galaxies and conclude that galaxies with different optical colours have fundamentally different dust emission properties. Results indicate that while blue galaxies are more luminous than red galaxies due to higher temperatures, the dust masses of the two samples are relatively similar. Dust mass is shown to correlate with stellar mass, although the dust-to-stellar mass ratio is much higher for low stellar mass galaxies, consistent with the lowest mass galaxies having the highest specific star formation rates.
We stack the 250\mum-to-$NUV$ luminosity ratio, finding results consistent with greater obscuration of star formation at lower stellar mass and higher redshift. 
Submillimetre luminosities and dust masses of all galaxies are shown to evolve strongly with redshift, indicating a fall in the amount of obscured star formation in ordinary galaxies over the last four billion years.}
\end{abstract}

\footnotetext[1]{\textit{Herschel} is an ESA space observatory with science instruments provided by European-led Principal Investigator consortia and with important participation from NASA.}
\footnotetext[2]{E-mail: ppxnb1@nottingham.ac.uk}

\renewcommand{\thefootnote}{\arabic{1}}

\section{Introduction}

Dust in galaxies represents only a tiny fraction of the mass density of the Universe\footnote{The cosmic dust mass density was estimated to be 0.0083 per cent of the baryon density at redshift 0.0 by \citet[][]{Driver2007}}, yet from an observational point of view it can provide some of the most important indications of the history of star formation. 
{This is possible because most stars form in dense clouds of gas and dust. The dust in these regions is heated as it absorbs ultraviolet (UV) radiation from the hot massive stars that form within, and re-radiates the energy as a modified blackbody (or greybody) with a characteristic temperature generally around $20-30$\,K. 
Measurement of the UV radiation from these massive stars is the most direct method of inferring the star-formation rate (SFR), although as with all SFR indicators this relies on estimating the rate of massive star formation from the UV and assuming an initial mass function (IMF) to infer the total SFR. Dust poses a problem to this method since the UV attenuation must be accounted for, and this will vary from one star forming region to another as well as being dependent on the inclination angle that the galaxy disk is observed at.
Observing the thermal emission of dust with far-infrared (FIR) and submillimetre (submm) telescopes provides a way to trace the UV radiation field in all parts of the interstellar medium (ISM) of a galaxy, since in general the ISM is optically thin to FIR wavelengths. Hence one can use the UV and FIR in tandem to measure the (massive) SFR in the unobscured and obscured phases respectively.

A further complication arises from the fact that dust also exists in the large-scale ISM, in regions not associated with star formation. Indeed this component forms the bulk of the dust mass in a galaxy, and in spiral galaxies can dominate the bolometric output in the FIR \citep{Helou1986,Dunne2001,Sauvage2005,Draine2007}. Because this ISM component is heated by the general interstellar radiation field, its FIR emission is not necessarily correlated with the SFR. The extent to which both the UV and FIR luminosities can be used to trace SFRs can only be understood by studying the full spectral energy distribution (SED) of galaxies from the UV to the FIR and submm.
}

We know that the mid/far-IR luminosity density of the Universe increases towards higher redshifts, as a result of increased star-formation activity and {increased dust content}
{\citep{Blain1999,Franceschini2001,Dunne2003,LeFloc'h2005}.}
\defcitealias{Dunne2011}{D11}
Recent analyses of the submm luminosity function (LF; \citealp{Eales2009,Eales2010,Dye2010}; \citealp{Dunne2011} [hereafter \citetalias{Dunne2011}]) with the Balloon-borne Large Aperture Submillimeter Telescope \citep[BLAST;][]{Devlin2009} and the \textit{Herschel Space Observatory} \citep{Pilbratt2010} reveal strong evolution up to a redshift of at least $\sim0.5$, so there must be a substantial increase in the numbers and/or luminosities of dusty star-forming galaxies as we look back in cosmic time.
In this paper we ask the question: what are the properties of dust in galaxies that are not selected to be dusty, and is there an evolution in their dust content with redshift equivalent to that seen in \textit{Herschel}-selected galaxies?

{Galaxies in the Universe comprise an extremely varied population, with a wide range of different properties. The galaxies that we will concentrate on in this paper are the quintessential Hubble tuning fork types, both spirals and ellipticals, that comprise the majority of galaxies selected in optical surveys \citep[e.g.][]{Driver2006}.
We make no prior selection with respect to dust content or FIR luminosity, but it may be expected that the typical galaxies sampled are quiescent in nature, and are not undergoing excessive starburst or nuclear activity (as in typical FIR-selected samples from \textit{IRAS} or \textit{Spitzer}). This sample may have more in common with the low redshift population in the \hatlas\ sample selected at 250\mum, which typically consists of optically luminous ($\Mr\lesssim-20$), blue ($NUV-r<4.5$) galaxies \citep[\citetalias{Dunne2011};][]{Dariush2011}; but unlike \hatlas\ this sample will not be biased towards dusty galaxies in any way.}

Most large statistical studies of the FIR/submm properties of FIR-faint galaxies selected by their stellar light have focused on high redshift samples selected in the NIR \citep{Zheng2006,Takagi2007,Serjeant2008,Greve2009,Marsden2009,Oliver2010a,Viero2010,Bourne2011}.
Studies of the FIR/submm properties of samples of normal galaxies at low/intermediate redshifts have been restricted to small sample sizes and most have therefore focussed more on individual galaxies than populations
\citep[e.g.][]{Popescu2002,Tuffs2002,Leeuw2004,Stevens2005,Vlahakis2005,Cortese2006,Stickel2007,Savoy2009,Temi2009}.
This is simply because deep submm imaging of large areas of sky is necessary to cover a large enough sample of low-redshift galaxies for statistical analysis. Until very recently, such data have not been available. 
Observations in the submm, over the Rayleigh-Jeans tail of the dust SED at $\gtrsim200\mum$, are crucial for constraining the mass of cold dust in the ISM of galaxies, since FIR studies using {\it IRAS} at $\lesssim100\mum$ were only able to constrain the more luminous but less massive contribution from warm dust in star-forming regions \citep{Dunne2001}.

The \textit{Herschel} Astrophysical Terahertz Large Area Survey \citep[\hatlas;][]{Eales2010a} is the first truly large area submm sky survey, and as such is ideal for this work. It is the largest open-time key project on the \textit{Herschel Space Observatory} and will survey 550\,deg$^2$ in five channels centred on 100, 160, 250, 350 and 500\mum, using the PACS \citep{Poglitsch2010} and SPIRE \citep{Griffin2010} instruments.
In this study we use SPIRE maps of the three equatorial fields in the Phase 1 Data Release, which cover 135\,deg$^2$ centred at R.A. of $9^{\rm h}$, $12^{\rm h}$ and $14.5^{\rm h}$ along the celestial equator.
{We are currently unable to use the \hatlas\ PACS maps for stacking due to uncertainties in the flux calibration at low fluxes, hence this will be pursued in a follow-up paper.
}

\begin{figure}
 \includegraphics[width=0.45\textwidth]{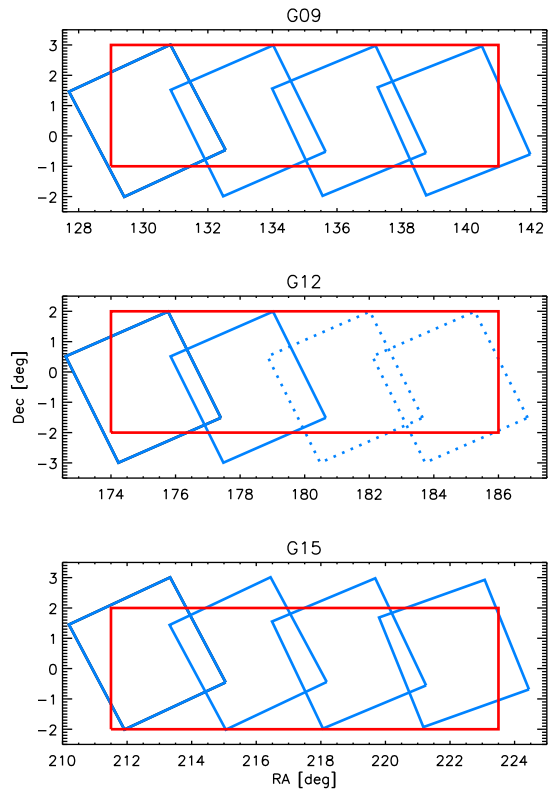}
\caption{{Comparison of the \hatlas\ (blue) and GAMA (red) coverage in the three equatorial fields. Dotted lines show the \hatlas\ tiles which are not covered in the current data; these will be included in a future release. The sample used in this work lies within the overlapping area between current \hatlas\ and GAMA coverage, approximately 126\,deg$^2$.}}
\label{fig:regions}
\end{figure}

For our galaxy sample we make use of UV, optical and near-infrared (NIR) data from the Galaxy and Mass Assembly (GAMA) survey \citep{Driver2009} which overlaps with the \hatlas\ equatorial fields at Dec.$>-1.0$\,\degree\ in the $9^{\rm h}$ field and Dec.$>-2.0$\,\degree\ in the other fields {(Fig.~\ref{fig:regions})}. The GAMA survey combines optical data from the Sloan Digital Sky Survey \citep[SDSS DR6;][]{Adelman-McCarthy2008}, NIR data from the UKIRT Infrared Deep Sky Survey (UKIDSS) Large Area Survey \citep[LAS DR4;][]{Lawrence2007}, and UV from the \textit{Galaxy Evolution Explorer} \citep[{\it GALEX};][]{Morrissey2005}, with redshifts measured with the Anglo-Australian Telescope and supplemented with existing redshift surveys (see \citealp{Driver2010} for further details).

In this paper we select galaxies { in the $r$-band, bin by their stellar properties derived from the UV-NIR,} and investigate their dust properties in the submm. We employ a stacking technique to recover median submm fluxes without being limited by detection limits in the \hatlas\ images. The optical data and sample selection are described in Section~\ref{sec:optdata}, and the submm imaging and stacking procedures are described in Section~\ref{sec:firdata}, although some more technical aspects of the stacking algorithm are left to Appendix~\ref{app:flux}. 
{In Section~\ref{sec:results} we present the results of stacking as a function of redshift, colour, mass and luminosity, and we discuss implications for the sources of the extragalactic background. In this Section we also derive rest-frame luminosities by inspecting the stacked SEDs, and investigate the effects of alternative binning schemes on our results.
Section~\ref{sec:disc} contains a more in-depth discussion of some of the implications of the stacking results with respect to dust luminosity, temperature and mass.} Final conclusions are summarised in Section~\ref{sec:conc}. Throughout this work we assume a cosmology of $\Omega_\Lambda=0.7$, $\Omega_\text{M}=0.3$, and $H_0=70$~km~s$^{-1}$~Mpc$^{-1}$. All celestial coordinates are expressed with respect to the J2000 epoch.

\section{Optical Data}
\label{sec:optdata}
\subsection{Sample selection}
\label{sec:sample}

{ We base our selection function on the GAMA `Main Survey' \citep{Baldry2010}, selecting} objects from the GAMA catalogue which are classified as galaxies by morphology and optical/NIR colours, and are limited in magnitude to $r_\text{petro}<19.8$ or ($z_\text{model}<18.2$ and $r_\text{model}<20.5$) or ($K_\text{model}<17.6$ and $r_\text{model}<20.5$).\footnote{{Model magnitudes are the best fit of an exponential and a de~Vaucouleurs fit as described by \citet{Baldry2010}.}} 
{ In fact only 0.3\% of the sample has $r_\text{petro} > 19.8$, so the sample is effectively $r$ selected. To simplify the selection function} we use the same selection in all fields, so we go below the GAMA `Main Survey' cut of $r_\text{petro}<19.4$ in the $9^{\rm h}$ { and $15^{\rm h}$ fields}.
For each galaxy we have matched-aperture Kron photometry in nine bands: $ugriz$ from SDSS and $YJHK$ from UKIDSS-LAS, plus $FUV$ and $NUV$ photometry from {\it GALEX}. More details of the GAMA photometry can be found in \citet{Hill2011}. All magnitudes used in this paper are corrected for galactic extinction using the reddening data of \citet{Schlegel1998} and are quoted in the AB system. The Kron magnitudes from \citet{Hill2011} are used for the colours described in this Section, however we use the Petrosian measurements for absolute magnitudes \Mr.
{We purposely do not apply dust corrections based on the UV-NIR SED or optical spectra, because we want to study dust properties as a function of empirical properties, excluding as much as possible any bias or prejudice to the expected dust content.}

\begin{figure}
 \includegraphics[width=0.45\textwidth]{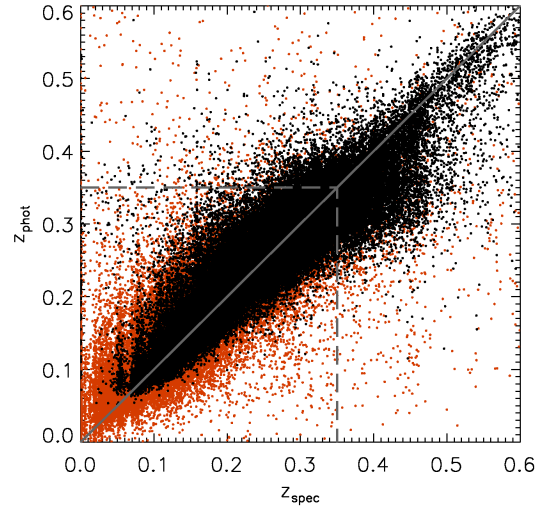}
\caption{Comparison of spectroscopic ($nQ\geq 3$) and photometric redshifts for the objects in our sample which have both. Black points have photometric redshifts with relative errors $<20\%$ while orange points have greater errors. Using this limiting error and a limiting redshift of 0.35 (dashed lines) ensures a reliable set of photometric redshifts for our purposes.}
\label{fig:zszp}
\end{figure}

\begin{figure}
 \includegraphics[width=0.45\textwidth]{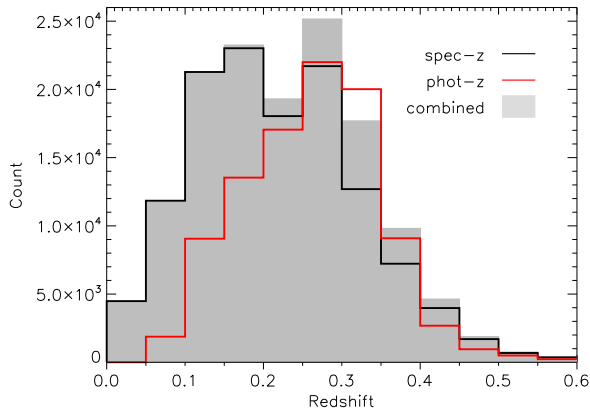}
\caption{Histogram of redshifts available in the catalogue: spectroscopic with $nQ\geq 3$ (black line); photometric with $\delta z/z\leq0.2$ (red line); all redshifts combined (grey shaded bars). {In constructing the grey histogram we take all the spectroscopic redshifts in the black histogram and add any additional photometric redshifts from the red.}}
\label{fig:zhist}
\end{figure}

We use spectroscopic redshifts from GAMA (year 3 data) where they are available and reliable (flagged with {\sc z\_quality}~$(nQ)\geq 3$). These are supplemented with photometric redshifts computed from the optical-NIR photometry using {\sc annz} \citep[for more details see][]{Smith2011a}. The comparison of photometric and spectroscopic redshifts is shown in Fig.~\ref{fig:zszp}. For this work we apply an upper limit in redshift of 0.35, {because the number of good spectroscopic and photometric redshifts drops rapidly beyond this point. This means that the redshift bins would have to be made wider to sample the same number of sources (hence diluting any sign of evolution); it also means that we only sample the brightest objects at the highest redshifts and their properties may not be comparable to the typically fainter objects sampled at lower redshift. We use photometric redshifts with relative errors $\delta z/z<20\%$ only, which excludes most of the poor matches in Fig.~\ref{fig:zszp}. In this way we exclude eight per cent of the whole sample at $z<0.35$ (seven per cent of $z_\text{phot}<0.35$), which means that the sample is still almost completely $r$-band limited. The limiting redshift error translates to a $20\%$ error on luminosity distance, a $40\%$ error on stellar mass, and an absolute magnitude error of $0.3$.
The criterion tends to exclude lower redshift objects, leading to a relative paucity of photometric redshifts at $z \lesssim 0.2$. This does not pose a problem since there is near complete spectroscopic coverage at these lower redshifts. Overall, 90 per cent of the redshifts we use are spectroscopic, although the photometric fraction does increase with redshift out to $z=0.35$.  The histograms of spectroscopic and photometric redshifts are shown in Fig.~\ref{fig:zhist}.
{ We tested the effect of random photometric redshift errors on our results by perturbing each photometric redshift by a random shift drawn from a Gaussian distribution with width $\sigma = \delta z$. After making these perturbations, we made the same cuts to the sample and repeated all the analysis, and found that all stacked results were robust, changing by no more than the error bars that we show.}

We note that a substantial number of photometric redshifts at $z>0.3$ appear to be biased low in Fig.~\ref{fig:zszp}. This explains why there appear to be more photometric redshifts than `all' redshifts at $0.3<z<0.35$ in Fig.~\ref{fig:zhist} -- i.e. some of those objects have spectroscopic redshifts which are greater than 0.35 and hence do not appear in the same bin in the `all redshifts' histogram. This issue could potentially affect the results in our highest redshift bin ($z>0.3$); the effect would be to contaminate that bin with galaxies from a slightly higher redshift, which may complicate any evolutionary trends seen across the $z=0.3$ boundary. We have chosen to leave the bin in our analysis because over 70 per cent of its galaxies have reliable spectroscopic redshifts, so the effect of a biased minority of photometric redshifts is considered to be small (and ultimately none of our conclusions hinge on this bin alone).}

In total we have a sample of 86,208 optically selected galaxies with good spectroscopic or photometric redshifts {within the 126~deg$^2$ overlapping area of the SPIRE masks and the GAMA survey}. We calculated $k$--corrections for the UV-NIR photometry using {\sc kcorrect v4.2} \citep{Blanton2007}, with the spectroscopic and photometric redshifts described above. The final component of the input catalogue is the set of stellar masses from \citet{Taylor2011}, which were computed by fitting \citet{Bruzual2003} stellar population models to the GAMA $ugriz$ photometry, assuming a \citet{Chabrier2003} { IMF}.\footnote{The NIR photometry was not used in deriving stellar masses due to problems fitting the UKIDSS bands as discussed by \citet{Taylor2011}.}
Altogether we have stellar masses estimated for 90 per cent of the sample. { The reason that 10 per cent are missing is that our sample reaches} deeper than the GAMA Main Survey in two of the fields: we use the same magnitude limit in all three fields so that we can sample as large a population as possible.

\subsection{Colour classifications and binning}
\label{sec:optcolours}
\begin{figure}
 \includegraphics[width=0.45\textwidth]{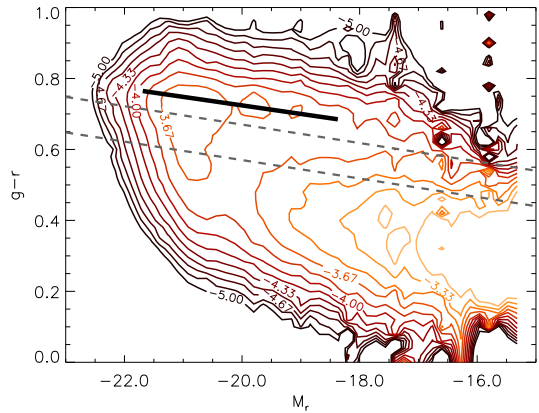}
\caption{Two-dimensional histogram (CMD) of rest-frame $g-r$ and \Mr\ for the full sample. Contours mark the log$_{10}$ of the histogram density function, { weighted with the $1/V_\text{max}$ method} to account for incompleteness as described in the text. {Contours are smoothed by a Gaussian kernel with width equal to 0.8 of the bin width (1/50 of the range in each axis).} The heavy black line is the red sequence fit given by equation~(\ref{eqn:g-r_fit}). The two dashed lines mark the boundaries between red/green and green/blue classifications respectively, which are given by equation~(\ref{eqn:g-r}).}
\label{fig:g-r}
\end{figure}

A simple way to divide the sample in terms of stellar properties is to make a cut in rest-frame optical colours. The bands which have good signal-to-noise data for the whole sample are the three central SDSS bands, hence the most reliable and complete optical colours to use are $g-r$, $g-i$, or $r-i$. We found very little difference between {the distributions of colours in any of these three alternatives; each appears equally effective at defining the populations of galaxy colours. We chose to use $g-r$ since these bands have the best signal-to-noise hence greatest depths,} and we plot the colour-magnitude diagram (CMD) in Fig.~\ref{fig:g-r}.

In this Figure the colour-magnitude space is sampled by a two-dimensional histogram in which the number density in each bin { is weighted by $\sum 1/V_\text{max}$ \citep{Schmidt1968}} to correct for the incompleteness of a flux-limited sample. To achieve this we weighted each galaxy by the ratio of the volume of the survey (the comoving volume at $z=0.35$) to the comoving volume enclosed by the maximal distance at which that galaxy could have been detected and included in the survey. To measure the latter we use the $r_\text{petro}$ limit which is the primary limiting magnitude of the sample (a negligible proportion of sources that were selected by $z$ or $K$ have fainter $r_\text{petro}$ magnitudes). Our redshift limit of 0.35 was also considered (so no $V_\text{max}$ is greater than the comoving volume at $z=0.35$).

\begin{figure}
 \includegraphics[width=0.4\textwidth]{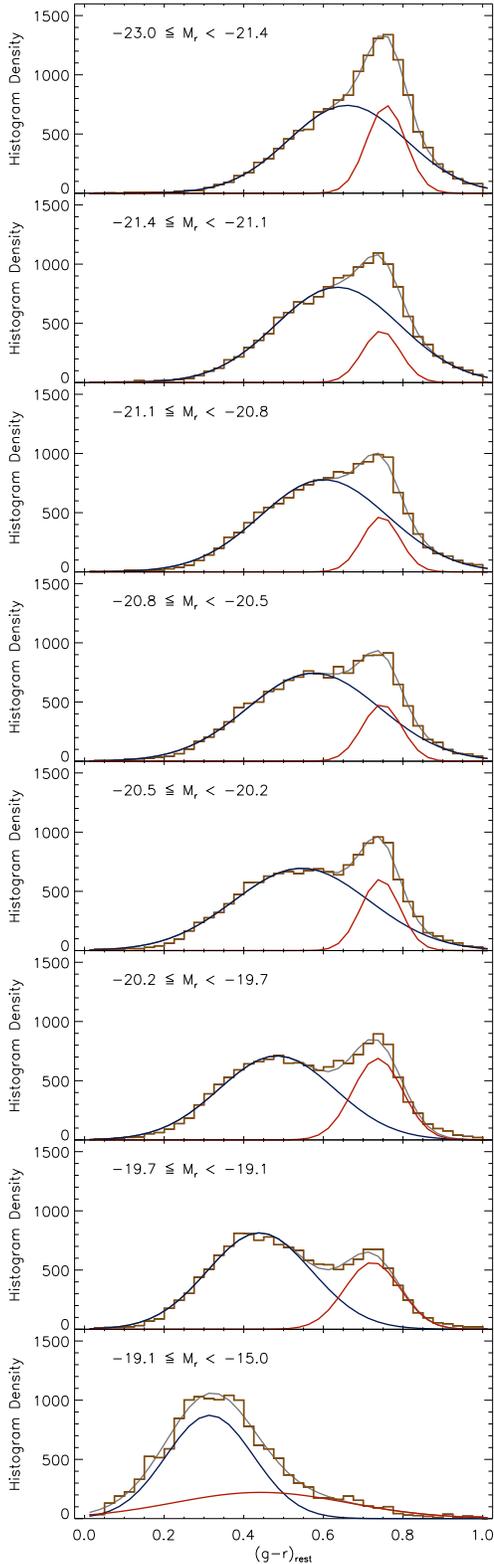}
\caption{Histograms of rest-frame $g-r$ colours split into eight \Mr\ bins between $-23$ and $-15$ { (weighted with the $1/V_\text{max}$ method)}. Overlaid are the two Gaussian functions which were simultaneously fitted to each histogram, representing the blue and red populations respectively, as well as the sum of the functions.}
\label{fig:g-r2}
\end{figure}

{It is now well established that the optical colours of galaxies fall into a bimodal distribution featuring a narrow `red sequence' and a more dispersed `blue cloud' \citep[etc]{Tresse1999,Strateva2001,Blanton2003,Kauffman2003,Baldry2004,Bell2004,Faber2007}.}
\Citet{Baldry2004} have shown that the optical CMD can be successfully modelled as the sum of two Gaussian functions in colour, which evolve with absolute magnitude and redshift. In Fig.~\ref{fig:g-r} we see a clear bimodality in $(g-r)_\text{rest}$ which can be modelled as a function of absolute magnitude \Mr\ (Petrosian) by splitting the distribution into eight bins between \Mr\ of $-15$ and $-23$, and computing the one-dimensional histogram of colours in each bin. These histograms were each fitted with the sum of two Gaussian functions, {shown in Fig.~\ref{fig:g-r2}. For convenience these functions can be thought of as representing two populations, one peaking on the red sequence and one in the blue cloud, although this interpretation has limited physical meaning.} A linear least-squares fit representing the red-sequence was obtained from the means and standard deviations of the red population as a function of absolute magnitude across the eight bins. {Note that the $\Mr\geq-19.1$ bin has effectively no red sequence, and has no contribution to the linear fit because the standard deviations were used as errors in the fitting.} We checked for any redshift dependency by splitting the population into three redshift bins as well as magnitude bins, but since no variation was found we use the red sequence fit to the eight magnitude bins with no redshift binning. This fit is shown as a heavy black line in Fig.~\ref{fig:g-r}, and is given by
\begin{equation}
 (g-r)_\text{rest} = 0.724 - 0.026(\Mr + 20).
\label{eqn:g-r_fit}
\end{equation}

\begin{figure}
 \includegraphics[width=0.45\textwidth]{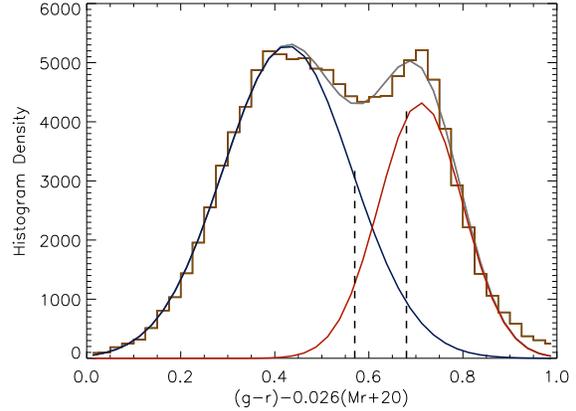}
\caption{Histogram of rest-frame, slope-corrected $g-r$ colours across $-23<\Mr<-18$ { (weighted with the $1/V_\text{max}$ method)}. Overlaid are the two Gaussian functions fitted to the histogram as well as the sum of the functions. The dashed vertical lines show the boundaries of the colour bins described in the text.}
\label{fig:g-r1}
\end{figure}

In order to divide red and blue populations as confidently as possible we examine the distribution of \mbox{$g-r$} colours across the range $-23<\Mr<-18$ in Fig.~\ref{fig:g-r1}. Here we plot the one-dimensional histogram of \mbox{$C_{gr} = (g-r)_\text{rest} - 0.026(\Mr+20)$}, thus removing the slope in the red sequence {to emphasise the bimodality.} 
{In Fig.~\ref{fig:g-r1}} we exclude $\Mr>-18$ because the red sequence becomes negligible at these faint luminosities and the distribution becomes dominated by the blue cloud, which hinders our two-component fitting {(note we do not make any absolute magnitude cut when stacking)}.
The distribution in Fig.~\ref{fig:g-r1} is fitted with the sum of two Gaussians: the red sequence has a mean of $\mu_\text{r}=0.71$ and standard deviation of $\sigma_\text{r}=0.09$; the blue cloud has $\mu_\text{b}=0.43$ and $\sigma_\text{b}=0.14$. To make a clean sample of red galaxies we make a cut at $C_{gr}>0.67$ (i.e. $\mu_\text{r}-0.5\sigma_\text{r}$). 
{{This cut was chosen to minimise the contribution of the `blue' functional fit, while also including the majority (55 per cent) of the area under the red fit. The fraction of this histogram at $C_{gr}>0.67$ that belongs to the blue function is seven per cent. It is recognised that the functional fits do not necessarily represent two distinct populations of galaxies, and this fraction does not imply a contamination of the red bin since all galaxies with $C_{gr}>0.67$ are empirically red. These cuts are largely arbitrary and the main purpose they serve is to separate the two modes of the colour distribution.}
Using similar arguments, we make a blue cut at $C_{gr}<0.57$ (i.e. $\mu_\text{b}+1\sigma_\text{b}$) which selects 86 per cent of the blue function; the fractional contribution of the red function in this bin is four per cent. 
The intermediate bin by its very nature is likely have a heterogenous composition including some galaxies close to the red sequence, others that are part of the blue cloud, and some proportion of genuine `green valley' galaxies \citep{Schiminovich2007,Martin2007}}. The relative contributions from each of these to the { intermediate (`green')} bin may vary as a function of redshift and absolute magnitude, which must be kept in mind when drawing any conclusions. However the red and blue bins will be dominated by completely different populations with respect to each other at all redshifts and absolute magnitudes, {which justifies the use of an intermediate bin to fully separate them}. The $g-r$ colour bins are summarised in equation~(\ref{eqn:g-r}):
\parbox{\columnwidth}{
\begin{align}
&\textrm{RED:} &0.67 + f(\Mr)&\,<\,(g-r)_\text{rest}\,<\,1.00&\notag \\
&\textrm{GREEN:} &0.57 + f(\Mr)&\,<\,(g-r)_\text{rest}\,<\,0.67 + f(\Mr) &\notag \\
&\textrm{BLUE:} & 0.00&\,<\,(g-r)_\text{rest}\,<\,0.57 + f(\Mr) & \notag 
\end{align}
\begin{equation}
 \textrm{where } f(\Mr)=-0.026(\Mr+20)=-0.026\Mr-0.52
\label{eqn:g-r}
\end{equation}
}
{These divide the sample into 41,350 blue, 17,744 green and 27,114 red galaxies.}

\begin{figure}
 \includegraphics[width=0.45\textwidth]{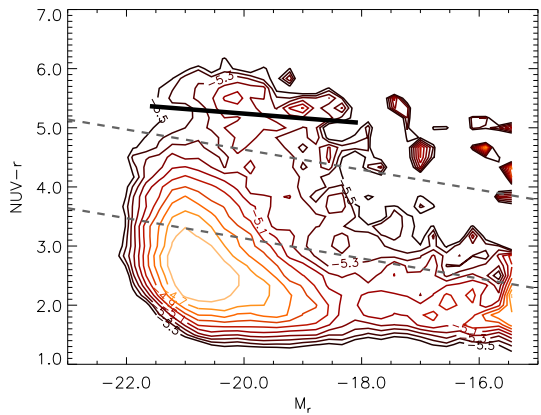}
\caption{Two-dimensional histogram (CMD) of rest-frame $NUV-r$ and \Mr\ for the full sample. Contours mark the log$_{10}$ of the histogram density function, { again using the $1/V_\text{max}$ method} to account for incompleteness as described in the text. {Contours are smoothed by a Gaussian kernel with width equal to one bin (1/40 of the range in each axis).} {The noise in the top right is a result of a small amount of data close to the $NUV$ detection limit having exceptionally high weights.} The heavy black line is the red sequence fit to our data given by equation~(\ref{eqn:NUV-r_fit}). The two dashed black lines mark the boundaries between red/green and green/blue classifications respectively, which are given by equation~(\ref{eqn:NUV-r}). }
\label{fig:NUV-r}
\end{figure}

A limitation of optical colours such as $g-r$ is the small separation in colour space between the red and blue populations. This would be improved by using a pair of bands which straddle the $4000$\AA\ spectral break, but the only Sloan band blueward of this is $u$, which has {poor signal-to-noise and therefore a relatively shallow magnitude limit}. It has been shown in the literature \citep[e.g.][]{Yi2005,Wyder2007,Cortese2009} that a UV-optical colour such as $NUV-r$ (or UV-NIR such as $NUV-H$) provides greater separation between red and blue and reveals a third population of galaxies in the green valley. A clear delineation of the populations would minimise contamination between the bins and should help to disambiguate trends in stacked results. 

In Fig.~\ref{fig:NUV-r} we plot the $NUV-r$ CMD (since the $r$-band data are deeper than $H$, and this gives an $NUV$-limited sample). We are unable to successfully fit the $NUV-r$ colour distribution in bins of \Mr\ using the simple double Gaussian function, due to a significant `green valley' excess. To overcome this we follow \citet{Wyder2007} by defining a clean red sample [$NUV-r>f(\Mr)-0.5$] and blue sample [$NUV-r<f(\Mr)-2.0$], where $f(\Mr)=1.73-0.17\Mr$ is the fit to a morphologically-selected red sequence by \citet{Yi2005}. To these subsets we attempt to fit Gaussian functions for the red and blue distributions respectively, again in eight bins of \Mr\ between $-15$ and $-23$. As before, we find a linear least-squares fit to the means of the red sequence, given by
\begin{equation}
 (NUV-r)_\text{rest} = 5.23 - 0.08(\Mr + 20).
\label{eqn:NUV-r_fit}
\end{equation}
The slope of this fit is somewhat flatter than the $-0.17$ found by \citet{Yi2005}, but the uncertainty is large due to the fact that we have only performed the Gaussian fit in each bin to the upper part of the red sequence. As in Fig.~\ref{fig:g-r1} we could subtract the slope and plot the histogram of the residual colour across all \Mr, but due to the uncertainty on the slope this does not give any extra benefit. We opt instead to simply adopt the colour cuts defined by \citet{Wyder2007} as boundaries between our blue, green and red samples:
\parbox{\columnwidth}{
\begin{align}
  &\textrm{RED:} &f(\Mr)-&0.5\,<\,(NUV-r)_\text{rest}\,<\,7.0 & \notag \\
  &\textrm{GREEN:} &f(\Mr)-&2.0\,<\,(NUV-r)_\text{rest}\,<\,f(\Mr)-0.5 \notag \\
  &\textrm{BLUE:} & &1.0\,<\,(NUV-r)_\text{rest}\,<\,f(\Mr)-2.0 \notag 
\end{align}
\begin{equation}
\textrm{where } f(\Mr)=1.73-0.17\Mr 
\label{eqn:NUV-r}
\end{equation}
}
{These divide the sample into 36,900 blue, 12,758 green and 3115 red galaxies.}
These numbers reveal two disadvantages of using the $NUV-r$ colour: that there are $NUV$ detections for only about 60 per cent of the sample, and that the $NUV$ selection naturally disfavours red $NUV-r$ colours leading to a smaller number of galaxies in the red bin. In contrast the $r$-band selection of the $g-r$ sample is relatively unbiased by the colour of the galaxy. 
However the differentiation between blue and green populations should be more successful using $NUV-r$ compared with $g-r$. Therefore both alternatives have their merits. {In Section~\ref{sec:results_g-r_NUV-r} we compare the results obtained using the two alternative colour cuts, but for the bulk of the paper we refer to the $g-r$ colour cuts unless otherwise stated.}

{In this paper we do not explicitly attempt to distinguish passive red galaxies from obscured, star-forming red galaxies; rather we focus on the submm properties as a function of { observed} optical colours. We therefore may expect a somewhat mixed population in the red (and green) bins, even using $NUV-r$. There are various ways one might attempt to overcome this -- applying dust corrections based on UV photometry or spectral line indices, or using the S\'{e}rsic index to predetermine the expected galaxy `type' -- however { we opt to avoid} biasing any of our results by any prior assumption about the nature of galaxies in each bin.\footnote{{The exception to this empirical approach is the assumption of the $k$--correction, which is necessary to make fair comparisons between redshift bins. The $k$--corrections are very well constrained by photometry in 5--11 bands and uncertainties are small compared with our colour bins.}} We leave any analysis that accounts for morphology or spectral properties to a follow-up paper.}

\section{Sub-Millimetre Data and Stacking}
\label{sec:firdata}
\subsection{Stacking into the SPIRE maps}
\label{sec:stacking}
For the submm imaging we use SPIRE images at 250, 350 and 500\mum\ of the three equatorial GAMA fields in \hatlas, which were made in a similar way to the science demonstration maps described by \citet{Pascale2010}. The fields consist of 53.25\,deg$^2$ at $9^{\rm h}$, 27.37\,deg$^2$ at $12^{\rm h}$ 53.93\,deg$^2$ at $14.5^{\rm h}$. Background subtraction was carried out using the {\sc nebuliser} routine developed by \citet{Irwin2010a} which effectively filters out the highly varying cirrus emission present in the \hatlas\ maps, {as well as extended background emission including the Sunyaev-Zel'dovich effect in clusters and unresolved clustered sources at high redshift.}

All sources are treated as point sources and { flux densities (hereafter `fluxes')} are measured in cut-outs of the map around each optical position, convolved with a point-response function (PRF). We account for sub-pixel scale positioning by interpolating the PRF from the point-spread function (PSF)\footnotemark\ at a grid of pixels offset from the centre by the distance between the optical source centre and the nearest pixel centre. {This convolution with the PRF is the standard technique to obtain the minimum-variance estimate of a point source's flux \citep{Stetson1987}. The PSFs at 250, 350 and 500\mum\ have full-widths at half maximum (FWHMs) equal to 18, 25 and 35~arcsec respectively.}
\footnotetext{The important distinction between the PSF and the PRF is that the PRF represents the discrete response function of the detector pixels to the continuous distribution of light (PSF) which reaches the detector from an astronomical point source.} 

We then measure fluxes in the maps at the positions in the optical catalogue, and use simplifying assumptions to account for blended sources which would otherwise lead to over-estimation of flux in the stacks. The procedure is described in detail in Appendix~\ref{app:flux}. {This automatically corrects stacked fluxes for the effect of clustered sources in the bins, with the caveat that sources not in the catalogue (i.e. below the optical detection limits) cannot be deblended.} We stack fluxes by dividing the sample into three colour bins (as described in the previous Section), then splitting each into five bins of redshift, then six bins of absolute magnitude. {Redshift and magnitude bins are designed to each contain an approximately equal number of objects;} in this way we ensure that the sample is evenly divided between the bins to maintain good number statistics in each. All three fields are combined in each stack. 
{ We choose to use the median statistic to represent the typical flux, since the median value with a suitable error bar is a robust representation of the distribution of fluxes in a given bin. Unlike the mean, the median is insensitive to bias due to exceptionally bright sources which are outliers from the general population \citep[e.g.][]{White2007}. }

We also measure the background in the maps, since although they have been sky-subtracted to remove the highly variable cirrus foreground emission, the overall background does not average to zero, and therefore has a significant contribution to stacked fluxes. In each map we create a sample of 100,000 random positions within the region covered by our optical catalogue, masking around each source in the main stacking catalogue with a circle of radius equal to the beam FWHM in order to avoid including the target positions in the background stack. We then perform an identical stacking analysis at these sky positions, but reject any positions where we measure a flux greater than $5\sigma$. This prevents a bias of our background measurement from sources detected in the SPIRE maps which have not been masked because they are not in the GAMA catalogue. The stacked flux measured in this way is a reasonable estimate of the average background flux in the corresponding map, and is subtracted from our stacked fluxes prior to further analysis. The average background levels are $3.5\pm0.1$, $3.0\pm0.1$ and $4.2\pm0.2$\,mJy\,beam$^{-1}$ in the 250, 350 and 500\mum\ bands respectively.

{ Fluxes measured in the SPIRE maps are calibrated for a flat $\nu S_\nu$ spectrum ($S_\nu \propto \nu^{-1}$), whereas thermal dust emission longward of the SED peak will have a slope between $\nu^0$ and $\nu^2$ depending on how far along the Rayleigh-Jeans tail a given waveband is. The SPIRE Observers' Manual\footnote{Available from \url{http://herschel.esac.esa.int/Docs/SPIRE/html/spire_om.html}} provides colour corrections suitable for various SED slopes, including the $\nu^2$ slope appropriate for bands on the Rayleigh-Jeans tail. This is a suitable description of the SED in each of the SPIRE bands at low redshift, and we therefore modify fluxes by the colour corrections for this slope: 0.9417, 0.9498 and 0.9395 in the three bands respectively. At increasing redshifts, however, a cold SED can begin to turn over in the observed-frame 250\mum\ band. From inspection of single-component SEDs fitted to our stacks we estimate that the SPIRE points in most of our bins fall on the Rayleigh-Jeans tail, although at the highest redshifts slopes can be as flat as $\nu^0$ at 250\mum\ and $\nu^{1.5}$ at 350\mum. The corresponding colour corrections are 0.9888 and 0.9630 respectively. We tried applying these corrections to the highest redshift bins and found no discernable difference to any of our stacked results, hence our results are not dependent on the colour correction assumed.}

\subsection{Simulations}

The stacking procedure that we used was tested on simulated maps to ensure that we could accurately measure faint fluxes when stacking in confused maps with realistic noise and source density. As described in Appendix~\ref{app:flux}, we were able to accurately reproduce median fluxes and correct errors, although fluxes of individual sources could be under-/over-estimated if they were blended with a fainter/brighter neighbouring source.

In addition we simulated various distributions of fluxes to test that the median measured flux is unbiased and representative when stacking faint sources close to and below the noise and confusion limits.
The results of these simulations indicate that the median can be biased in the presence of noise { \citep[see also][]{White2007}}. We show details of the simulations in Appendix~\ref{app:sims}. In summary, we assume a realistic distribution of fluxes described by
$dN/dS\propto S^{-2}$, $S_\text{min}<S<S_\text{max}$; $R=\log_{10}(S_\text{max}/S_\text{min})=1.3$,
and for this we estimate the amount of bias in the measured median as a function of the true median flux, and correct our stacked fluxes for this bias. Correction factors are all in the range $0.6 - 1.0$, and the effect is greatest for low fluxes ($\lesssim 10$\,mJy). { If we consider the true median to be representative of the typical source in any bin, then relative to this, the bias in the measured median is always less than or equal to the `bias' in the mean resulting from extreme values (as we explain in Appendix~\ref{app:sims}).}
We tested the sensitivity of the results to assumptions about the flux distribution, and found that although the level of bias does depend on the limits and slope of the distribution, all of our measured trends remain significant and all conclusions remain valid for any reasonable choice of distribution. This is equally true if we make no correction to the measured median.

We also tested the correlations found in the data by simulating flux distributions with various dependencies on redshift, absolute magnitude and colour. We first made simulations in which fluxes were randomised with no built-in dependencies but with the same scatter as in the real data, and saw that stacked results were equal in every bin (as expected). 
We tried simulations in which fluxes varied with redshift as a non-evolving LF (simply varying as the square of the luminosity distance, modified by the $k$--correction), with realistic scatter; and also as an evolving LF (log flux increasing linearly with redshift at a realistic rate), also with scatter.
We also allowed flux to vary linearly with optical colour and logarithmically with \Mr, again including realistic scatter. In all simulations we found that we were able to recover the input trends by stacking.

Finally we simulated fluxes in the three SPIRE bands to produce a randomised distribution of submm colours ($S_{250}/S_{350}$ etc) with scatter similar to that in the data but with no correlations built in. Results showed that no artificial correlations were introduced by stacking. 

These results indicate that the correlations detected in the real data (described in the following chapter) should be true representations of the intrinsic distributions in the galaxy population, and are not artifacts created by the stacking procedure.

\subsection{Errors on SPIRE fluxes}
\label{sec:errors}
{Errors on stacked fluxes are calculated in two different ways. Firstly we estimate the instrumental and confusion noise on each source and propagate the errors through the stacking procedure. We estimate the instrumental noise by convolving the variance map at the source position with the same PRF used for the flux measurement. To this we add in quadrature a confusion noise term, as in \citet{Rigby2010}. Since fluxes are measured by filtering the map with a kernel based on the PSF (see Appendix~\ref{app:flux}), we need to use the confusion noise as measured in the PSF-filtered map. \Citet{Pascale2010} measured confusion noises of 5.3, 6.4 and 6.7~mJy per beam in the PSF-filtered \hatlas\ Science Demonstration Phase (SDP) maps. We estimate that the confusion noise is at a similar level in our maps after PSF-filtering, by comparing the total variance in random stacks on the background to the average instrumental noise described above. Hence we combine these values of confusion noise with the measured instrumental noise of each source. In each stack the mean of these measured noises, divided by the square root of the number of objects stacked, and combined with the error on background subtraction, gives the `measurement error' ($\sigma_\text{N}$) in Table~\ref{tab:flux}.
}

We also use a second method to calculate errors on stacked fluxes (as well as other stacked quantities), which is the $1\sigma$ error on the median described by \citet{Gott2001}, as used in \citet{Dunne2009} and \citet{Bourne2011}. This is calculated from the distribution of {flux} values in the bin and so automatically takes into account measurement errors as well as genuine variation within the bin resulting from the underlying population from which it is drawn. 
{Briefly, the method sorts the $N$ values in a bin, assigning each a unique rank $r$ between 0 and 1. In the limit of large $N$, the expectation value of the \emph{true} median of the population sampled is $\langle r \rangle = 0.5$, and its standard deviation is $\langle r^2 - \langle r \rangle^2 \rangle ^{1/2} = 1/\sqrt{4N}$. If the measurement at rank $r$ is $m(r)$, then the median measurement is $m(0.5)$, which gives the expectation value of the true median of the population sampled. The error on this expectation value is then
\begin{equation}
\dfrac{m(0.5 + 1/\sqrt{4N}) - m(0.5 - 1/\sqrt{4N})}{2}.
\end{equation}
This formula gives the `statistical error' ($\sigma_\text{S}$) in Table~\ref{tab:flux}. { These values are typically three to four times larger than the measurement error $\sigma_\text{N}$, indicating that the uncertainty resulting from the spread of intrinsic fluxes in a bin is greater than the combined noise in the map at all the positions in the stack.
}}

\section{Results}
\label{sec:results}

\subsection{Stacked fluxes}
\label{sec:fluxevo}
The results of stacking SPIRE fluxes as a function of $g-r$ colour, redshift and absolute magnitude \Mr\ are given in Table~\ref{tab:flux}. Secure detections are obtained at 250 and 350\mum\ in most bins, although many bins { have low signal-to-noise} at 500\mum. Note that the signal-to-noise ratios in the Table are based on the measurement error reduced by $\sqrt{N}$ (i.e. $\sigma_\text{N}$), since this strictly represents the noise level (instrumental plus confusion) which we compare our detections against. When talking about errors in all subsequent analysis we will refer to the statistical uncertainty on the median ($\sigma_\text{S}$) because this takes into account both instrumental noise and the distribution of values in the bin, both of which are contributions towards the uncertainty on the median stacked result.

{Table~\ref{tab:flux} also contains the results of Kolmogorov--Smirnov (KS) tests which were carried out on each stack to test the certainty that the stacked flux represents a signal from a sample of real sources and is not simply due to noise. This was done by comparing the distribution of measured fluxes in each bin with the distribution of fluxes in the background sample for the same SPIRE band. These background samples (described in Section~\ref{sec:stacking}) were placed at a set of random positions in the map, after masking around the positions of input sources. If any of our stacks did not contain a significant signal from real sources then the KS test would return a high probability that the distribution of fluxes is drawn from the same population as the background sample. The great majority of our bins were found to have an extremely small KS probability, meaning that we can be confident that the signals measured are real. The highest probability is 0.04, for a 500\mum\ stack in the highest-redshift, red-colour bin. 
A small sample of the bins are explored in more detail in Appendix~\ref{app:figs}, where we show stacked postage-stamp images and histograms on which the KS test was carried out.}
 
\begin{figure*}
 \includegraphics[width=\textwidth]{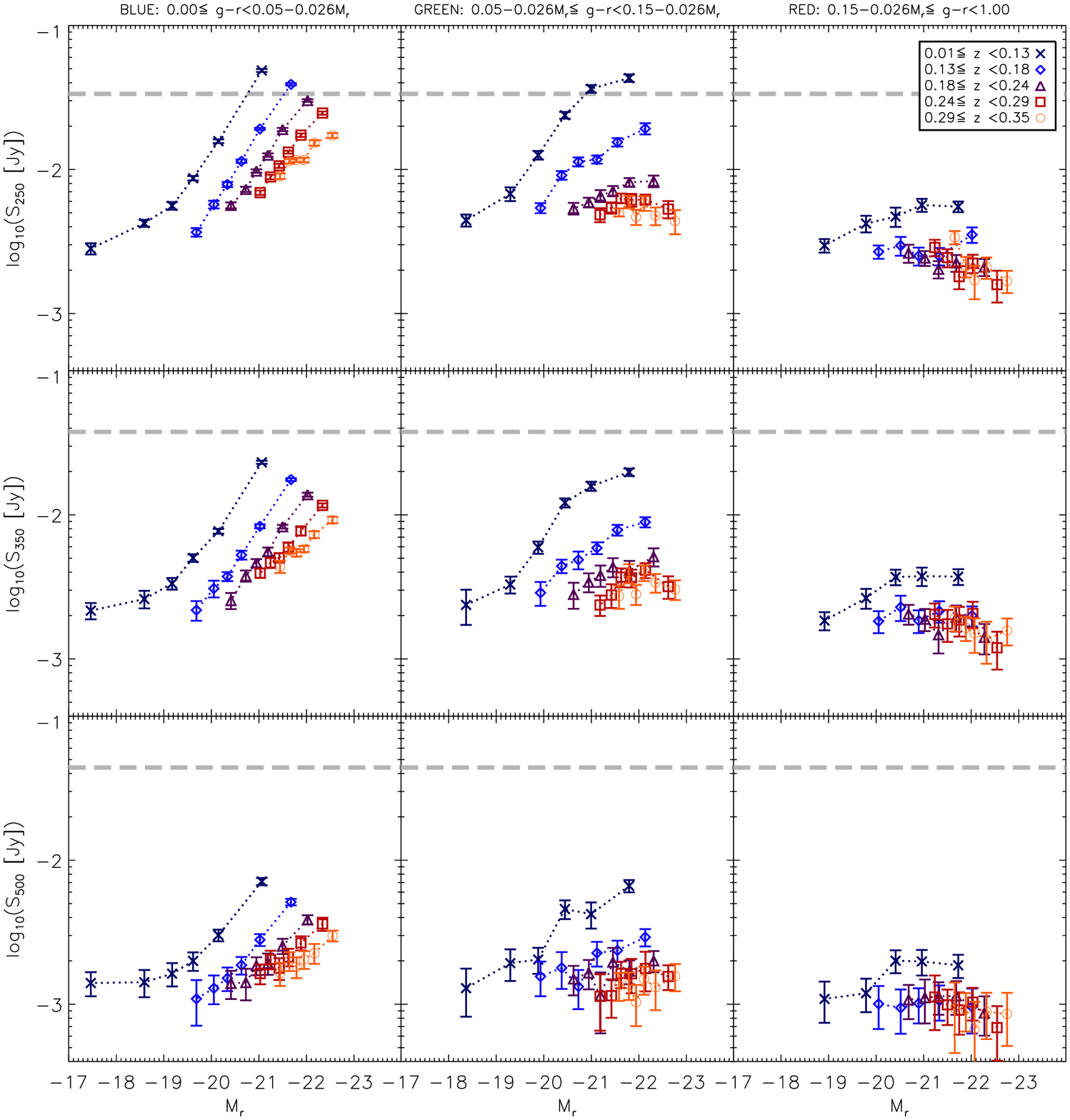}
\caption{Stacked SPIRE fluxes (not $k$--corrected) as a function of $g-r$ colour, redshift and absolute magnitude \Mr. Top: 250\mum, middle: 350\mum, bottom: 500\mum. Galaxies are binned by optical colour from blue to red (shown in panels from left to right) and by redshift (shown by plot symbols) and stacked fluxes in each bin are plotted against \Mr. Error bars are the statistical $1\sigma$ errors in the bins as described in Section~\ref{sec:errors}, and also include errors due to background subtraction. The horizontal dashed lines at 33.5, 39.5 and 44.0~mJy in 250, 350 and 500\mum, represent the $5\sigma$ point-source detection limits as measured in the PSF-convolved Phase 1 maps.}
\label{fig:flux}
\end{figure*}

Stacked fluxes at 250, 350 and 500\mum\ in the observed frame are plotted in Fig.~\ref{fig:flux}, showing the dependence on \Mr\ and $g-r$ at different redshifts. The majority of the bins have stacked fluxes well below the $5\sigma$ point-source detection limits shown as horizontal lines on the Figure. In all three bands there is a striking difference between the submm fluxes of blue, green and red galaxies, and a strong correlation with \Mr\ in the low-redshift bins of blue and green galaxies. These trends unsurprisingly indicate that the red galaxies tend to be passive and have lower dust {luminosities} than blue, and are generally well below the detection threshold in all SPIRE bands. They also show that submm flux varies little across the range of \Mr\ in red galaxies, while in blue galaxies it correlates strongly with \Mr\ such that only the most luminous $r$-band sources have fluxes above the 250\mum\ detection limit -- {a point noted by \citet{Dariush2011} and \citetalias{Dunne2011}}. The variation with redshift is also very different between the three optical classes, with the fluxes of blue galaxies diminishing with redshift more rapidly than those of red galaxies. {Fluxes in the green bin initially fall more rapidly with increasing redshift than those in the blue, but at $z>0.18$ they resemble those in the red bin and evolve very little}. This is potentially due to a change in the nature of the galaxies classified as green at different redshifts, {which is unsurprising since this bin contains a mixture of different galaxy types in the overlapping region between the blue cloud and red sequence. It is likely that the relative fractions of passive, relatively dust-free systems and dusty star-forming systems in this bin will change with redshift, as the star-formation density of the Universe evolves over this redshift range \citep[etc]{Lilly1996,Madau1996}.}
The evolutionary trends discussed in this Section can be better explored by deriving submm luminosities, which first requires a model for $k$--correcting the fluxes, {as we will { discuss} in Section~\ref{sec:firkcorrs}.}

At this point it is worth considering some potential sources of bias in different bins in case they might impact on the apparent trends. For example, it is reasonable to expect that certain classes of galaxy are more likely than others to inhabit dense environments: in particular redder galaxies, and more massive galaxies, are known to be more clustered \citep[e.g.][]{Zehavi2010}. While we do account for blending in the flux measurements (see Appendix~\ref{app:flux}) this is limited to the blending of sources within the catalogue. As we move to higher redshifts the catalogue becomes more incomplete, and it becomes more likely that the clustered galaxies will be blended with some unseen neighbour. We make no correction for this effect, but we expect the contamination to be small for the following reasons: The input sample is complete down to below the knee in the optical LF at $z<0.3$ (see next section); we therefore account for the blending with most of the galaxies in the same redshift range. Contaminating flux would have to come from relatively small galaxies which are not likely to contribute a large amount of flux. Moreover, the blending corrections are on average very small in comparison with the trends that we observe (see Appendix~\ref{app:flux} Table~\ref{tab:clustercorr}), so a small additional blending correction should not significantly alter our conclusions.

\begin{sidewaystable*}\scriptsize
\centering
\caption{Results of stacking in bins of $g-r$ colour, redshift, and absolute magnitude (\Mr). Columns are as follows: colour bin $C$ = B (blue), G (green), R (red); median redshift $z$ in bin (approximate $z$ bin boundaries are 0.01, 0.12, 0.17, 0.22, 0.28, 0.35); median \Mr; count $N$ in the bin. The following columns for each of the three SPIRE bands: background-subtracted flux $S$ in mJy; signal-to-noise ratio $S/\sigma_\text{N}$; measurement error $\sigma_\text{N}$ (mJy) computed from the mean variance of positions in the stack, divided by $\sqrt{N}$, and including the error on background subtraction; statistical error on the median flux ($\sigma_\text{S}$, mJy) following \citet{Gott2001}; KS probability that the distribution of fluxes in each bin is the same as that at a set of random positions; median $k$--correction $K'=K(z)/(1+z)$ in the bin.}
\begin{tabular}{ c r l r l l l l l l l l l l l l l l l l l l }
\hline
  \multicolumn{1}{c}{$C$} &
  \multicolumn{1}{c}{$z$} &     
  \multicolumn{1}{c}{$M_r$} &
  \multicolumn{1}{c}{$N$} &
  \multicolumn{1}{c}{$S_{250}$} &
  \multicolumn{1}{c}{SNR$_{250}$} &
  \multicolumn{1}{c}{$\sigma_{\text{N},250}$} &
  \multicolumn{1}{c}{$\sigma_{\text{S},250}$} &
  \multicolumn{1}{c}{KS$_{250}$} &
  \multicolumn{1}{c}{$S_{350}$} &
  \multicolumn{1}{c}{SNR$_{350}$} &
  \multicolumn{1}{c}{$\sigma_{\text{N},350}$} &
  \multicolumn{1}{c}{$\sigma_{\text{S},350}$} &
  \multicolumn{1}{c}{KS$_{350}$} &
  \multicolumn{1}{c}{$S_{500}$} &
  \multicolumn{1}{c}{SNR$_{500}$} &
  \multicolumn{1}{c}{$\sigma_{\text{N},500}$} &
  \multicolumn{1}{c}{$\sigma_{\text{S},500}$} &
  \multicolumn{1}{c}{KS$_{500}$} &
  \multicolumn{1}{c}{$K'_{250}$} &
  \multicolumn{1}{c}{$K'_{350}$} &
  \multicolumn{1}{c}{$K'_{500}$} \\
\hline
  B & 0.11 & -21.1 & 1567 & 48.69 &  286.4  & 0.17  & 1.15 &     0 & 23.15  &   178.1 &   0.13  & 0.58 &     0 &  7.11 &  59.3   &    0.12  & 0.41 &     0 &     0.77  &    0.71  &    0.67 \\
  B & 0.10 & -20.1 & 1568 & 15.72 &  104.8  & 0.15  & 0.46 &     0 &  7.70  &    70.0 &   0.11  & 0.30 &     0 &  3.01 &  30.1   &    0.10  & 0.28 &     0 &     0.77  &    0.71  &    0.68 \\
  B & 0.10 & -19.6 & 1567 & 8.69  &   62.1  & 0.14   & 0.31 &     0 & 5.02   &    45.6 & 0.11    & 0.31 &     0 & 2.00  &  20.0   &  0.10    & 0.29 & 3E-39 &    0.77  &    0.72  &    0.68 \\
  B & 0.09 & -19.2 & 1568 & 5.60  &   43.1  & 0.13   & 0.34 &     0 & 3.33   &    33.3 & 0.10    & 0.31 &     0 & 1.63  &  16.3   &  0.10    & 0.30 & 5E-21 &    0.79  &    0.74  &    0.70 \\
  B & 0.08 & -18.6 & 1567 & 4.24  &   35.3  & 0.12   & 0.24 &     0 & 2.60   &    26.0 & 0.10    & 0.36 & 9E-43 & 1.42  &  14.2   &  0.10    & 0.30 & 1E-17 &    0.82  &    0.78  &    0.75 \\
  B & 0.04 & -17.5 & 1567 & 2.82  &   23.5  & 0.12   & 0.25 &     0 & 2.16   &    21.6 & 0.10    & 0.28 & 1E-26 & 1.40  &  14.0   &  0.10    & 0.27 & 7E-14 &    0.90  &    0.87  &    0.85 \\
[1ex]
  B & 0.15 & -21.7 & 1300 & 39.05 &  216.9 &  0.18 & 0.73 &     0 & 17.58  &   125.6  &   0.14  & 0.39 &     0 &  5.12 &  42.7   &    0.12  & 0.27 &     0 &     0.70  &    0.62  &    0.57 \\
  B & 0.15 & -21.0 & 1300 & 19.14 &  112.5 &  0.17 & 0.32 &     0 &  8.34  &    64.2  &   0.13  & 0.28 &     0 &  2.80 &  25.4   &    0.11  & 0.27 &     0 &     0.69  &    0.62  &    0.57 \\
  B & 0.15 & -20.6 & 1300 & 11.42 &   71.4 &  0.16 & 0.31 &     0 &  5.26  &    43.8  &   0.12  & 0.37 &     0 &  1.87 &  17.0   &    0.11  & 0.26 & 1E-25 &     0.70  &    0.62  &    0.57 \\
  B & 0.15 & -20.3 & 1300 & 7.88  &   52.5  & 0.15   & 0.32 &     0 & 3.73   &   33.9  & 0.11    & 0.27 &     0 & 1.52  & 13.8   &   0.11   & 0.28 & 1E-14 &     0.70  &    0.62  &    0.57 \\
  B & 0.15 & -20.1 & 1300 & 5.71  &   40.8 &  0.14  & 0.35 &     0 & 3.08   &    28.0  & 0.11    & 0.41 & 6E-45 & 1.29  &  11.7  &   0.11   & 0.29 & 7E-14 &     0.70  &    0.63  &    0.58 \\
  B & 0.14 & -19.7 & 1299 & 3.67  &   28.2  & 0.13   & 0.25 &     0 & 2.18   &   19.8  & 0.11    & 0.34 & 8E-27 & 1.09  &  9.9   &   0.11   & 0.38 & 2E-06 &     0.72  &    0.65  &    0.60 \\
[1ex]
  B & 0.21 & -22.0 & 1295 & 30.11 &  167.3 &  0.18 & 0.61 &     0 & 13.80  &   98.6   &   0.14  & 0.44 &     0 &  3.86 &   32.2  &    0.12  & 0.28 &     0 &     0.63  &    0.53  &    0.47 \\
  B & 0.21 & -21.5 & 1295 & 18.94 &  111.4 &  0.17 & 0.49 &     0 &  8.27  &   63.6   &   0.13  & 0.26 &     0 &  2.54 &   23.0  &    0.11  & 0.31 &     0 &     0.63  &    0.53  &    0.47 \\
  B & 0.21 & -21.2 & 1295 & 12.59 &   78.7 &  0.16 & 0.35 &     0 &  5.55  &   46.3   &   0.12  & 0.38 &     0 &  1.92 &    17.5 &    0.11  & 0.33 & 4E-22 &     0.63  &    0.53  &    0.48 \\
  B & 0.21 & -20.9 & 1295 & 9.75  &   65.0  & 0.15   & 0.27 &     0 & 4.63   &  38.6   & 0.12    & 0.30 &     0 & 1.85  &  16.8  &   0.11   & 0.27 & 6E-23 &     0.63  &    0.53  &    0.48 \\
  B & 0.20 & -20.7 & 1295 & 7.29  &   48.6  & 0.15   & 0.30 &     0 & 3.77   &  34.3   & 0.11    & 0.33 &     0 & 1.41  &  12.8  &   0.11   & 0.35 & 7E-14 &     0.63  &    0.54  &    0.49 \\
  B & 0.19 & -20.4 & 1294 & 5.64  &   40.3  & 0.14   & 0.27 &     0 & 2.54   &  23.1   & 0.11    & 0.33 & 3E-34 & 1.40  &  12.7  &   0.11   & 0.31 & 5E-12 &     0.65  &    0.56  &    0.50 \\
[1ex]
  B & 0.27 & -22.3 & 1323 & 24.65 &  145.0 &  0.17 & 0.64 &     0 & 11.63  &   89.5   &   0.13  & 0.39 &     0 &  3.61 &   32.8   &   0.11   & 0.38 &     0 &    0.57  &    0.46  &    0.40 \\
  B & 0.27 & -21.9 & 1324 & 17.31 &  108.3 &  0.16 & 0.50 &     0 &  7.72  &   64.3   &   0.12  & 0.52 &     0 &  2.66 &   24.2   &   0.11   & 0.30 & 4E-44 &    0.57  &    0.46  &    0.40 \\
  B & 0.27 & -21.6 & 1324 & 13.19 &   82.4  & 0.16  & 0.30 &     0 &  5.97  &  49.8   &   0.12  & 0.42 &     0 &  2.13 &  19.4    &   0.11   & 0.30 & 7E-33 &    0.57  &    0.46  &    0.40 \\
  B & 0.27 & -21.4 & 1324 & 10.59 &   70.6  & 0.15  & 0.34 &     0 &  5.03  &  41.9   &   0.12  & 0.37 &     0 &  1.78 &  16.2    &   0.11   & 0.31 & 1E-17 &    0.57  &    0.46  &    0.40 \\
  B & 0.26 & -21.2 & 1324 & 8.88  &   59.2  & 0.15   & 0.33 &     0 & 4.65   &  42.3   & 0.11    & 0.38 &     0 & 2.05  &  18.6   &  0.11    & 0.30 & 2E-27 &    0.57  &    0.46  &    0.40 \\
  B & 0.26 & -21.0 & 1323 & 6.90  &   49.3  & 0.14   & 0.24 &     0 & 3.94   &  35.8   & 0.11    & 0.29 &     0 & 1.63  &  14.8   &  0.11    & 0.26 & 1E-20 &    0.58  &    0.47  &    0.41 \\
[1ex]
  B & 0.32 & -22.6 & 1377 & 17.22 &  107.6 &  0.16 & 0.67 &     0 &  9.22  &   76.8   &   0.12  & 0.48 &     0 &  2.99 &  27.2    &   0.11   & 0.26 &     0 &    0.53  &    0.41  &    0.34 \\
  B & 0.32 & -22.2 & 1377 & 15.30 &   95.6  & 0.16  & 0.67 &     0 &  7.30  &  60.8   &   0.12  & 0.40 &     0 &  2.26 &  20.5    &   0.11   & 0.36 & 2E-34 &    0.53  &    0.41  &    0.34 \\
  B & 0.32 & -21.9 & 1377 & 11.64 &   77.6  & 0.15  & 0.43 &     0 &  5.80  &  48.3   &   0.12  & 0.29 &     0 &  2.05 &  18.6    &   0.11   & 0.29 & 2E-29 &    0.52  &    0.41  &    0.34 \\
  B & 0.32 & -21.8 & 1377 & 11.65 &   77.7  & 0.15  & 0.39 &     0 &  5.45  &  49.5   &   0.11  & 0.32 &     0 &  1.82 &  18.2    &   0.10   & 0.30 & 3E-23 &    0.53  &    0.41  &    0.34 \\
  B & 0.31 & -21.6 & 1377 & 11.41 &   76.1  & 0.15  & 0.41 &     0 &  5.66  &  51.5   &   0.11  & 0.28 &     0 &  2.00 &  18.2    &   0.11   & 0.30 & 2E-30 &    0.53  &    0.42  &    0.35 \\
  B & 0.30 & -21.4 & 1377 & 8.99  &   59.9  & 0.15   & 0.40 &     0 & 4.37   &  39.7   & 0.11    & 0.43 &     0 & 1.64  &  16.4    & 0.10     & 0.30 & 7E-17 &   0.54  &    0.43  &    0.36  \\
\hline
  G & 0.11 & -21.8 & 452  & 43.13 &  139.1 &  0.31 & 2.57 &     0 & 19.79  &   82.5   &   0.24  & 1.20 &     0 &  6.64 &  31.6     &  0.21    & 0.65 &     0  &  0.77  &    0.71  &    0.67  \\
  G & 0.10 & -21.0 & 452  & 36.34 &  121.1 &  0.30 & 2.18 &     0 & 15.82  &   65.9   &   0.24  & 1.16 &     0 &  4.22 &  21.1     &  0.20    & 0.87 & 1E-37  &  0.78  &    0.72  &    0.68  \\
  G & 0.10 & -20.4 & 452  & 23.80 &   82.1 &  0.29 & 1.30 &     0 & 12.13  &   55.1   &   0.22  & 0.88 &     0 &  4.58 &  22.9     &  0.20    & 0.68 & 8E-29  &  0.78  &    0.72  &    0.68  \\
  G & 0.10 & -19.9 & 452  & 12.58 &   46.6 &  0.27 & 0.87 &     0 &  5.95  &   29.8   &   0.20  & 0.58 &     0 &  2.04 &  11.3     &  0.18    & 0.42 & 1E-10  &  0.77  &    0.71  &    0.68  \\
  G & 0.10 & -19.3 & 452  & 6.78  &   27.1 &  0.25  & 0.74 &     0 & 3.28   &   17.3   & 0.19    & 0.44 & 7E-23 & 1.93  &  10.2    & 0.19     & 0.48 & 6E-08  &  0.78  &    0.72  &    0.69  \\
  G & 0.06 & -18.4 & 452  & 4.44  &   19.3 &  0.23  & 0.43 & 3E-37 & 2.37   &   13.2   & 0.18    & 0.65 & 4E-12 & 1.29  &  7.2     & 0.18     & 0.47 & 1E-04  &  0.86  &    0.82  &    0.79  \\
[1ex]
  G & 0.15 & -22.1 & 569  & 19.23 &  76.9  &  0.25 & 1.71 &     0 &  8.91  &   46.9   &   0.19  & 0.72 &     0 &  2.92 &  18.3    &   0.16   & 0.40 & 3E-24 &    0.70  &    0.62  &    0.57 \\
  G & 0.15 & -21.5 & 570  & 15.47 &  61.9  &  0.25 & 1.04 &     0 &  7.85  &   41.3   &   0.19  & 0.67 &     0 &  2.36 &  14.8    &   0.16   & 0.40 & 7E-18 &    0.70  &    0.62  &    0.57 \\
  G & 0.15 & -21.1 & 570  & 11.74 &  48.9  &  0.24 & 0.76 &     0 &  5.89  &   32.7   &   0.18  & 0.56 &     0 &  2.27 &  13.4    &   0.17   & 0.44 & 1E-14 &    0.70  &    0.62  &    0.57 \\
  G & 0.16 & -20.7 & 570  & 11.31 &  49.2  &  0.23 & 0.82 &     0 &  4.87  &   28.6   &   0.17  & 0.69 & 4E-41 &  1.33 &   8.3     &  0.16    & 0.40 & 3E-09 &   0.69  &    0.61  &    0.57  \\
  G & 0.15 & -20.4 & 570  & 9.12  &  39.7  &  0.23  & 0.63 &     0 & 4.41   &   25.9   & 0.17    & 0.47 & 2E-34 & 1.79  &  11.1    & 0.16     & 0.51 & 1E-07 &   0.70  &    0.62  &    0.57  \\
  G & 0.14 & -19.9 & 569  & 5.40  &  25.7  &  0.21  & 0.43 &     0 & 2.88   &   18.0   & 0.16    & 0.54 & 3E-20 & 1.56  &   9.8    & 0.16     & 0.42 & 1E-07 &   0.72  &    0.64  &    0.60  \\
\hline\end{tabular}

\label{tab:flux}
\end{sidewaystable*}
\begin{sidewaystable*}\scriptsize
\centering
\caption*{Table~\ref{tab:flux} continued}
\begin{tabular}{ c r l r l l l l l l l l l l l l l l l l l l }
\hline
  \multicolumn{1}{c}{$C$} &
  \multicolumn{1}{c}{$z$} &     
  \multicolumn{1}{c}{$M_r$} &
  \multicolumn{1}{c}{$N$} &
  \multicolumn{1}{c}{$S_{250}$} &
  \multicolumn{1}{c}{SNR$_{250}$} &
  \multicolumn{1}{c}{$\sigma_{\text{N},250}$} &
  \multicolumn{1}{c}{$\sigma_{\text{S},250}$} &
  \multicolumn{1}{c}{KS$_{250}$} &
  \multicolumn{1}{c}{$S_{350}$} &
  \multicolumn{1}{c}{SNR$_{350}$} &
  \multicolumn{1}{c}{$\sigma_{\text{N},350}$} &
  \multicolumn{1}{c}{$\sigma_{\text{S},350}$} &
  \multicolumn{1}{c}{KS$_{350}$} &
  \multicolumn{1}{c}{$S_{500}$} &
  \multicolumn{1}{c}{SNR$_{500}$} &
  \multicolumn{1}{c}{$\sigma_{\text{N},500}$} &
  \multicolumn{1}{c}{$\sigma_{\text{S},500}$} &
  \multicolumn{1}{c}{KS$_{500}$} &
  \multicolumn{1}{c}{$K'_{250}$} &
  \multicolumn{1}{c}{$K'_{350}$} &
  \multicolumn{1}{c}{$K'_{500}$} \\
\hline
  G & 0.21 & -22.3 & 619  & 8.30  &  39.5  &  0.21  & 0.77 &     0 & 5.12   &   32.0   & 0.16    & 0.74 & 3E-38 & 2.00  &  13.3    & 0.15     & 0.34 & 2E-17  &  0.63  &    0.53  &    0.48  \\
  G & 0.21 & -21.8 & 620  & 8.17  &  38.9  &  0.21  & 0.55 &     0 & 4.11   &   25.7   & 0.16    & 0.68 & 2E-36 & 1.70  &  11.3    & 0.15     & 0.32 & 3E-10  &  0.63  &    0.54  &    0.48  \\
  G & 0.21 & -21.5 & 620  & 7.09  &  33.8  &  0.21  & 0.58 &     0 & 4.34   &   27.1   & 0.16    & 0.66 & 3E-35 & 1.96  &  12.3    & 0.16     & 0.50 & 8E-12  &  0.63  &    0.53  &    0.48 \\
  G & 0.21 & -21.2 & 619  & 6.56  &  31.2  &  0.21  & 0.62 &     0 & 3.80   &   23.8   & 0.16    & 0.64 & 2E-29 & 1.14  &   7.6    & 0.15     & 0.52 & 2E-05  &  0.63  &    0.53  &    0.48 \\
  G & 0.21 & -20.9 & 620  & 5.90  &  28.1  &  0.21  & 0.46 &     0 & 3.41   &   21.3   & 0.16    & 0.51 & 6E-27 & 1.65  &  10.3    & 0.16     & 0.38 & 8E-11  &  0.63  &    0.54  &    0.48  \\
  G & 0.20 & -20.6 & 619  & 5.36  &  26.8  &  0.20  & 0.50 &     0 & 2.80   &   17.5   & 0.16    & 0.58 & 8E-20 & 1.50  &   9.4    & 0.16     & 0.35 & 4E-07  &  0.64  &    0.55  &    0.50  \\
[1ex]
  G & 0.27 & -22.6 & 662  & 5.31  &  27.9  &  0.19  & 0.73 &     0 & 3.18   &   21.2   & 0.15    & 0.57 & 2E-23 & 1.56  &  11.1    & 0.14     & 0.31 & 8E-09  &  0.57  &    0.46  &    0.40  \\
  G & 0.27 & -22.1 & 663  & 6.19  &  31.0  &  0.20  & 0.50 &     0 & 4.13   &   25.8   & 0.16    & 0.49 & 5E-34 & 1.77  &  11.8    & 0.15     & 0.54 & 1E-10  &  0.57  &    0.46  &    0.40  \\
  G & 0.27 & -21.8 & 662  & 6.14  &  30.7  &  0.20  & 0.63 &     0 & 3.65   &   24.3   & 0.15    & 0.52 & 6E-26 & 1.63  &  10.9    & 0.15     & 0.44 & 4E-09  &  0.57  &    0.46  &    0.40  \\
  G & 0.27 & -21.6 & 663  & 6.31  &  31.6  &  0.20  & 0.49 &     0 & 3.69   &   23.1   & 0.16    & 0.56 & 2E-31 & 1.63  &  10.9    & 0.15     & 0.34 & 8E-09  &  0.57  &    0.46  &    0.40  \\
  G & 0.27 & -21.4 & 662  & 5.42  &  27.1  &  0.20  & 0.49 &     0 & 2.78   &   18.5   & 0.15    & 0.52 & 4E-17 & 1.15  &   7.7    & 0.15     & 0.34 & 3E-04  &  0.57  &    0.46  &    0.40  \\
  G & 0.26 & -21.2 & 662  & 4.85  &  25.5  &  0.19  & 0.54 &     0 & 2.37   &   15.8   & 0.15    & 0.38 & 2E-18 & 1.14  &   7.6    & 0.15     & 0.48 & 2E-05  &  0.58  &    0.47  &    0.41  \\
[1ex]
  G & 0.32 & -22.8 & 653  & 4.39  &  23.1  &  0.19  & 0.84 &     0 & 3.03   &   20.2   & 0.15    & 0.47 & 7E-24 & 1.57  &  11.2   &  0.14    & 0.34 & 9E-10    &  0.53 &     0.41 &     0.35  \\
  G & 0.32 & -22.4 & 653  & 4.78  &  25.2  &  0.19  & 0.67 &     0 & 3.38   &   22.5   & 0.15    & 0.50 & 9E-25 & 1.31  &   8.7    & 0.15     & 0.40 & 8E-08  &  0.53  &    0.41  &    0.35  \\
  G & 0.32 & -22.1 & 653  & 5.84  &  29.2  &  0.20  & 0.68 &     0 & 3.96   &   24.8   & 0.16    & 0.50 & 1E-32 & 1.49  &   9.9    & 0.15     & 0.33 & 1E-09  &  0.53  &    0.41  &    0.34  \\
  G & 0.32 & -21.9 & 653  & 4.68  &  24.6  &  0.19  & 0.57 &     0 & 2.81   &   18.7   & 0.15    & 0.45 & 5E-22 & 1.03  &   6.9    & 0.15     & 0.33 & 4E-04  &  0.53  &    0.41  &    0.34  \\
  G & 0.31 & -21.8 & 653  & 5.97  &  29.9  &  0.20  & 0.59 &     0 & 3.95   &   24.7   & 0.16    & 0.60 & 8E-29 & 1.37  &   9.1    & 0.15     & 0.30 & 6E-07  &  0.53  &    0.41  &    0.35  \\
  G & 0.30 & -21.6 & 653  & 5.04  &  26.5  &  0.19  & 0.32 &     0 & 2.73   &   18.2   & 0.15    & 0.50 & 4E-20 & 1.35  &   9.0    & 0.15     & 0.30 & 6E-07  &  0.54  &    0.43  &    0.36  \\
\hline                                                                                                                                                                             
  R & 0.10 & -21.7 & 991  & 5.52  &  34.5  &  0.16  & 0.47 &     0 & 3.72   &   31.0   & 0.12    & 0.47 & 2E-43 & 1.86  &  15.5   &  0.12    & 0.34 & 3E-24 &    0.77  &    0.71  &    0.67 \\
  R & 0.10 & -21.0 & 992  & 5.66  &  35.4  &  0.16  & 0.59 &     0 & 3.75   &   28.8   & 0.13    & 0.55 & 5E-43 & 1.98  &  16.5   &  0.12    & 0.40 & 2E-17 &    0.77  &    0.71  &    0.68 \\
  R & 0.10 & -20.4 & 991  & 4.71  &  29.4  &  0.16  & 0.72 &     0 & 3.70   &   28.5   & 0.13    & 0.46 & 3E-39 & 2.00  &  16.7   &  0.12    & 0.36 & 9E-17 &    0.77  &    0.72  &    0.68 \\
  R & 0.10 & -19.8 & 992  & 4.21  &  28.1  &  0.15  & 0.56 &     0 & 2.65   &   22.1   & 0.12    & 0.42 & 2E-23 & 1.19  &   9.9    & 0.12     & 0.31 & 4E-07  &  0.78  &    0.72  &    0.68 \\
  R & 0.08 & -18.9 & 991  & 2.97  &  21.2  &  0.14  & 0.32 & 6E-45 & 1.84   &   15.3   & 0.12    & 0.27 & 1E-16 & 1.09  &   9.1    & 0.12     & 0.35 & 2E-05  &  0.82  &    0.78  &    0.74 \\
[1ex]                                                                                                                                                                                   
  R & 0.15 & -22.0 & 1177 & 3.52  &  27.1  &  0.13  & 0.44 &     0 & 2.00   &   20.0   & 0.10    & 0.32 & 3E-22 & 0.96  &  9.6    &  0.10    & 0.33 & 2E-11 &    0.70  &    0.62  &    0.57 \\
  R & 0.15 & -21.3 & 1177 & 2.50  &  19.2  &  0.13  & 0.35 & 2E-42 & 2.15   &   19.5   & 0.11    & 0.37 & 2E-23 & 1.06  &  9.6    &  0.11    & 0.29 & 4E-12 &    0.70  &    0.62  &    0.57 \\
  R & 0.15 & -20.9 & 1177 & 2.51  &  19.3  &  0.13  & 0.36 & 3E-38 & 1.84   &   16.7   & 0.11    & 0.33 & 3E-15 & 1.01  &  9.2    &  0.11    & 0.28 & 6E-08 &    0.70  &    0.62  &    0.57 \\
  R & 0.15 & -20.5 & 1177 & 2.96  &  22.8  &  0.13  & 0.44 &     0 & 2.29   &   20.8   & 0.11    & 0.46 & 9E-21 & 0.94  &  8.5    &  0.11    & 0.32 & 9E-06 &    0.70  &    0.62  &    0.57 \\
  R & 0.14 & -20.1 & 1177 & 2.68  &  20.6  &  0.13  & 0.28 & 5E-44 & 1.83   &   16.6   & 0.11    & 0.32 & 2E-16 & 1.01  &  9.2    &  0.11    & 0.33 & 4E-06 &    0.71  &    0.64  &    0.59 \\
[1ex]                                                                                                                                                                            
  R & 0.21 & -22.3 & 1110 & 2.11  &  16.2  &  0.13  & 0.29 & 3E-28 & 1.41   &   14.1   & 0.10    & 0.34 & 3E-13 & 0.87  &   7.9    & 0.11     & 0.27 & 7E-10 &   0.63  &    0.54  &    0.48  \\
  R & 0.21 & -21.7 & 1111 & 2.25  &  17.3  &  0.13  & 0.30 & 9E-32 & 1.94   &   17.6   & 0.11    & 0.40 & 5E-17 & 1.14  &  10.4    & 0.11     & 0.30 & 1E-07 &   0.63  &    0.54  &    0.48  \\
  R & 0.21 & -21.3 & 1110 & 2.03  &  15.6  &  0.13  & 0.28 & 8E-27 & 1.47   &   13.4   & 0.11    & 0.38 & 3E-11 & 1.13  &  10.3    & 0.11     & 0.29 & 7E-06 &   0.63  &    0.54  &    0.48  \\
  R & 0.20 & -21.0 & 1111 & 2.43  &  18.7  &  0.13  & 0.29 & 9E-35 & 1.89   &   17.2   & 0.11    & 0.33 & 1E-16 & 1.11  &  10.1    & 0.11     & 0.38 & 2E-06 &   0.63  &    0.54  &    0.48  \\
  R & 0.19 & -20.7 & 1110 & 2.63  &  18.8  &  0.14  & 0.38 & 5E-42 & 2.05   &   18.6   & 0.11    & 0.32 & 4E-19 & 1.08  &   9.8    & 0.11     & 0.29 & 4E-06 &   0.64  &    0.55  &    0.50  \\
[1ex]                                                                                                                                                                              
  R & 0.27 & -22.5 & 1002 & 1.59  &  12.2  &  0.13  & 0.40 & 3E-15 & 1.19   &   10.8   & 0.11    & 0.35 & 1E-07 & 0.69  &  6.3     & 0.11     & 0.28 & 5E-05 &   0.57  &    0.46  &    0.40 \\
  R & 0.27 & -22.0 & 1002 & 2.23  &  15.9  &  0.14  & 0.33 & 6E-28 & 2.07   &   17.3   & 0.12    & 0.42 & 4E-15 & 1.04  &  8.7     & 0.12     & 0.26 & 4E-07 &   0.57  &    0.46  &    0.40 \\
  R & 0.27 & -21.7 & 1003 & 1.80  &  12.9  &  0.14  & 0.33 & 2E-20 & 1.86   &   16.9   & 0.11    & 0.43 & 8E-14 & 0.91  &  7.6     & 0.12     & 0.30 & 2E-04 &   0.57  &    0.46  &    0.40 \\
  R & 0.27 & -21.5 & 1002 & 2.44  &  17.4  &  0.14  & 0.35 & 2E-31 & 1.75   &   14.6   & 0.12    & 0.44 & 2E-12 & 0.99  &  8.3     & 0.12     & 0.27 & 2E-06 &   0.57  &    0.46  &    0.40 \\
  R & 0.26 & -21.2 & 1002 & 2.87  &  20.5  &  0.14  & 0.38 & 6E-42 & 2.04   &   17.0   & 0.12    & 0.39 & 6E-17 & 1.12  &  9.3     & 0.12     & 0.46 & 1E-04 &   0.58  &    0.47  &    0.41 \\
[1ex]                                                                                                                                                                               
  R & 0.32 & -22.8 & 903  & 1.68  &  12.0  &  0.14  & 0.30 & 1E-16 & 1.57   &   13.1   & 0.12    & 0.33 & 9E-10 & 0.86  &  7.2     & 0.12     & 0.34 & 2E-05 &   0.53  &    0.41  &   0.34  \\
  R & 0.32 & -22.3 & 904  & 2.09  &  14.9  &  0.14  & 0.36 & 2E-22 & 1.37   &   11.4   & 0.12    & 0.44 & 1E-10 & 0.89  &  7.4     & 0.12     & 0.31 & 6E-05 &   0.53  &    0.41  &   0.34  \\
  R & 0.32 & -22.1 & 903  & 1.70  &  12.1  &  0.14  & 0.45 & 1E-18 & 1.50   &   12.5   & 0.12    & 0.40 & 6E-09 & 0.67  &  5.2     & 0.13     & 0.37 & 4E-02 &   0.52  &    0.41  &   0.34  \\
  R & 0.32 & -21.9 & 904  & 2.12  &  15.1  &  0.14  & 0.35 & 2E-26 & 1.64   &   13.7   & 0.12    & 0.31 & 4E-11 & 0.92  &  7.7     & 0.12     & 0.27 & 3E-04 &   0.53  &    0.41  &   0.35  \\
  R & 0.30 & -21.7 & 903  & 3.38  &  22.5  &  0.15  & 0.36 & 8E-45 & 1.93   &   16.1   & 0.12    & 0.38 & 3E-13 & 0.94  &  7.8     & 0.12     & 0.48 & 1E-03 &   0.54  &    0.42  &   0.36  \\
\hline\end{tabular}

\end{sidewaystable*}

\subsection{Contamination from lensing}
\label{sec:lens}
{There is a risk} that the submm fluxes of some galaxies in the sample are contaminated by flux from lensed background sources, via galaxy-galaxy lensing. This is especially likely in a submm survey as a result of the negative $k$--correction and steep evolution of the LF \citep{Blain1996,Negrello2007}. These factors conspire to make submm sources detectable up to very high redshifts, therefore providing an increased probability for one or more foreground galaxies to intrude in the line of sight and magnify the flux via strong gravitational lensing. The strong potential for detecting lensed high redshift sources in \hatlas\ was conclusively demonstrated by \citet{Negrello2010}. 
In this study we target low redshift sources selected in the optical, but our sample will inevitably include some of the foreground lenses whose apparent fluxes are likely to be boosted by flux from the lensed background sources.
The flux magnification is likely to be greatest for massive spheroidal lenses, as a result of their mass distribution \citep{Negrello2010}. This could pose a problem for our red bins, where spheroids will be mostly concentrated. To make matters worse, measured fluxes are lowest in our red bins which means that even a small lensing contamination of order 1\,mJy could significantly boost the flux.

{ We can make an estimate of the lensing contribution to stacked fluxes by considering the predicted number counts of lenses from \citet{Lapi2011}, which are based on the amplification distribution of strong lenses (amplification factors $\geq 2$) from \citet{Negrello2007}. Integrating these counts gives a total of 470 lensed submm sources per square degree, and integrating their fluxes per square degree gives the total surface brightness of lensed sources shown in the first line of Table~\ref{tab:lens}. However, the counts are not broken down by redshift, and only those at $z<0.35$ will contribute to our stacks. 
It is not trivial to predict what fraction of strong lenses are in this redshift range, but recent results from \hatlas\ can provide us with the best estimate that is currently possible. \Citet{Gonzalez-Nuevo2011} created a sample of 64 candidate strong lenses from the \hatlas\ SDP by selecting sources with red SPIRE colours which have no reliable SDSS IDs, or have SDSS IDs with redshifts inconsistent with the submm SED. After matching to NIR sources in the VISTA Kilo-degree INfrared Galaxy survey (VIKING; Sutherland,~W. et~al. in preparation), they reduced this sample to 33 candidates with photometric redshifts for both the lens (using the NIR photometry) and source (using SPIRE and PACS photometry).  This sample, the \hatlas\ Lensed Objects Selection (HALOS), is unique in being selected in the submm, enabling the selection of candidate lenses over a much larger redshift range than other lens samples to date (their lenses had photometric redshifts $\lesssim 1.8$, while other surveys were confined to $z<1$). HALOS therefore provides the best observational measurement of the lens redshift distribution.

Seven of the 33 HALOS candidates have lens redshifts $<0.35$. \citeauthor{Gonzalez-Nuevo2011} removed two of these from their final sample because the lenses were at $z<0.2$, and they considered lenses at such low redshifts to have low probability both on theoretical grounds and on the evidence of previous surveys \citep{Browne2003, Oguri2006, Faure2008}. However, we do not want to risk underestimating the number of low redshift lenses, so we conservatively include those two in our analysis. The fraction of lenses at $z<0.35$ is therefore $7/33 = 21^{+9}_{-5}\,\%$, 
using binomial techniques to estimate the 1-$\sigma$ confidence interval \citep{Cameron2011}. Using these results to scale the total lensed flux from all redshifts, we obtain the contribution from lenses at $z<0.35$, as shown in Table~\ref{tab:lens}. 
Assuming that all these low redshift lenses fall in the red bin of our sample, we can compare these fluxes to the total stacked flux of our red bins as shown in Table~\ref{tab:lens}, which indicates that about 10, 20 and 30 per cent of the 250, 350, and 500\mum\ fluxes respectively comes from high redshift sources lensed by the targets. This may be a slight overestimate since some of the lenses may fall in the other bins; however \citet{Auger2009} showed that 90 per cent of lenses are massive early type galaxies. Any lensing contribution to the blue or green bins would be negligible compared to the fluxes measured in those bins. 

The lensed flux is divided between the redshift bins of the red sample in a way that is determined by the product of the lens number distribution $n_l(z)$ and the lens efficiency distribution $\Phi(z)$. The numbers $n_l(z)$ are given by \citet{Gonzalez-Nuevo2011}, while the efficiency depends on the geometry between source, lens and observer. We estimate $\Phi(z)$ from the HALOS source and lens redshift distributions using the formula of \citet{Hu1999}, and compute the lens flux distribution from the product of total lensed flux, $n_l(z)$ and $\Phi(z)$. Comparing this to the total flux of red galaxies in each redshift bin, we compute the fractional contamination from lensed flux as shown in Table~\ref{tab:lens}. Errors on the lensed flux per redshift bin are dominated by the Poisson error on the normalisation of $n_l(z)$, which is simply the Poisson error on the count of 33 lenses. {The relative error on the lensed flux is therefore $\sqrt{33}/33=0.17$. The error on the stacked red galaxy fluxes is dominated by the 7\%\ flux calibration error (Pascale E., et al. in preparation), hence the errors on the fractions in Table~\ref{tab:lens} are given by the quadrature sum of 7\%\ and 17\%, which is 19\%.}
%
Using the fractions derived above we can remove the estimated lensed contribution to stacked fluxes in each redshift bin of the red sample. The effect of subtracting this fraction from the fluxes of red galaxies is minor in comparison to the trends described in Section~\ref{sec:fluxevo}. The effect on other derived results will be discussed later in the paper. 
}

\begin{table}
\caption{ Total surface brightness of lensed sources from the \citet{Lapi2011} counts model, and estimated contribution from the low redshift population of lenses assuming the lens redshift distribution from HALOS \citep{Gonzalez-Nuevo2011}. This is compared to the total surface brightness of red galaxies ($g-r$ colour) at $z<0.35$ from our stacks. We then estimate the fraction of the flux in each redshift bin of the red sample that comes from lensed background sources.}
\begin{tabular}{l c c c}
\hline
 & 250\mum & 350\mum & 500\mum \\
\hline
 & \multicolumn{3}{c}{Total surface brightness Jy\, deg$^{-2}$}  \\
All lensed flux & $1.09$ & $1.34$ & $1.22$ \vspace{3pt} \\
Lenses at $z<0.35$ & $0.23^{+0.09}_{-0.06}$ & $0.28^{+0.12}_{-0.07}$ & $0.26^{+0.11}_{-0.07}$ \\
Red galaxies & $2.6\pm0.5$ & $1.6\pm0.2$ & $0.8\pm0.1$ \vspace{3pt} \\
\hline
 & \multicolumn{3}{c}{Lensed flux/red galaxy flux by $z$ bin}  \\
$0.01<z<0.12$ &     $0.00\pm0.00$    &  $0.00\pm0.00$   &  $0.00\pm0.00$ \\
$0.12<z<0.17$ &     $0.06\pm0.01$    &  $0.12\pm0.02$   &  $0.21\pm0.04$ \\
$0.17<z<0.22$ &     $0.11\pm0.02$    &  $0.20\pm0.04$   &  $0.33\pm0.06$ \\
$0.22<z<0.28$ &     $0.16\pm0.03$    &  $0.27\pm0.05$   &  $0.44\pm0.08$ \\
$0.28<z<0.35$ &     $0.20\pm0.04$    &  $0.35\pm0.07$   &  $0.58\pm0.11$ \\
\hline
\end{tabular} 
\label{tab:lens}
\end{table}

\subsection{Resolving the cosmic IR background}
\label{sec:cib}
A useful outcome of stacking on a well-defined population of galaxies such as the GAMA sample is that we can easily measure the integrated flux from this population and infer how much it contributes to the cosmic infrared background \citep[CIB;][]{Puget1996,Fixsen1998}. The cosmic background at FIR/submm wavelengths makes up a substantial fraction of the integrated radiative energy in the Universe \citep{Dole2006}, although the sources of this radiation are not fully accounted for. For example, \citet{Oliver2010} calculated that the HerMES survey resolved only $15\pm4\%$ of the CIB into sources detected with SPIRE at 250\mum, down to a flux limit of 19\,mJy. A greater fraction can be accounted for using $P(D)$ fluctuation analysis to reach below the detection limit of the map: in HerMES, \citet{Glenn2010} resolved $64\pm16\%$ of the 250\mum\ CIB into SPIRE sources with $S_{250}>2$\,mJy. Stacking on 24\mum\ sources has also proved successful, utilising the greater depth of 24\mum\ maps from \textit{Spitzer}-MIPS to determine source catalogues for stacking at longer wavelengths. Stacking into BLAST, \Citet{Bethermin2010} resolved $48\pm27\%$ of the 250\mum\ CIB into 24\mum\ sources with $S_{250}>6.2\,$mJy while \citet{Marsden2009} resolved $83\pm21\%$ into sources with $S_{24}>15\,\mu$Jy. { However, these BLAST measurements included no corrections for clustering; the authors claimed that the effect was negligible, although this observation may appear to conflict with similar analyses in the literature \citep{Negrello2005,Serjeant2008,Serjeant2010a,Bourne2011}}.

Similarly we can stack the GAMA sample to estimate what fraction of the CIB at 250, 350 and 500\mum\ is produced by optically detected galaxies at low redshifts.
{ To do this we measure the {\it sum} of measured fluxes in each bin and scale} by a completeness correction to obtain the total flux of all $r<19.8$ galaxies at $z<0.35$. The correction accounts for two levels of incompleteness. The first is the completeness of the original magnitude-limited sample: \citet{Baldry2010} estimate that the GAMA galaxy sample (after star-galaxy separation) is $\gtrsim 99.9\%$ complete. 
The second completeness is the fraction of the catalogue for which we have good spectroscopic or photometric redshifts { (i.e. spectroscopic {\sc z\_quality}~$\geq 3$ or photometric $\delta z/z<0.2$; see Section~\ref{sec:sample})}.
This fraction is 91.9\%; however we have only included galaxies with redshifts less than 0.35, which comprise 86.8\% of the good redshifts. We cannot be sure of the redshift completeness at $z<0.35$ (accounting for both spectroscopic and photometric redshifts) so we simply assume that we have accounted for 91.9\% of these, to match the redshift completeness of the full sample.\footnote{This may be a slight underestimate of the redshift completeness at $z<0.35$, in which case we would overestimate the total corrected flux by a maximum of 8.7\%.}
Finally we scale by the fraction of galaxies at $z<0.35$ that are within the overlap region between the SPIRE mask and the GAMA survey, which is 72.4\%. The combined correction factor is 
$\eta=1/(0.999 \times 0.919 \times 0.724) = 1.504$.
The corrected flux is converted into a radiative intensity (nW\,m$^{-2}$\,sr$^{-1}$) by dividing by the GAMA survey area (0.0439\,sr).
We compare this to the CIB levels expected in the three SPIRE bands \citep{Glenn2010} -- these are calculated by integrating the CIB fit from \citet{Fixsen1998} over the SPIRE bands. We find that the optical galaxies sampled by GAMA { account for $\lesssim 5\%$ of the background} in the three bands (see Table~\ref{tab:cib}). In the Table we also show the percentage of the CIB produced by galaxies at $z<0.28$, since in this range the catalogue is complete down to below the knee of the optical LF at $M_r^\star = -21.4$ \citep[Petrosian magnitude, $h=0.7$;][]{Hill2011}.

\begin{table}
\caption{Total intensities of $r_\text{petro}<19.8$ galaxies from stacking at 250, 350 and 500\mum, in comparison to the corresponding CIB levels from \citet{Fixsen1998}. We show the intensity as a percentage of the CIB for the full stack, and for the $z<0.28$ subset which is complete in $M_r$ down to $M_r^\star = -21.4$ \citep{Hill2011}. We also show the contributions of the individual redshift bins and $g-r$ colour bins.
{ All contributions from red galaxies have been corrected for the lensed flux contamination using the fractions in Table~\ref{tab:lens}. }
All errors include our statistical error bars from stacking, { the error on the lensing correction (where applicable) and a 7\% flux calibration error (Pascale, E. et al. in preparation).}}
\begin{tabular}{ l c c c }
\hline 
 & 250\mum & 350\mum & 500\mum \\
\hline 
 & \multicolumn{3}{c}{Intensity nW\,m$^{-2}$\,sr$^{-1}$} \\ 
CIB  &
      $10.2 \pm       2.3$ &       $5.6 \pm       1.6$ &       $2.3 \pm      0.6$ \\
Total Stack &
     $0.508 \pm     0.036$ &      $0.208 \pm    0.015$ &     $0.064 \pm    0.005$ \\
$0.01<z<0.28$ &
     $0.428 \pm     0.030$ &      $0.173 \pm    0.012$ &     $0.054 \pm    0.004$ \\
\hline 
 & \multicolumn{3}{c}{\% of CIB} \\ 
Total Stack &
      $4.98 \pm      0.39$ &       $3.71 \pm      0.30$ &       $2.79 \pm      0.22$ \\
$0.01<z<0.28$ & 
      $4.19 \pm      0.34$ &       $3.08 \pm      0.26$ &       $2.33 \pm      0.19$ \\ \\
$0.01<z<0.12$ &
      $1.57 \pm      0.17$ &       $1.11 \pm      0.13$ &      $0.81 \pm     0.09$ \\
$0.12<z<0.17$ &
      $0.97 \pm     0.10$ &      $0.71 \pm     0.08$ &      $0.53 \pm     0.06$ \\
$0.17<z<0.22$ &
     $0.85 \pm     0.08$ &      $0.63 \pm     0.07$ &      $0.49 \pm     0.05$ \\
$0.22<z<0.28$ &
     $0.80 \pm     0.08$ &      $0.63 \pm     0.07$ &      $0.50 \pm     0.05$ \\
$0.28<z<0.35$ &
     $0.78 \pm     0.08$ &      $0.63 \pm     0.07$ &      $0.46 \pm     0.05$ \\ \\
Blue &
      $3.02 \pm      0.26$ &       $2.34 \pm      0.21$ &       $1.67 \pm      0.15$ \\
Green &
      $1.12 \pm     0.10$ &      $0.84 \pm     0.08$ &      $0.64 \pm     0.06$ \\
Red &
     $0.83 \pm     0.08$ &      $0.64 \pm     0.07$ &      $0.48 \pm     0.05$ \\
\hline
\end{tabular}

\label{tab:cib}
\end{table}

\subsection{The submm SED and {\it k}--corrections}
\label{sec:firkcorrs}

Monochromatic luminosities (W\,Hz$^{-1}$) at rest-frame 250, 350 and 500\mum\ can be calculated using equation~(\ref{eqn:lum}), in which $S_{\nu}$ is the SPIRE flux in Jy, $K(z)$ is the $k$--correction at redshift $z$, and $D_\text{L}$ the corresponding luminosity distance (m). The $(1+z)$ on the denominator is the bandwidth correction, which together with the $k$--correction converts an observed 250\mum\ flux to the flux at rest-frame 250\mum.
\begin{equation}
L_\nu =10^{-26}\;\dfrac{4\,\pi\, D_\text{L}^2\, S_\nu K(z)}{(1+z)}
\label{eqn:lum}
\end{equation}
$K$--corrections are obtained by assuming that the SED emitted by dust at a temperature $T_\text{dust}$ is governed by a greybody of the form $\nu^\beta B(\nu,T_\text{dust})$ (where $B$ is the Planck function). The $k$--correction for this SED is given by
\begin{equation}
K(z)=\left(\dfrac{\nu_\text{o}}{\nu_\text{e}}\right)^{3+\beta} \dfrac{e^{h\nu_\text{e}/kT_\text{dust}}-1}{e^{h\nu_\text{o}/kT_\text{dust}}-1}.
\label{eqn:kcorr}
\end{equation}
where $\nu_\text{o}$ is the observed frequency in the 250, 350 or 500\mum\ band, $\nu_\text{e} = (1+z)\,\nu_\text{o}$ is the rest-frame (emitted) frequency, $k$ is the Boltzmann constant and $h$ the Planck constant.

Since we cannot fit an SED to individual galaxies we instead examine the ratios between stacked fluxes in each bin. { Fluxes in the red bins are first corrected to remove the contribution from lensing as discussed in Section~\ref{sec:lens}.} The colour-colour diagrams in Fig.~\ref{fig:ccdiags} show { the resulting flux ratios in the observed frame in each of} the five redshift bins, alongside a selection of models, which are plotted with a range of temperatures increasing from left to right. We try both a single greybody and a two component model, but there is little to choose between them in these colours, since the SPIRE bands are at long wavelengths at which the SED is dominated by the cold dust, {with little contribution from transiently heated small grains or hot dust}. We therefore adopt a single component for simplicity. 
{
The scatter in the data is large, as are the errors on the 350/500\mum\ flux ratio. Moreover, with all our data points on the longward side of the SED peak we are unable to resolve the degeneracy between the dust temperature and the emissivity index ($\beta$). This is shown by the close proximity of the $\beta=1$ and $\beta=2$ models on Fig.~\ref{fig:ccdiags}, which overlap in different temperature regimes.
With these limitations we are forced to assume a constant value of $\beta$ across all our bins. We choose a value of 2.0, which has been shown to be realistic in this frequency range \citep[e.g.][]{Dunne2001,James2002,Popescu2002,Blain2003,Leeuw2004,Hill2006,Paradis2009}. 
For comparison the Planck Collaboration found an average value of $\beta=1.8\pm0.1$ by fitting SEDs to data at 12\mum--21\,cm from across the Milky Way \citep[also references therein]{PlanckCollaboration2011b}.\footnote{{A further issue with fitting SEDs is that $\beta$ may vary with frequency. For example \citet{Paradis2009} analysed data on the Milky Way from 100\mum\ to 3.2\,mm and showed that $\beta$ was generally steeper at $100-240\mum$ than $550-2100\mum$. This is an effect that we cannot take any account of without many more photometric points on the SED, but it could have some effect on our fitted temperatures and therefore luminosities.}}

Under the assumption of a constant $\beta$ (whatever its value) the dust temperatures implied by Fig.~\ref{fig:ccdiags} take a wide range of values across the various bins (between 11 and 22\,K for $\beta=2$). This is not just random scatter; red galaxies tend towards colder temperatures than blue and green, while blue galaxies in some redshift bins show a trend towards lower temperatures at brighter \Mr. 
For the purposes of $k$--corrections we can estimate the temperature more accurately by fitting greybody SED models to the three data points at the emitted frequencies given by the observed frequency scaled by $1+z$, using the median redshift in the bin.\footnote{{The temperature fits and trends mentioned here are discussed further in Section~\ref{sec:temps}.}}
{In general one must be careful when using stacked fluxes in this way to examine the SED, since when stacking many galaxies with different SEDs, the ratios between the stacked fluxes can be unpredictable and not representative of the individual galaxy SEDs. In this case however we believe we can be fairly confident of the results because we bin the galaxies in such a way that we should expect the SEDs within each bin to be similar, and so inferred dust temperatures and other derived parameters should be accurate.}

The best-fit temperatures range between $12-28$\,K, with a median value of 18.5\,K. Using $\beta=1.5$ instead, the temperatures are increased by a factor $1.2-1.6$, ranging from $13-46$\,K with a median of 23.0\,K.
We can compare this median value to temperatures derived from single-component fits in the literature. For example, \citet{Dye2010} derived a median isothermal temperature of 26\,K ($\beta=1.5$) for the detected population in the \hatlas\ science demonstration data, in agreement with the BLAST sample of \citet{Dye2009}. 
The value { of} 26\,K is within the range of our temperatures using $\beta=1.5$, and only slightly {higher} than the median.  
{Higher temperatures were found by \citet{Hwang2010} in their PEP/HerMES/SDSS sample of 190 local galaxies : they reported median temperatures rising as function of IR luminosity, from around 26\,K at $10^9\,\text{L}_\odot$ to 32\,K at $10^{11}\,\text{L}_\odot$ and 40\,K at $10^{12}\,\text{L}_\odot$ ($\beta=1.5$).  
{These temperatures may be higher because \citeauthor{Hwang2010} required a detection shortward of the SED peak (i.e. in an {\it AKARI}-FIS  or {\it IRAS} band) for galaxies to be included in their sample. Fitting a single greybody to an SED which contains both a cold ($\lesssim 20$\,K) and a warm ($\gtrsim 30$\,K) component \citep{Dunne2001} may give results that are not comparable to ours, which fit only the cold component.
On the other hand, \citet{Smith2011b} fitted greybodies with $\beta=1.5$ to the \hatlas\ 250\mum-selected sample of low redshift galaxies matched to SDSS, and found a median temperature of $22.5\pm5.5$\,K (similar to our result), and unlike \citeauthor{Hwang2010} they found no evidence for a correlation with luminosity.
}

{The \citet{PlanckCollaboration2011a} compiled a sample of around 1700 local galaxies by matching the {\it Planck} Early Release Compact Source Catalogue and the Imperial {\it IRAS} Faint Source Redshift Catalogue, and fitted SEDs to data between $60-850$\mum\ using both single-component fits with variable $\beta$, and dual-component fits with fixed $\beta=2$. In their single-component fits they found a wide range of temperatures ($15-50$\,K) with median $T=26.3$\,K and median $\beta=1.2$. This median temperature is consistent with the {\it Herschel} and BLAST results, and the low value of $\beta$ is likely { to be} due to the inclusion of shorter wavelength data. The authors state that the two-component fit is statistically favoured in most cases; these fits indicate cold dust temperatures mostly between $\sim10-22$\,K, consistent with the range in our data.}
}

In any case we do not necessarily expect to find the same dust temperatures in an optically selected sample as in a submm selected sample. 
For the purposes of $k$--corrections this is relatively unimportant, {at least at the low redshifts covered in this work}. The choice between cold~$T$/high~$\beta$ and hot~$T$/low~$\beta$ makes very little difference to {monochromatic} luminosities, as crucially they both fit the data. Likewise the {\it range} of temperatures has little effect on $k$--corrections: using the median fitted temperature of 18.5\,K in all bins gives essentially the same results as the using the temperature fitted to each bin separately. To remove the effect of the variation between models, we carry out all analysis of monochromatic luminosities using the median temperature of $T_\text{dust}=18.5$\,K and $\beta=2.0$ to derive $k$--corrections using equation~(\ref{eqn:kcorr}) (except where stated otherwise). 
The implications of the fitted SEDs on the physical properties of galaxies in the sample will be discussed in Section~\ref{sec:disc}. First we will concentrate on the observational results of the stacking which are not dependent on the model used to interpret the submm fluxes.

\begin{figure*}
 \includegraphics[width=\textwidth]{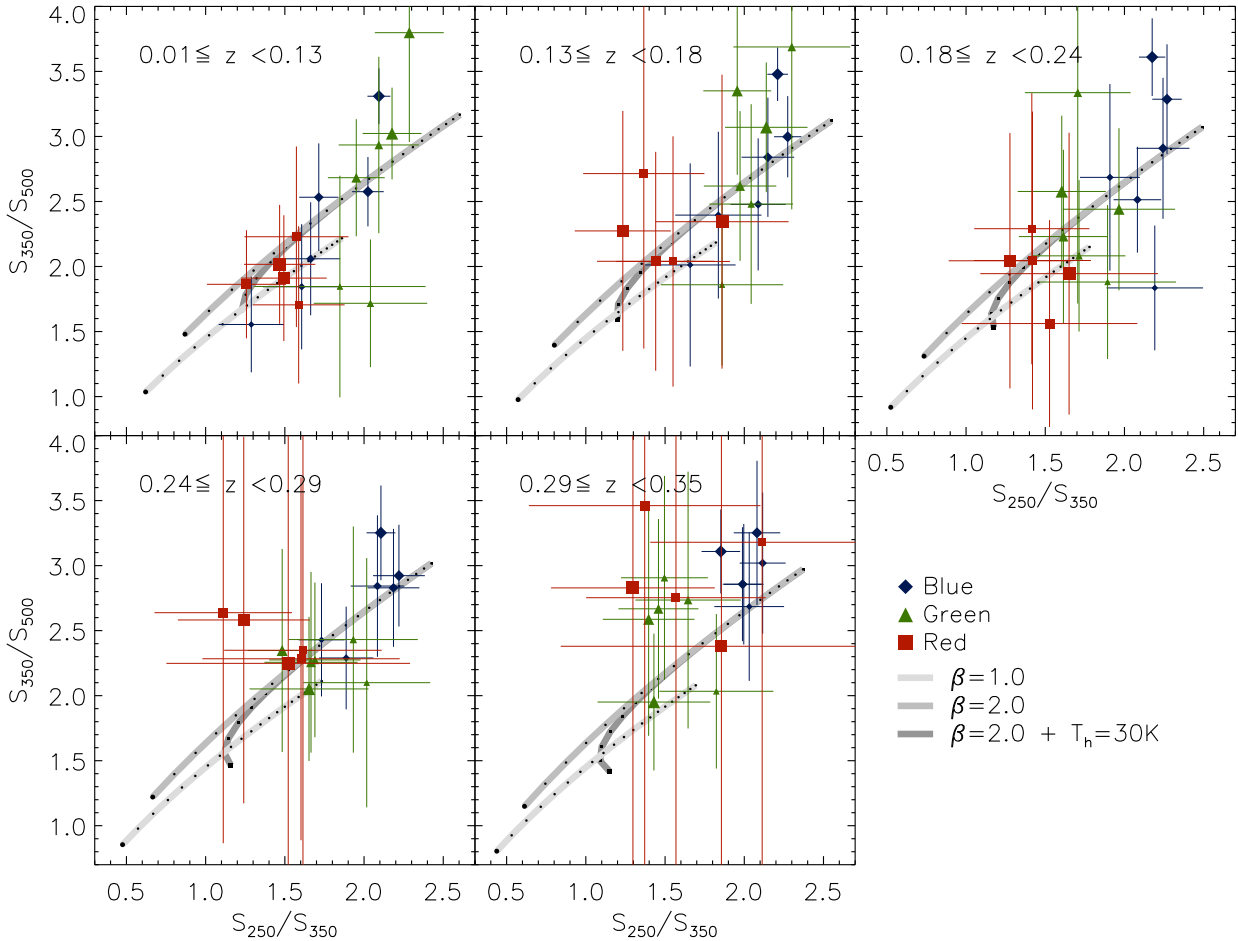}
\caption{Colour-colour diagram of the observed-frame SPIRE fluxes. The plot is divided into five redshift bins, in which the data from that bin are plotted along with the colours expected from various models as they would be observed at the median redshift in the bin. Data are divided into the three $g-r$ colour bins, denoted by symbols and colours, and six \Mr\ bins denoted by the size of the data point (larger=brighter). { Data points and error bars in the red bins include the lensing correction and its uncertainty.} Three families of models are shown: two consist of isothermal SEDs with either $\beta=1$ or $\beta=2$, and various dust temperatures; the third is a two-component SED with $\beta=2$, warm dust temperature $T_w=30$\,K, with a cold/hot dust ratio of 100. Each model is given a range of (cold) dust temperatures; the dots along the lines indicate 1\,K increments from 10\,K (lower left) to 30\,K. Choosing a single component model with $\beta=2$ leads to a range of temperatures between 13 and 22\,K. }
\label{fig:ccdiags}
\end{figure*}

\subsection{Luminosity evolution}
\label{sec:lumevo}
To calculate stacked luminosities we apply equation~(\ref{eqn:lum}) to each measured flux and stack the results. The error on the stacked value is again calculated using the \citet{Gott2001} method. {Note that this method is not the same as applying equation~(\ref{eqn:lum}) to the median flux and median redshift of each bin, since luminosity is a bivariate function of both flux and redshift.} { Fluxes of sources in the red bin are corrected for the fractional contributions from lensing given in Table~\ref{tab:lens}, as explained in Section~\ref{sec:lens}. The 1-$\sigma$ errors on these corrections are included in the luminosity errors.}
Results in Fig.~\ref{fig:lums} show a generally strong correlation between luminosities in the $r$-band and all three submm bands, as is the expected trend across such a broad range, but the dependence is not on \Mr\ alone. This becomes obvious when comparing the data points with the grey line, which shows the linear least-squares fit to the results from the lowest-redshift blue galaxies (the line is the same in each panel from left to right). In the blue galaxies, there may be a slight flattening of the correlation for the brightest galaxies and/or the higher redshifts, but this effect is much stronger in the green galaxies, which are intermediate between the blue and red samples.  For the red galaxies the correlation disappears entirely but for the faintest bin at low redshift. The luminosities of the red galaxies all lie below the grey line, showing that red galaxies emit less in the submm than blue or green galaxies of the same \Mr, {strongly suggesting that they are dominated by a more passive population than green and blue galaxies.} { These trends are greater than the uncertainties on the lensing correction}.

Apart from this colour dependence there is also a significant increase in submm luminosity with redshift for green and red galaxies of the same $r$-band luminosity. This evolution appears to occur at all \Mr, without being particularly stronger for either bright or faint galaxies, but it is especially strong for red galaxies. {This may indicate a transition in the make-up of the red population, with obscured star-forming galaxies gradually becoming more dominant over the passive population as redshift increases. Such a scenario might be expected as we look back to earlier times towards the peak of the universal star-formation history. {One problem with this explanation is that we might expect an increase in obscured star-formation to be accompanied by an increase in the dust temperatures at higher redshifts, which we found no evidence for in the SPIRE colours (Section~\ref{sec:firkcorrs}).

Meanwhile the green sample shows similarities with the blue at low redshift and low $r$-band luminosity, but at high redshifts and stellar masses the luminosity dependence on \Mr\ is flatter and more similar to that of red galaxies. This could be due to a shift in the dominant population of the green bin, between blue-cloud-like galaxies and red-sequence-like galaxies at different redshifts and \Mr.}

\begin{figure*}
 \includegraphics[width=\textwidth]{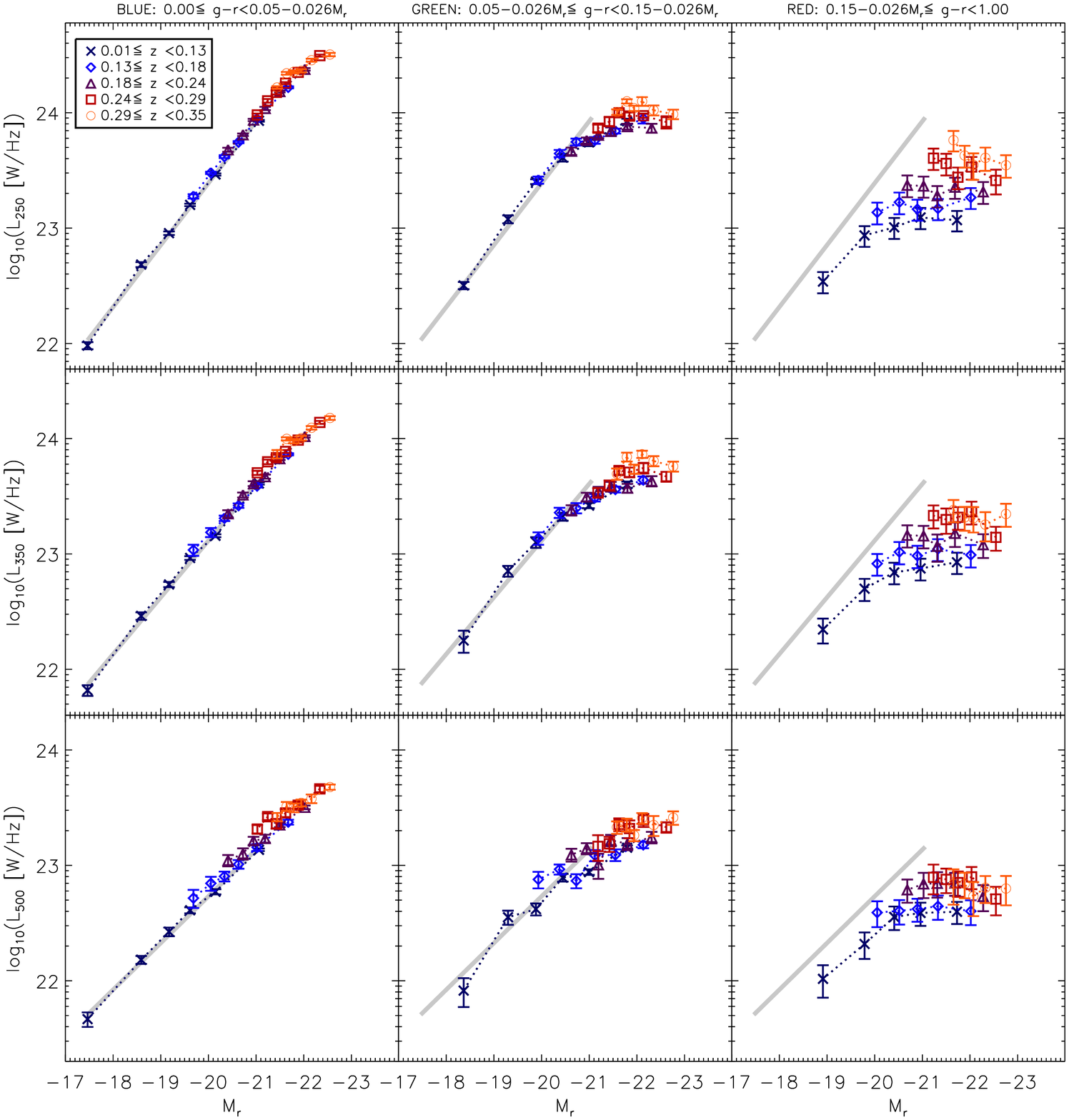}
\caption{Stacked SPIRE luminosities as a function of $g-r$ colour, redshift and absolute magnitude \Mr. Layout as in Fig.~\ref{fig:flux}. Error bars are the statistical $1\sigma$ errors in the bins as described in Section~\ref{sec:errors}. { Fluxes in the red bin have been corrected for the lensing contribution as described in Section~\ref{sec:lens}, and error bars include the associated uncertainty.} The thick grey line is the same from left to right, and is the linear least-squares fit to the results for the lowest-redshift blue galaxies.}
\label{fig:lums}
\end{figure*}

\subsubsection{UV-optical versus optical Colours}
\label{sec:results_g-r_NUV-r}
Splitting the sample by the $NUV-r$ colour index provides a slightly different sampling regime and reduces contamination between the colour bins because the red and blue populations are better separated (see Section~\ref{sec:optcolours}). It therefore offers a useful test of the robustness of the results of stacking by $g-r$. 
Figure~\ref{fig:lums_NUV-r} shows that stacked 250\mum\ luminosities follow the same trends with colour, redshift and \Mr\ as in the $g-r$ stacks (350 and 500\mum\ results are similar). 
This supports the interpretation that the three colour bins sample intrinsically different populations in terms of the dust properties. The red sample in either $NUV-r$ or $g-r$ appears to be dominated by passive galaxies at low redshifts at least, but the emission from dust increases by a factor of { around} 10 over the redshift range.

Errors are slightly larger in this sample, particularly in the red bin, because we are limited to the 52,773 galaxies with $NUV$ detections. Since results appear to be independent of the colour index used, we opt to use the more complete $r$-limited sample of 86,208 sources with $g-r$ colours for all subsequent analysis.

\begin{figure*}
 \includegraphics[width=\textwidth]{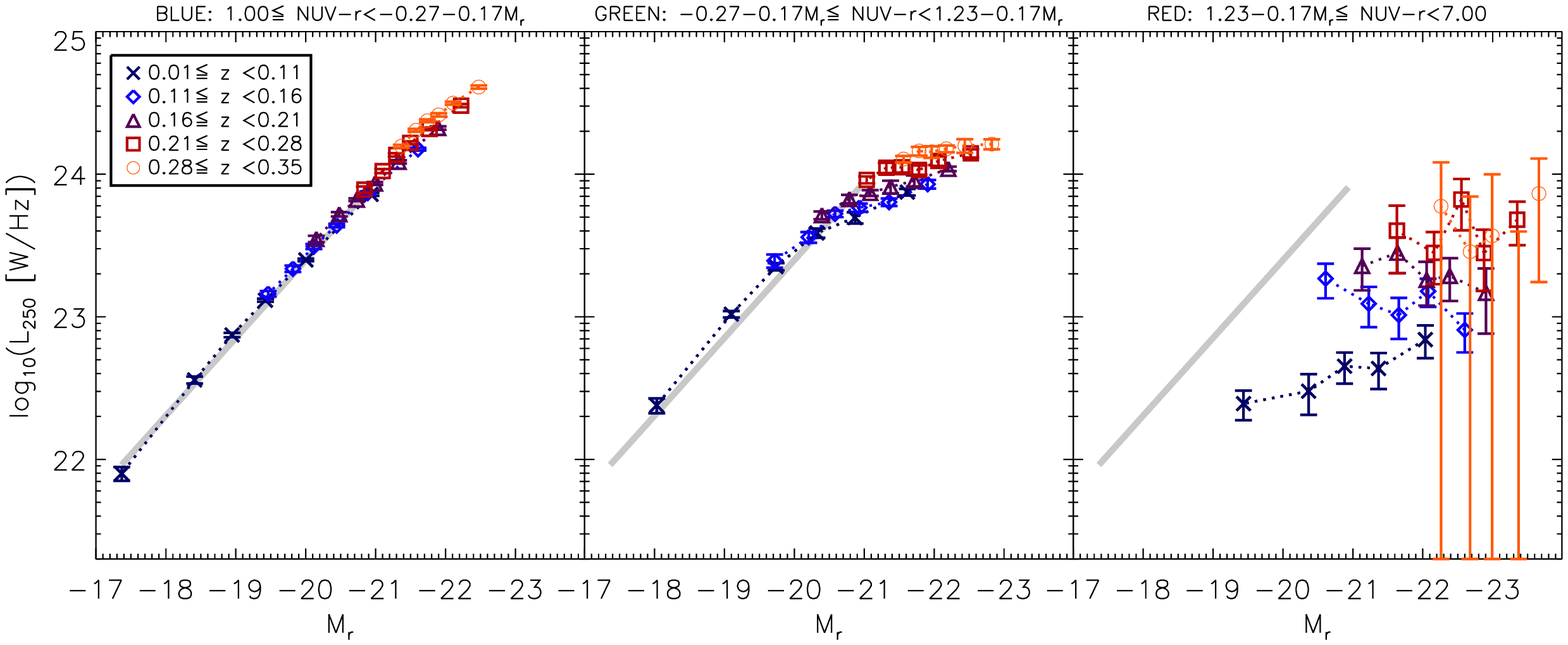}
\caption{Stacked 250\mum\ luminosity as a function of $NUV-r$ colour, redshift and \Mr. Error bars are the statistical $1\sigma$ errors in the bins as described in Section~\ref{sec:errors}. { Data and errors in the red bin incorporate the correction for lensing.} The thick grey line is the same from left to right, and is the linear least-squares fit to the results for the lowest-redshift blue galaxies.}
\label{fig:lums_NUV-r}
\end{figure*}

\subsubsection{Stellar mass versus absolute magnitude}
An alternative to dividing the sample by \Mr\ is to use the stellar masses which were calculated from the GAMA $ugriz$ photometry by \citet{Taylor2011} assuming a \citet{Chabrier2003} IMF.
Stellar mass is a simple physical property of the galaxy so may reveal more about intrinsic dependencies; on the other hand it depends much more on the models used to fit the optical SED than \Mr, which is only subject to a small $k$--correction and the assumed cosmology (for a given redshift).
{ Relative errors on stellar masses are dominated by systematics, but are small ($\Delta \log M_\text{star} \sim 0.1$; \citealp{Taylor2011}). We confirmed that our results are robust to these errors by repeating all analysis after making random perturbations to the stellar masses, where the size of each perturbation was drawn from a Gaussian distribution with width $\sigma=\Delta \log M_\text{star}$. No results were systematically affected by these perturbations, and random deviations in stacked values were smaller than the error bars. 
}

Figure~\ref{fig:lums_bymass} shows that the results of stacking by $g-r$ colour and stellar mass differ slightly from the results of stacking by \Mr\ (Fig.~\ref{fig:lums}). Again we found very little difference from these results when we stacked by $NUV-r$ colour and stellar mass.
The results of stacking by mass seem to differ most in the blue bin. {Whereas there was little luminosity evolution at fixed \Mr, these results show evolution at fixed stellar mass. Furthermore this evolution is dependent on stellar mass, suggesting that smaller galaxies tend to evolve more rapidly. }
{The samples in Figures~\ref{fig:lums} and \ref{fig:lums_bymass} are slightly different since stellar masses were only available for 90 per cent of the full sample; however we know that this is not responsible for the discrepancy since repeating the stacking by \Mr\ with the stellar mass sample gives identical results (the stellar mass incompleteness does not vary between bins).}
The difference arises because \Mr\ does not directly trace stellar mass, which leads to a mixing of galaxies of different masses within a given \Mr\ bin. 
This is unavoidable since we must split the sample in three ways (colour, redshift and mass/magnitude) because dust luminosity varies strongly as a function of all of these. We split the sample by colour first, then by redshift and finally divide into mass or magnitude bins, but each bin can still contain a relatively broad range of redshifts. Within each bin there will be a strong degeneracy between redshift and \Mr, {simply because \Mr\ is a strong function of redshift. In Fig.~\ref{fig:mass_mr} we plot stellar mass against \Mr\ with the points colour-coded by redshift, showing that redshifts increase steadily from left to right, with { decreasing} \Mr.} A narrow range in \Mr\ would select a narrow range of redshifts, while a similarly narrow range in mass selects a much broader range of redshifts.

The effect of this on the blue bins in Figures~\ref{fig:lums} and \ref{fig:lums_NUV-r} is a tendency for the data points of different redshift bins to lie along the same relation of $L_{250}$ as a function of \Mr. The degeneracy is (partially) broken when splitting by stellar mass, thus separating out the trends with redshift and with mass in Fig.~\ref{fig:lums_bymass}. This effect is much less noticeable in the red bin simply because the redshift evolution is much stronger while the mass dependency is very weak in the red sample.

\begin{figure*}
 \includegraphics[width=\textwidth]{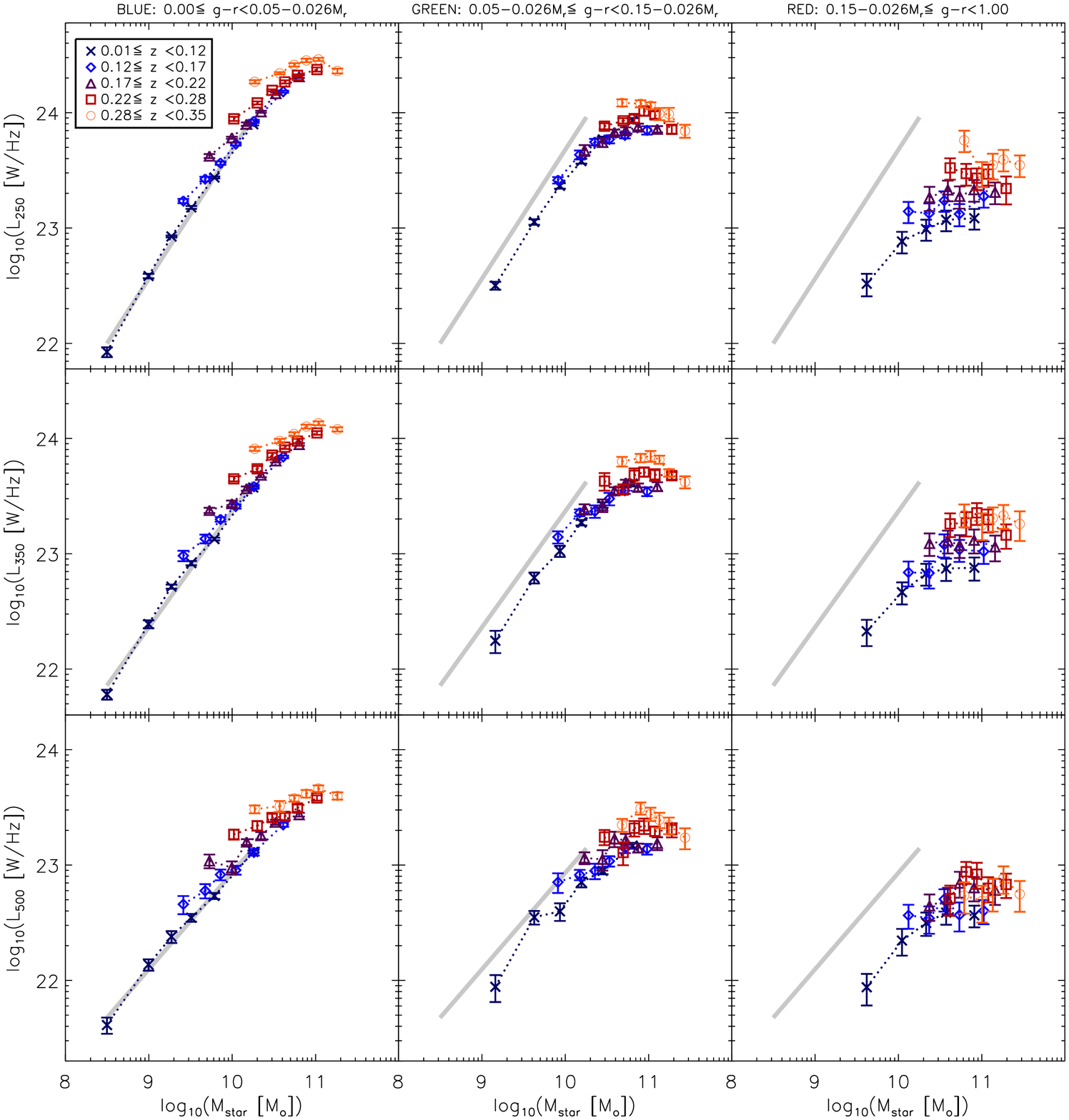}
\caption{Stacked SPIRE luminosities as a function of $g-r$ colour, redshift and stellar mass. Error bars are the statistical $1\sigma$ errors in the bins as described in Section~\ref{sec:errors}. { Data and errors in the red bin incorporate the correction for lensing.} The thick grey line is the same from left to right, and is the linear least-squares fit to the results for the lowest-redshift blue galaxies.}
\label{fig:lums_bymass}
\end{figure*}

\begin{figure}
\includegraphics[width=0.5\textwidth]{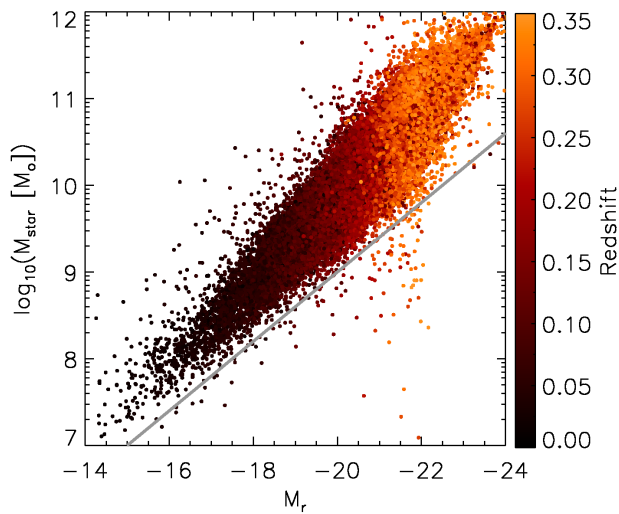}
\caption{Stellar masses of the sample as a function of absolute magnitude and coloured by redshift, showing the degeneracy between \Mr\ and $z$ resulting from an $r$-band selection. { To guide the eye, the slope of the grey line indicates direct proportionality between mass and luminosity, i.e. $\log_{10}M_\text{star} = -0.4\Mr + C$ (for this line $C=1.0$). The spread in the data perpendicular to this line reveals the broad range of mass-to-light ratios which is responsible for the differences between stacking by \Mr\ and by $M_\text{star}$}}
\label{fig:mass_mr}
\end{figure}

\section{Discussion}
\label{sec:disc}

\subsection{Dust temperatures and SED fitting}
\label{sec:temps}
In Section~\ref{sec:firkcorrs} we stated that assumptions about the dust temperature and $\beta$ had negligible effect on the $k$--corrections to SPIRE fluxes at these low redshifts. Hence we chose to use the same SED model to compute monochromatic luminosities, assuming a single-component greybody with $T_\text{dust}=18.5$\,K and $\beta=2.0$.
However if we want to infer properties of the full IR SED these considerations are much more important.

We fitted single-component SEDs with $\beta=2.0$ to the stacked SPIRE fluxes in each bin, shifting the observed wavelengths by $(1+z)$ using the median redshift in the corresponding bin. 
{ Fluxes in the red bin were first corrected for the predicted lensing contamination as described in Section~\ref{sec:lens}. The effect of this is to increase the fitted temperatures in the red bin by around $1-3$\,K, which is small compared with the range of temperatures observed, although the errors on temperatures are significantly increased.}
{The fitting was carried out using the IDL routine {\sc mpfitfun},\!\footnote{{\sc mpfitfun} available from Craig Markwardt's IDL library:\\ \url{http://cow.physics.wisc.edu/~craigm/idl/idl.html}} which performs Levenberg-Marquardt least-squares fitting to a general function. Best-fit values of the free parameters (temperature, normalisation) are returned with formal $1\sigma$ errors computed from the covariance matrix.} Some examples of the fits are plotted in Appendix~\ref{app:figs}, showing a range of fitted temperatures.
{The derived temperatures depend on the assumption of a fixed emissivity parameter ($\beta$). Varying this as a function of optical colour and/or stellar mass could to some extent account for the variation in submm colours, which we interpret as a temperature variation. However the variation in $\beta$ would need to be severe ($\Delta\beta >1$) to fully account for the trends in the stacked colours in Fig.~\ref{fig:ccdiags}. It therefore seems likely that { the cause for these variations is either temperature alone or a combination of $\beta$ and temperature.}}

Figure~\ref{fig:fittdust} shows the results of all the temperature fits as a function of colour, stellar mass and redshift. There are strong deviations in some bins from the value that we have been using, and these show strong dependence on colour and stellar mass. In the blue bin we see that dust temperature is tightly correlated with stellar mass in all redshift bins, with a peak at around $6\times10^{10}\text{M}_\odot$, but galaxies of higher masses appear to have colder dust.
The temperature distribution in the the green bin is less well correlated and very noisy, but again there is evidence that the warmest galaxies are towards the middle of the mass range. There is no clear evolution with redshift.
As with the luminosity results, the red bin appears very different from the other two, and there is no evidence for the temperature to increase with stellar mass. {The most important result would seem to be that temperatures are generally much lower in the red than the blue bins. Over the mass range in which the bins overlap ($3\times10^9<M_\text{star}<3\times10^{11}$), the mean (standard deviation) of the temperatures in the blue bins is { 22.7\,(2.9)\,K, compared with 19.4\,(2.8)\,K in the green and 16.1\,(2.9)\,K in the red. 
The difference in the means is statistically significant, since an unpaired $t$-test gives a probability of $10^{-12}$ that the means of the red and blue bins are the same. The $t$-test assumes that the errors on each of the measurements are independent, but this is not true since the errors on the red stacks are dominated by the error on the lensing correction. The mean temperature error of the red stacks is 3.1\,K, which means that including the lensing uncertainty, the blue and red mean temperatures are only different at the $2.1\sigma$ level.} 
%
The incidence of colder dust in redder galaxies may be explained by the relationship between dust temperature and the intensity of the interstellar radiation field (ISRF). The colour temperature of the ISRF is directly related to the stellar population. Old stars produce less UV flux, and so heat the dust less, which causes galaxies dominated by older stars to have cooler dust. Such a temperature differential is therefore consistent with the notion of red galaxies being passive.
{
Meanwhile, colder dust temperatures in the least massive blue galaxies is consistent with these galaxies having more extended dust disks in comparison with their stellar disks, since the ISRF becomes weaker at greater galactocentric radii. Evidence for an extended dust disk in at least one low mass system has been reported by \citet{Holwerda2009}, although larger samples would be needed to judge whether this is a widespread phenomenon.
}

It is possible when fitting the SPIRE bands that the SED shape could be biased by differential effects between the three bands. In particular the 500\mum\ band is the most affected by confusion and blending, as well as being the noisiest, and is also potentially subject to contamination from other emission mechanisms, including synchrotron from within the galaxies, and extended radiation from the Sunyaev-Zel'dovich effect in galaxy clusters (although we note the latter should have been removed in background subtraction). To check for such bias, we also tried fitting only the 250 and 350\mum\ data, and found that the derived temperatures and trends were not significantly different.

%

\begin{figure*}
 \includegraphics[width=\textwidth]{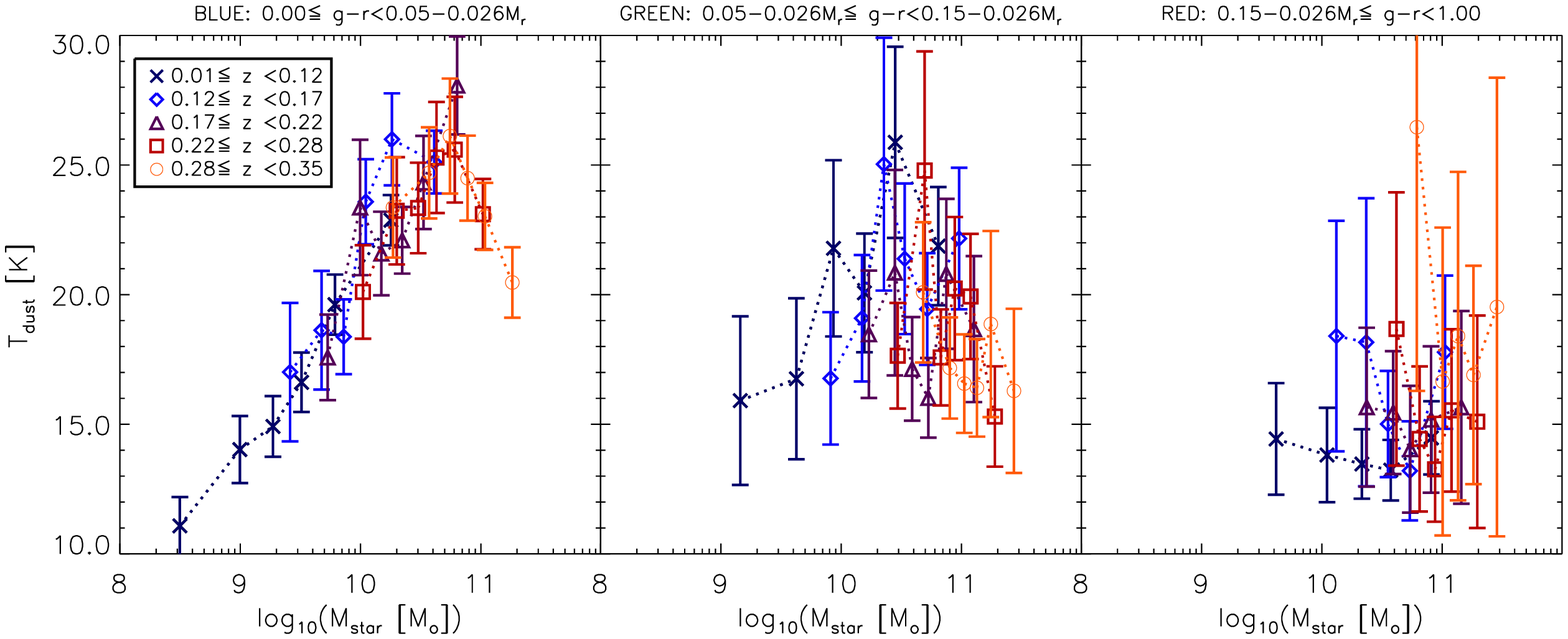}
\caption{Results of fitting single-component greybodies with $\beta=2.0$ to the observed (not $k$--corrected) fluxes in each bin to estimate dust temperatures. { Fluxes in the red bin were corrected for lensing before fitting.} Error bars are the $1\sigma$ errors on fitted temperatures computed by {\sc mpfitfun}.}
\label{fig:fittdust}
\end{figure*}

\subsubsection{Bolometric luminosities}

It can be useful to consider the total IR luminosities $L_\text{TIR}$ ($8-1000\mum$) of galaxies as this allows some comparison between observations at different IR wavelengths and between data and the predictions of models. In all cases this entails making assumptions about the shape of the SED, which must be interpolated -- and indeed extrapolated -- from the limited photometric data available. In this particular case we are limited to just three SPIRE bands, all of which lie longward of the peak in the SED and as such do little to constrain the warmer end of the SED at $\lambda \lesssim 100\mum$. This is why they are well fitted by single-component SEDs, representing a single component of cold dust. In contrast, the TIR luminosity is highly sensitive to emission from the hotter components of dust, especially the $\gtrsim 30$\,K dust associated with H{\sc ii} regions, which is heated by UV radiation from hot young stars. 

\defcitealias{Chary2001}{CE01}
Bearing in mind these limitations, we nevertheless consider it useful to make some attempt at estimating the TIR luminosities representative of our stacked samples. Since our sample is thought to be dominated by normal star-forming and quiescent galaxies, we need to choose an appropriate IR SED template. A commonly used set of templates is that of \citet[hereafter \citetalias{Chary2001}]{Chary2001}. These templates are based on libraries of mid- and far-IR templates representing a range of SED types (from normal spirals to ULIRGs\footnote{Ultraluminous IR galaxies; $L_\text{TIR}>10^{12}\text{L}_\odot$.}) fitted to data on $\sim 100$ local galaxies at 6.7, 12, 15, 25, 60, 100 and 850\mum.\footnote{\citetalias{Chary2001} templates were obtained from \url{http://www.its.caltech.edu/~rchary/}}
From these we select the most appropriate template for each stack by computing chi-squared between each of the templates and our rest-frame { (lensing corrected)} SPIRE luminosities, and assign to each stack the $L_\text{TIR}$ of the template with the minimum chi-squared. Results are shown in Fig.~\ref{fig:fitldust}(a). {Errors on $L_\text{TIR}$ were estimated with Monte-Carlo simulations using the $1\sigma$ errors on the SPIRE luminosities and re-fitting the templates 200 times to obtain the $1\sigma$ error bar on the template $L_\text{TIR}$.}

The \citetalias{Chary2001} templates are of limited value for our sample because they are fitted to {\it IRAS} and SCUBA data for a relatively small sample of local galaxies. The necessity for {\it IRAS} detections means that the galaxies in their sample may have been biased towards hotter SEDs, and may not be representative of the larger population sampled in this work. As an alternative we can compare the results of using the \citetalias{Chary2001} templates with a set of templates modelled on the \hatlas\ SDP source catalogue \citep{Smith2011b}. There is a danger that the opposite bias is active here, since the templates are based on sources selected at 250\mum, which are more likely to have cold SEDs. However by comparing the \hatlas\ $L_{250}$ LF from \citetalias{Dunne2011} with the range of $L_{250}$ of optical galaxies (Figures~\ref{fig:lums}--\ref{fig:lums_bymass}) we see that the luminosity ranges spanned by the two surveys are remarkably similar, implying that the SEDs of \hatlas\ sources could provide a reasonable representation of an optically selected sample.
We use a single template based on the mean of all \hatlas\ SED models from \citet{Smith2011b}. In Fig.~\ref{fig:fitldust}(b) we show the results of fitting this template to our stacked SPIRE luminosities, minimising chi-squared to obtain the correct normalisation and integrating the SED from 8--1000\mum\ to obtain $L_\text{TIR}$ {(errors were estimated from Monte-Carlo simulations using the $1\sigma$ errors on the SPIRE luminosities in the same way as for the \citetalias{Chary2001} templates).}

The results of the two sets of templates are strikingly different, with the \citetalias{Chary2001} models suggesting significantly higher luminosities, reaching the level of `luminous IR galaxies' (LIRGs; $L_\text{TIR} > 10^{11} \text{L}_\odot$) at $z>0.22$ or $M_\text{star}\gtrsim 2\times10^{10} \text{M}_\odot$. The \hatlas\ templates are much colder so give much more moderate luminosities, with around five times lower bolometric luminosity for the same $L_{250}$. With only the SPIRE data to constrain the SED we cannot conclusively say that either set of templates is better suited to describing the optical sample, although for the reasons outlined above we believe that the \hatlas\ templates are more likely to be suitable. The addition of data points at shorter wavelengths, from PACS in the FIR and {\it WISE} in the MIR, would permit a much more accurate derivation of the bolometric luminosity{;} we leave this for a future study.

\begin{figure*}
\subfloat[CE01 templates]{\includegraphics[width=\textwidth]{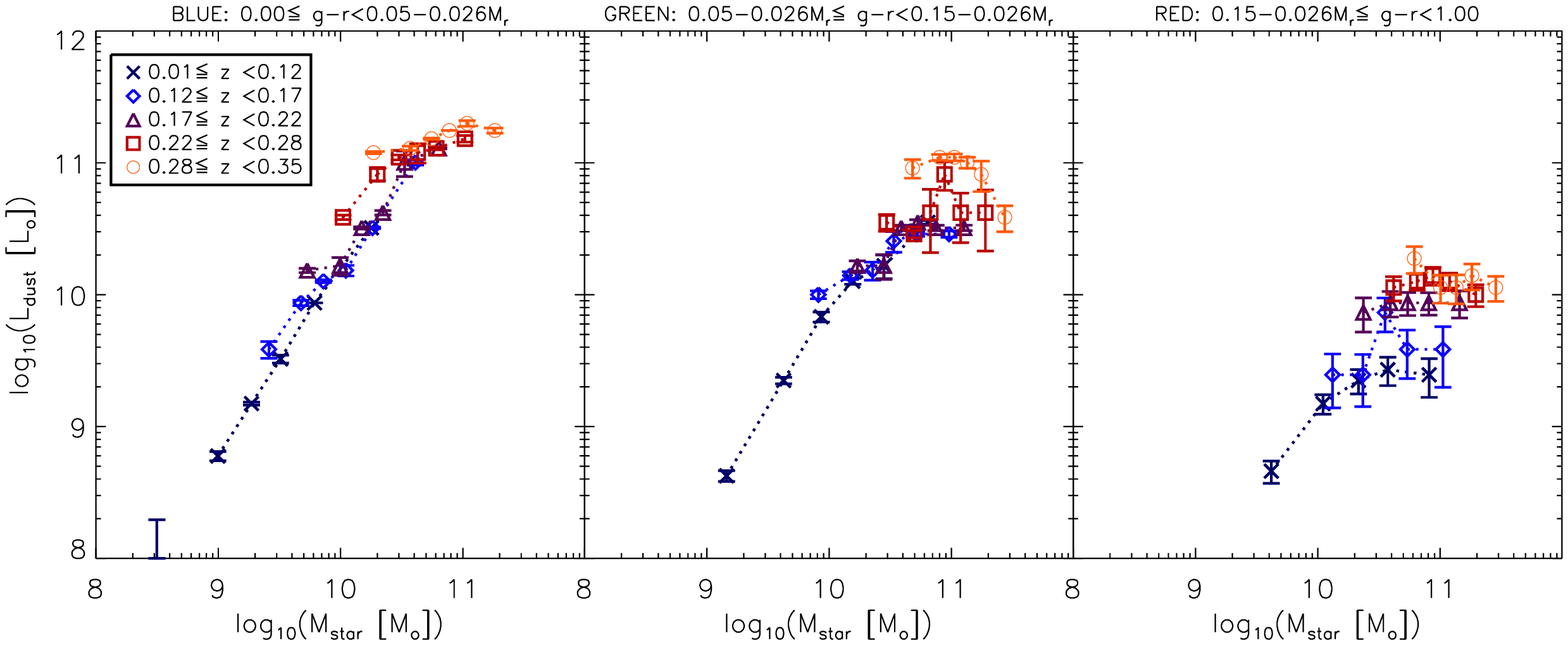}} \quad
\subfloat[\hatlas\ templates]{\includegraphics[width=\textwidth]{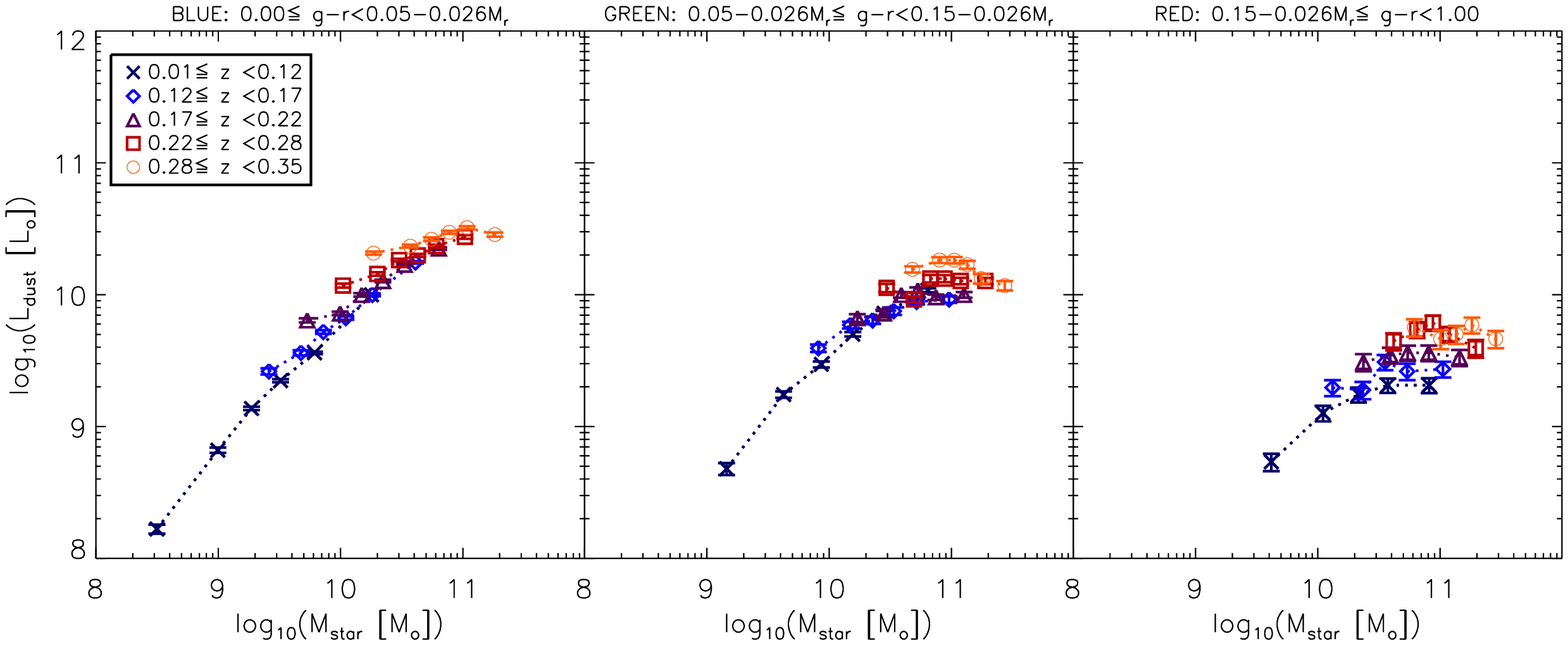}} \quad

\caption{Integrated $L_\text{TIR}$ (8-1000\mum) of (a) \citet{Chary2001} and (b) \hatlas\ \citep{Smith2011b} templates fitted to the stacked SPIRE luminosities in each bin as described in the text. {Error bars are $1\sigma$ errors estimated from Monte-Carlo simulations using the $1\sigma$ errors on the SPIRE luminosities.} {Fluxes in the red bin have been corrected for the lensing contribution as described in Section~\ref{sec:lens}, and error bars include the associated uncertainty.} Note that in panel (a) the first point in the blue sample (i.e. lowest mass, lowest redshift bin) had SPIRE luminosities that fall below all of the CE01 templates, and so the luminosity of the faintest template is used as an upper limit.}
\label{fig:fitldust}
\end{figure*}

Both parts of Fig.~\ref{fig:fitldust} show trends in the bolometric luminosities that are similar to those seen in the monochromatic SPIRE luminosities. That is to say there is a clear evolution with redshift and that this is much stronger in the red than the blue sample. 
We can quantify this evolution in the form $L(z)\propto(1+z)^\alpha$ for the mass range in which the bins overlap. We fit this function in log-space by chi-squared minimisation, using the IDL routine {\sc linfit} to fit the bins which fall in the mass range $M_\text{star}=2-7\times 10^{10}\,\text{M}_\odot$. We find that $L_\text{TIR}$ from the \hatlas\ SEDs { evolves with $\alpha=4.1\pm0.2$ in the blue sample; $\alpha=1.3\pm0.3$ in the green; and $\alpha=6.9\pm1.0$ in the red}. The same evolution is also seen in $L_{250}(z)$ in the same mass bins: this is { described by $\alpha=4.0\pm0.2$ (blue); $\alpha=1.1\pm0.4$ (green); and $\alpha=7.7\pm1.6$ (red).} It appears counter-intuitive that the intermediate green bin should show the least evolution. The likely reason for this is the aforementioned possibility for the green sample to probe different populations at different redshifts (see Sections~\ref{sec:lumevo} and \ref{sec:fluxevo}). Any sign of genuine luminosity evolution would be counteracted by sampling a less luminous population (e.g. with more passive red sequence galaxies in the green bin) at higher redshifts. { We must however note that any evolution in the submm SED of any of these samples (blue, green or red) would render these single template fits unreliable.}  
}

The evolution in $L_\text{TIR}$ of normal galaxies was also observed by \citet{Oliver2010a} for a large sample selected in the optical and NIR with redshifts between $0-2$. Dividing their sample by redshift, stellar mass and optical class (each derived from broadband SED fitting of \citealp{Rowan-Robinson2008}) they stacked into 70 and 160\mum\ {\it Spitzer} images and showed that both `blue' galaxies (with spiral-like SEDs) and `red' galaxies (with elliptical-like SEDs) increased in specific IR luminosity (i.e. $L_\text{TIR}/M_\text{star}$) as a function of redshift. 
{The evolution in specific IR luminosity of all galaxies in their sample increased as $(1+z)^{4.4\pm0.3}$ (independent of stellar mass), which is nearly identical to our result for blue galaxies. When they split the sample into red and blue colours, they also found that red galaxies evolved more strongly, but only with the index $5.7\pm2.5$ compared with our $6.9\pm1.0$. Their blue sub-sample evolved with the index $3.4\pm0.3$ compared with our $4.1\pm0.2$. The agreement is not exact but general trends with redshift and colour are certainly compatible between the two samples.}
{Assuming a correlation between IR luminosity and SFR (e.g. \citealp{Kennicutt1998}),} we can also draw parallels with other {\it Spitzer}-stacking \citep[such as][]{Magnelli2009, Damen2009, Damen2009a} as well as radio-stacking studies \citep{Dunne2009,Pannella2009,Karim2010}, all of which have shown similar dependence of (specific) SFR on stellar mass and redshift in NIR-selected massive galaxies covering larger redshift ranges (up to $z \sim 3$). {These studies have variously reported redshift evolution in specific SFR with indices ($\alpha$) ranging from 3.4 to 5.0, all comparable with the luminosity evolution of our blue sample. It is unsurprising that our blue sample generally agrees with other samples selected by rest-frame optical light with no regard to colour, since our blue selection is by far the largest of our colour bins, comprising 50 per cent of our sample.}

It is well reported in the literature that there is strong evolution in the IR LF out to at least $z=1$ { (\citealp{Saunders1990,Blain1999,Pozzi2004,LeFloc'h2005,Eales2009,Eales2010,Dye2010,Gruppioni2010,Rodighiero2009}; \citetalias{Dunne2011}; \citealp{Goto2011,Sedgwick2011})}. 
This requires an increase in the luminosity of the brightest ($\gtrsim L_\text{IR}^\star$) galaxies, leading to an increase in the numbers of LIRGs and ULIRGs at higher redshifts. Our results show that this evolution at low redshifts also occurs in ordinary galaxies well below the LIRG threshold; these are the galaxies that dominate the number density. Such an evolution in the IR luminosities of all galaxies leads naturally to an evolution in the characteristic luminosity $L^\star$. Exactly this sort of evolution in normal (i.e. non-merging) star-forming galaxies is predicted by the semi-analytic model of \citet{Hopkins2010}, essentially as a result of an evolving gas fraction and using the Schmidt-Kennicutt law \citep{Kennicutt1998}. \citetalias{Dunne2011} also show that an evolving gas fraction is required to explain the luminosity evolution in the \hatlas\ sample, based on the chemical evolution model of Gomez,~{ H.\,L.} et al. (in preparation).

\subsection{The cosmic spectral energy distribution}
{
Having discussed evolution in the IR luminosity density of the Universe, it is natural to consider the local luminosity density (at $z=0$), since this provides a reference point for similar measurements at higher redshifts.
In Section~\ref{sec:cib} we calculated the integrated intensity of low redshift galaxies and showed that they contribute a small fraction of the CIB at submm wavelengths. Building on this result, we can estimate the $z=0$ cosmic SED at submm wavelengths, i.e. the integrated luminosity of all galaxies at $z=0$. To do this we make use of the completeness-corrected integrated intensities in the range $0.01<z<0.12$ in Table~\ref{tab:cib}, but apply $k$- and $e$-corrections to account for the redshifted wavelengths and luminosity evolution respectively. We divide by the comoving volume of the redshift bin ($V_c$) to obtain the luminosity per unit comoving volume:
\begin{equation}
\nu L_\nu= 4\pi d_L^2~\nu I_\nu~\dfrac{K(z)}{1+z}~\dfrac{4\pi}{V_c}~e(z)
\label{eqn:csed}
\end{equation}
where $I_\nu$ is in W\,m$^{-2}$\,sr$^{-1}$\,Hz$^{-1}$, $d_L$ is in m, $V_c$ is in Mpc$^{3}$, 
and the luminosity $\nu L_\nu$ is expressed in units of ${\rm W}\,{\rm Mpc}^{-3}\,h_{70}~(h_{70}=H_0/70\,{\rm km}\,{\rm s}^{-1}\,{\rm Mpc}^{-1})$. Note that while $V_c=0.523\,{\rm Gpc}^3$ represents the total comoving volume of the bin, $d_L=382.5$\,Mpc is the luminosity distance of the median redshift in the bin, $\langle z\rangle=0.084$.
We use the median values of $K(z)/(1+z)$ for this redshift bin: 0.8094 (250\mum), 0.7586 (350\mum), and 0.7259 (500\mum); and for the evolution at all submm wavelengths we take the fitted function $L_{250}(z) \propto (1+z)^{\alpha}$ from Section~\ref{sec:temps}, assuming that the shape of the rest-frame SED does not evolve with redshift (which is supported by our non-evolving temperature results). We assume $\alpha=4.0\pm0.2$, the value derived for the blue sample, since these represent roughly half of all galaxies; the evolution for all galaxies may be slightly stronger (red galaxies seem to evolve more strongly, although the green sample evolve less) but an index around 4 is consistent with other results in the literature (see discussion in Section~\ref{sec:temps}). The evolution from $z=0-0.084$ (the median redshift of the bin) is therefore a factor $(1+0.084)^{4.0}=1.38$, hence the correction is $e(z)=0.72$. 

We thus calculate the luminosities of the cosmic SED at $z=0$ to be $3.9\pm0.3\times10^{33}$, $1.5\pm0.1\times10^{33}$ and $4.3\pm0.3\times10^{32}$\,W\,Mpc$^{-3}\,h_{70}$ at 250, 350 and 500\mum\ respectively. Errors are dominated by the 7\% calibration error on the SPIRE fluxes, which is correlated across the three bands. These results agree very closely with the submm luminosities predicted from GAMA data by Driver,~S. et~al. (in preparation), by calculating the total energy absorbed by dust in the UV-NIR and assuming it is reprocessed as FIR emission with templates from \citet{Dale2002}. 
{They are also close to the pre-Herschel-era prediction of \citet{Serjeant2005}, based on modelling the SEDs of IRAS sources with SCUBA submm measurements. Using equation~(7) of that paper, the predicted luminosities at 250, 350 and 500\mum\ are $4.52\times10^{33}$, $1.43\times10^{33}$ and $3.48\times10^{32}$\,W\,Mpc$^{-3}\,h_{70}$ respectively. Our measurements are within $3\sigma$ of these values, although they arguably suggest that the slope of the cosmic SED may be a little shallower than the prediction.
These measurements are independent of the SEDs and temperatures assumed (since $k$--corrections are small) and of the lensing assumptions (the lensed flux is negligible at $z\lesssim0.1$).
}}

\subsection{Evolution of dust masses}
Dust mass is a quantity which we can expect to constrain much more accurately than $L_\text{TIR}$ using the SPIRE luminosities, because the cold component that they trace is thought to dominate the total dust mass \citep[and references therein]{Dunne2001}. We therefore estimate the mass of the cold dust component described by our greybody fits and use this as a proxy for the total dust mass, assuming that any warmer components have a negligible contribution to the mass.

The dust mass as a function of temperature $T_\text{dust}$ is estimated from the 250\mum\ flux using equation~(\ref{eqn:mdust}):
\begin{equation}
 M_\text{dust}=\dfrac{S_{250}\; D_\text{L}^2\; K(z)}{\kappa_{250}\, B(\nu_{250},T_\text{dust})\, (1+z)}
\label{eqn:mdust}
\end{equation}
We use a dust mass absorption coefficient at 250\mum\ of $\kappa_{250}=0.89\, \text{m}^2\text{kg}^{-1}$ \citepalias[and references therein]{Dunne2011}.

Dust mass results depend strongly on the temperatures assumed {(although not as strongly as the bolometric luminosities). They are therefore subject to our assumption of constant emissivity index ($\beta$), as well as the assumption of a constant absorption coefficient ($\kappa$). Any variation of $\kappa$ with redshift, stellar mass, optical colour or indeed with dust temperature or dust mass itself could alter the trends that we see in dust mass. 
}

In Fig.~\ref{fig:mdust_bymass} we show the dust masses derived using the fitted temperatures from Fig.~\ref{fig:fittdust}. The dust mass is seen to range from around $2\times10^6$ to $8\times10^7 \text{M}_\odot$ across the sample. In all bins dust and stellar mass are correlated, but this weakens slightly with increasing redshift and/or stellar mass. 
There is a definite evolution towards higher dust masses with increasing redshift. {Following the evolutionary form $M_\text{dust}(z)\propto(1+z)^\alpha$ we fit data in the range $M_\text{star}=2-7\times10^{10}\,\text{M}_\odot$ with { slopes of $\alpha=3.9\pm1.7$ for the blue sample; $\alpha=3.0\pm3.3$ for the green; and $\alpha=6.8\pm4.6$ for the red.}
The slopes of the evolution are thus similar to the evolution in luminosities (as might be expected from the lack of evolution in temperature). { Given the error bars we cannot claim to detect evolution in the dust masses of red galaxies, despite having a significant detection of evolution of their luminosities. The reason for this is that the uncertainty of the lensing contribution increases the uncertainty in the fitted temperatures. If we assume that dust temperatures of the red sample do not evolve (as they do not for the blue and green samples) then the evolving luminosities must result from dust mass evolution. However this may not be a valid assumption if the composition of the red sample changes with redshift.}

{ 
If we ignore the highest redshift red bin, which has the largest errors due to lensing, then
the results seem to suggest that the difference in dust mass between the red and blue galaxies is weaker than the difference in luminosity, for a given stellar mass and redshift. The dependence of luminosity on colour therefore appears to be driven more by the temperature than the mass of the dust. We note that if the same temperatures were used in deriving the dust mass in every bin} then the dust masses would be directly proportional to $L_{250}$ and would follow the trends seen in Fig.~\ref{fig:lums_bymass}.
This shows the vital importance of having photometry at multiple points along the SED, without which it would be impossible to constrain the SED shape and inferences about dust mass evolution would have to assume a constant temperature.

Significantly, this analysis suggests that the cold dust masses of red and blue galaxies are { not strikingly discrepant} in the stellar mass range $\sim 1 \times 10^{10} - 2 \times 10^{11} \text{M}_\odot$. 
Previous studies fitting two-component dust models to normal spiral galaxies in the local Universe have derived ranges of \emph{cold} dust masses comparable to our sample: $3\times10^5-1\times10^8$ \citep{Popescu2002}; $2\times10^7-1\times10^9$ \citep{Stevens2005}; and $4\times10^6-6\times10^7\,\text{M}_\odot$ \citep{Vlahakis2005}. Dust masses measured from fits to the cold dust in local ellipticals (which should be akin to our red sample) are often a little lower: $2\times10^5-2\times10^6$ \citep{Leeuw2004}; {$4\times10^4-5\times10^7$ \citep{Temi2004};} $2\times10^4-2\times10^7$ \citep{Savoy2009}. However, both \citet{Vlahakis2005} and \citet{Stickel2007} reported little or no significant difference in the typical dust masses of galaxies of different Hubble types (including spheroidals, spirals and irregulars), although Stickel et al. reported that spheroidal and irregular types reached significantly lower dust masses. One particular issue noted by Vlahakis et al. was the possibility of contamination of their 850\mum\ fluxes with synchrotron emission, which would lead them to overestimate a few of their dust masses. 
We note the caveat that in contrast to our unbiased sample, the references in this paragraph were studies of individual galaxies selected variously with {\it IRAS} or ISOPHOT in the FIR, or SCUBA in the submm, so the range of dust masses sampled would not have been complete {(with the exception of the optically selected sample of \citealp{Popescu2002}).}

{\Citet{Popescu2002} observed colder dust in \emph{later} Hubble types (which might be expected to be the bluest galaxies). Their sample of spirals would probably reside entirely within our blue bin so such a trend would not be apparent between our colour bins if it does not extend beyond the blue cloud. However there is a correlation between Hubble type and mass; later types have lower stellar masses, so our observation of a strong correlation between stellar mass and dust temperature in blue galaxies is entirely consistent with the findings of Popescu et al.
}

{
Meanwhile, results from the {\it Herschel} Reference Survey \citep[HRS;][]{Boselli2010a} indicate that early type galaxies (E+S0+S0a) detected by {\it Herschel} in a volume-limited sample of the local Universe have dust masses in the range $1\times 10^5 - 2\times 10^7\,\text{M}_{\odot}$ \citep{Smith2012} -- although they only detected 34 per cent of ellipticals and 61 per cent of S0's. Their sample have $NUV-r$ colours that place them in our `red' bin, yet their dust masses are much lower than the typical dust masses that we find for red galaxies. This is perhaps due in part to the higher derived temperatures of the HRS sample: $T_\text{dust}=16-32$\,K; with a mean of 24\,K in comparison with our mean of 16.1\,K for the red sample (both assuming $\beta=2$). Smith et al. concluded that the dust masses of S0's were around 10 times lower than those of the HRS spirals, while those of ellipticals were 10 times lower again (for the same stellar mass), which seems to contrast with our results.
}

{\Citet{Rowlands2011} studied early type galaxies detected in \hatlas\ and found dust masses mostly between $2\times10^{7}$ and $2\times10^{8}\,\text{M}_\odot$, with a mean dust mass similar to that of spirals ($5.5\times10^7\,\text{M}_\odot$). However, the stellar mass distributions of spirals and early types were very different. The $NUV-r$ colours of the \citeauthor{Rowlands2011} spiral sample lie mostly within our blue bin, and the early types mostly within our green bin. Their redshift range is similar to ours ($z<0.3$), and the mean dust and stellar masses of their spirals/early types lie within the locus of our blue/green bins in Fig.~\ref{fig:mdust_bymass}. 
However, derived dust masses depend on the temperature assumed. The cold dust temperatures fitted by \citeauthor{Rowlands2011} range from $15$ to $25$\,K, while the temperatures in our blue/green samples for the same stellar mass range ($>10^{10}\,\text{M}_\odot$) are between $15$ and $28$\,K. The correspondence is not exact but the ranges are similar so average dust mass results should be comparable. For reference, changing the temperature from 15 to 25\,K results in a drop in the derived dust mass by a factor 5, which is similar to the range of dust masses across the redshift range in Fig.~\ref{fig:mdust_bymass}.
{ \citet{Rowlands2011} compared their \hatlas-detected early types with a control sample of undetected early-types selected to have a matching distribution of redshifts and $r$-band magnitudes. They} concluded that the detected early types were unusually dusty compared with the control sample, and could be undergoing a transition from the blue cloud to the red sequence. It seems likely that those objects do indeed comprise the top end of the dust mass distribution (at a given stellar mass and redshift), although they do not appear to be exceptional outliers when compared with the median dust masses in our sample. This is surprising when we consider that \citeauthor{Rowlands2011} found the detected early types to have around 10 times as much dust as typical early types: why are they not also outliers compared to typical {\it red} galaxies? The answer {may} be that typical red galaxies are generally dustier than typical early types, { which} supports the notion (as discussed in earlier sections) that our red sample is comprised of a mixture of different populations. The red bin is likely to contain most of the early type galaxies in the sample volume, and if these are relatively dust-poor then there must be a substantial population of red dusty galaxies boosting the median dust masses in the red bin.
{
This could also explain the discrepancy between our red sample and the HRS early types \citep{Smith2012}. 

There is also the possibility of an environmental factor in the offset between the HRS results and our own. Many of the galaxies in the HRS sample reside in the Virgo cluster, while most of the galaxies in GAMA will be in lower density environments. The lower dust masses of HRS early types compared with GAMA red galaxies could be due to early types in clusters being dominated by passive red-sequence systems, while red galaxies in lower density environments are more likely to be dusty. Such a division is indeed suggested by the higher detection rate of early types outside of Virgo in HRS, compared to those inside \citep{Smith2012}. It is however unclear whether such an effect could be strong enough to fully explain the discrepancy we find.

On the other hand, if the lensing contamination is slightly greater than we have predicted, then the derived dust masses of our red sample could be a lot lower; however this is only likely to affect the higher redshift bins due to the weak lensing efficiency at lower redshifts. The \citeauthor{Rowlands2011} sample is less likely to be biased by strong lensing than our sample is, because they excluded detections. The HRS results are unlikely to be biased by lensing because their fluxes were much higher than the red galaxies in our sample, and the low dust masses and high temperatures they derive (relative to our own) argue against their results being biased by lensing. 
}

\subsubsection{Dust-to-stellar mass ratios}

In Fig.~\ref{fig:mdms_bymass} we plot the ratio of dust to stellar mass across the sample, which shows several interesting features. Firstly there is in general a strong correlation with stellar mass: the more massive galaxies have smaller dust-to-stellar mass ratios. The correlation appears steeper for red galaxies of the highest masses in each redshift bin, but this is not so obvious in the blue and green samples which do not reach to quite such high stellar masses. Nevertheless it seems not unreasonable to observe that the most massive red galaxies, many of which will be passively evolving giant ellipticals, have especially low dust-to-stellar mass ratios. 

{ It is worth pointing out the potential for a negative correlation between $M_\text{dust}/M_\text{star}$ and $M_\text{star}$ to be produced artificially in binned data. This can occur if there is a large range of stellar masses with large errors, even when there is no correlation between $M_\text{dust}$ and $M_\text{star}$, since a bin that selects data with low $M_\text{star}$ also selects those with high $M_\text{dust}/M_\text{star}$. In general the slope of the measured correlation could be affected by this artificial phenomenon; however we can be sure in this case that the trends are real because they can also be discerned in the median $M_\text{dust}$ values in Fig.~\ref{fig:mdust_bymass}, and in any case our stellar mass errors are small ($\Delta \log M_\text{star} \sim 0.1$; \citealp{Taylor2011}).}

The { aforementioned} redshift evolution is very apparent in Fig.~\ref{fig:mdms_bymass}, and although we naturally select higher stellar masses at higher redshift, the dust masses in the sample rise more rapidly resulting in an increasing dust-to-stellar mass ratio with redshift. {Using the $(1+z)^\alpha$ model once again we find that the evolution is consistent with the slopes derived for the dust mass evolution.}
This evolution in dust mass echoes the results of D11, who found a strongly evolving dust mass function (DMF) in \hatlas\ sources up to $z\sim0.4$, as well as results from other surveys reaching higher redshifts \citep{Eales2009,Eales2010}. {Using dust masses from SED fitting by \citet{Smith2011b},} \citetalias{Dunne2011} showed that the characteristic dust mass ($M_\text{dust}^\star$) of the \hatlas\ DMF increases from $3.8\times10^7\,\text{M}_\odot$ at $z<0.1$ to around $2.1\times10^8\,\text{M}_\odot$ at $z\sim0.35$ (although they note that this does not measure the true evolution because there is an accompanying fall in the characteristic density $\phi^\star$). This range is indicated by the shaded region in Fig.~\ref{fig:mdust_bymass}.
In all of the bins, our typical dust masses reach lower than the minimum mass sampled in equivalent redshift slices in the \citetalias{Dunne2011} DMF, but the fact that we see evolution indicates that galaxies of a given stellar mass shift up the DMF at increasing redshifts. We see this happening in galaxies of all colours and stellar masses, indicating that the evolution in the DMF is the result of changing dust masses in all galaxies, both passive and star-forming. 
The evolution (around a factor $3-4$) we see is similar to that seen by \citetalias{Dunne2011}, who fitted two-component dust masses temperatures using a detailed physically-motivated SED model. 

It is also interesting to compare $M_\text{dust}/M_\text{star}$ in our results with the detected \hatlas\ galaxies in \citetalias{Dunne2011}. The \hatlas\ galaxies were found to have higher dust-to-stellar mass ratios than predicted by models, ranging from $2\times 10^{-3}$ at $z<0.1$ to $7\times 10^{-3}$ at $z\sim 0.3$ (this range is shaded in Fig.~\ref{fig:mdms_bymass}). Our results are typically lower but the strong dependence on stellar mass means that they span a very wide range; many of the blue galaxies in the lower mass bins have much higher dust/stellar mass ratios than the \hatlas\ sample for the same redshifts. It is perhaps not surprising that we sample a much wider range than the \hatlas\ sources since our sample selection criteria are independent of dust content.
{ The dust/stellar mass results have implications for understanding the dust production mechanism, as we can show by comparing the results from a chemical evolution model with the parameters obtained for the \hatlas\ galaxies (Gomez,~H.\,L. et al. in preparation). The models \citep[based on the framework in][]{Morgan2003} show that values of $M_\text{dust}/M_\text{star}>10^{-3}$ cannot be achieved with a purely stellar source of dust. Even including dust production in supernovae ejecta \citep[e.g.][]{Rho2008,Dunne2009a,Gomez2011a,Matsuura2011}, models require the condensation efficiency in the ejecta to be close to 100 per cent to reach the high values of $M_\text{dust}/M_\text{star}\sim 10^{-2}$ seen both in the \hatlas\ detected sample and in the lowest-mass blue galaxies in our sample. However, as discussed by \citetalias{Dunne2011} and Gomez et al. (in preparation), such high dust yields from supernovae are difficult to produce, leading to the invocation of alternative explanations such as dust grain growth in the ISM \citep{Draine1979, Dwek1980, Draine1990, Edmunds2001} or a top-heavy IMF \citep[e.g.][]{Harayama2008}.
The models also indicate that the low mass galaxies with high $M_\text{dust}/M_\text{star}$ are less efficient at turning their gas into stars (compared to high mass sources with low $M_\text{dust}/M_\text{star}$). In other words, low mass systems have a longer star formation timescale, so although they produce less dust mass per year from stars, there are more metals and dust in the ISM for longer, in comparison to massive galaxies (Gomez et al. in preparation).
}

{
{ Moving on to higher redshifts,} \Citet{Santini2010} showed that the dust/stellar mass ratios of ordinary low redshift galaxies are much lower than those of high-redshift submm galaxies (SMGs) from the \textit{Herschel}-PEP survey. Our $M_\text{dust}/M_\text{star}$ values from stacking are consistent with their sample of low-redshift spirals from the SINGS survey (with typical stellar masses of $\sim10^{11}\text{M}_\odot$). Their dust masses were derived by fitting GRASIL models \citep{Silva1998} to photometry spanning the FIR/submm SED, and are consistent with the value of $\beta=2$ that we have assumed. Our results therefore support the conclusion of \citeauthor{Santini2010} that high-redshift SMGs have much higher dust content (by a factor $\sim30$) than local spiral galaxies. This offset is also consistent with the comparison between low-redshift \hatlas\ sources in \citetalias{Dunne2011} and high-redshift SCUBA SMGs in \citet{Dunne2003}. It has now become clear that the SMGs detected in early submm surveys are exceptionally dusty systems in comparison with the low redshift galaxy population.
}
%

There is evidence that the dust/stellar mass ratio correlates with specific SFR \citep[Gomez,~{ H.\,L.} et al. in preparation]{Cunha2010a,Smith2011b}, so these results imply that the least massive galaxies at low redshift are the most actively star-forming, and that all galaxies become more active towards higher redshifts. At $z\sim0.3$, galaxies with $2\times10^{10}$~M$_{\odot}$ of stars have the same $M_\text{dust}/M_\text{star}$ as galaxies a quarter of that size do at $z\lesssim0.1$. This is consistent with the picture of downsizing, in which the specific star formation rates of high mass galaxies peak earlier in the Universe than those of low-mass galaxies \citep{Cowie1996}. 
}

\begin{figure*}
 \includegraphics[width=\textwidth]{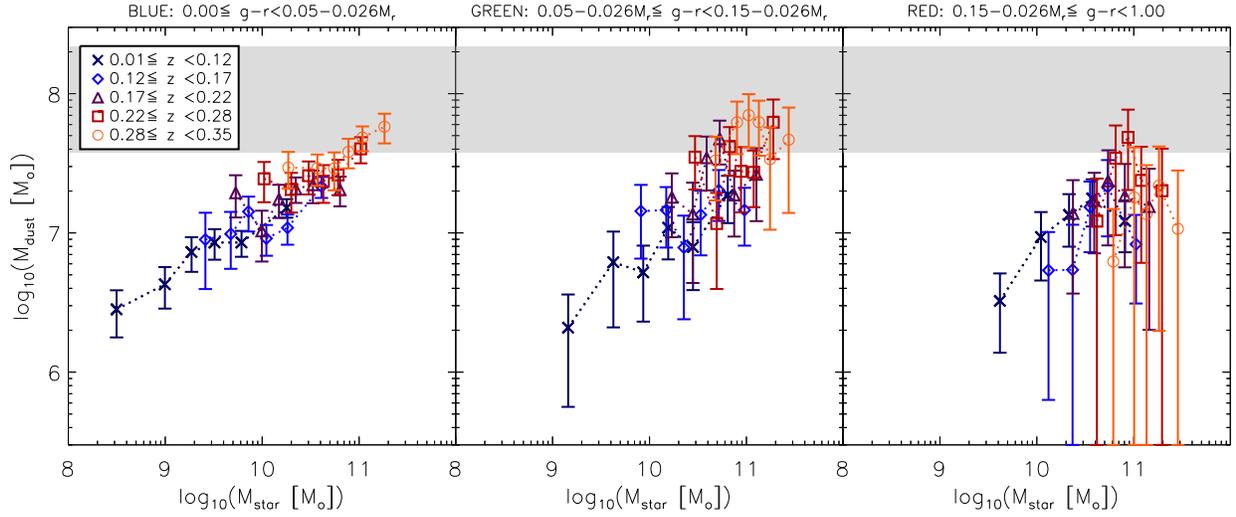}
\caption{Stacked dust mass as a function of $g-r$ colour, redshift and stellar mass. Dust mass is derived from equation~(\ref{eqn:mdust}) using the fitted dust temperatures from Fig.~\ref{fig:fittdust}. Error bars include the statistical $1\sigma$ errors in the bins as described in Section~\ref{sec:errors}, with an additional contribution due to the error on the fitted temperature. { The lensing contribution has been removed from the red bins, and error bars include the associated uncertainty.} The shaded region shows the range of characteristic dust masses measured by \citetalias{Dunne2011}, which evolve from $3.8\times10^7\,\text{M}_\odot$ at $z<0.1$ to $2.2\times10^8\,\text{M}_\odot$ at $z\sim0.35$.}
\label{fig:mdust_bymass}
\end{figure*}

\begin{figure*}
 \includegraphics[width=\textwidth]{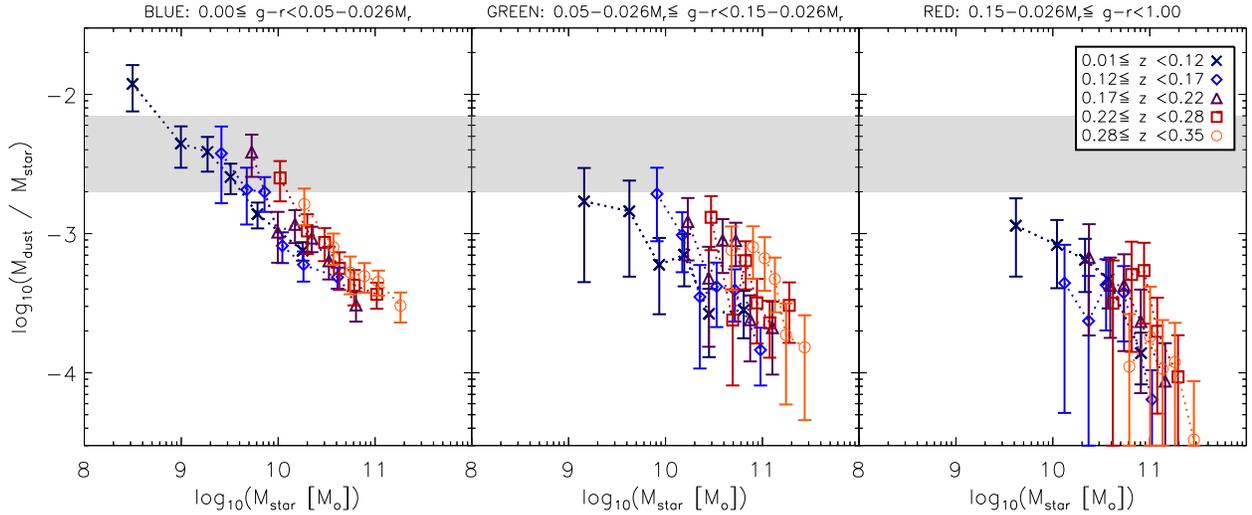}
\caption{Stacked dust mass per unit stellar mass as a function of $g-r$ colour, redshift and stellar mass. Error bars include the statistical $1\sigma$ errors in the bins as described in Section~\ref{sec:errors}, with an additional contribution due to the error on the fitted temperature. The shaded region shows the range from \citetalias{Dunne2011}, from $2\times 10^{-3}$ at $z<0.1$ to $7\times 10^{-3}$ at $z\sim 0.35$.}
\label{fig:mdms_bymass}
\end{figure*}

\subsection{Obscuration}
\label{sec:obsc}
A further step that we can take towards understanding the nature of the galaxies is to investigate the relative luminosities at UV and submm wavelengths, which can give information about the fraction of star-formation that is obscured by dust in the galaxies \citep[e.g.][]{Buat2010, Wijesinghe2011}. The submm luminosity represents the energy absorbed and re-radiated by dust. {Naively, we might expect that this energy originated as UV radiation from young stars, hence the ratio of submm light to UV light detected is directly related to the fraction of UV light which is obscured by dust. If UV light is assumed to come primarily from star-forming regions, this is a measure of the ratio of obscured to unobscured star-formation. However both the UV and the submm radiation could also trace populations unrelated to star-formation: there are open questions as to how much UV radiation can be produced by evolved stars \citep[and references therein]{Chavez2011} as well as how much of the dust probed at submm wavelengths is heated by old stars in the galaxy \citep{Bendo2010,Boselli2010,Law2011}. {Detailed radiative transfer calculations \citep[e.g.][]{Popescu2011}, which have been used to make detailed predictions for a few well-studied spiral galaxies,  can be used in the future to analyse statistical samples of galaxies to address this question in a quantitative way. For the moment, however, the generalisation to galaxy populations as a whole is uncertain.}

In Fig.~\ref{fig:stackedIRUV_bymass} we stack $L_{250}/L_{NUV}$ for the $NUV$-detected sample. Since we require $NUV$ detections for this, we use the $NUV-r$ colour which is likely to be a cleaner colour separation, and we use $L_{250}$ instead of $L_\text{TIR}$ because the monochromatic luminosity is not model-dependent.
Simulations showed that stacking this ratio is robust even for small 250\mum\ fluxes with low signal-to-noise, since the $NUV$ fluxes all have reasonably high signal-to-noise (on the contrary stacking $L_{NUV}/L_{250}$ was found to be unreliable in simulations since this quantity diverges as the 250\mum\ flux approaches zero).
{ As before, we correct 250\mum\ fluxes for the expected contribution from lensing as described in Section~\ref{sec:lens}.
}}

{Some of the results implied by Fig.~\ref{fig:stackedIRUV_bymass} are not trivial to explain, and should be treated with caution since there is a strong bias introduced by the UV selection.}
It appears that the obscuration increases with increasing stellar mass for blue galaxies, but there appears to be a decrease with redshift, at least for stellar masses $\lesssim5\times10^{10}\,\text{M}_\odot$. This contrasts with the increase in 250\mum\ luminosity with redshift, which would imply that while the obscured SFR increases with redshift up to $z=0.3$, the unobscured SFR (UV luminosity) must increase faster for the relative obscuration to fall. {However it is likely that these observations are affected by selection bias: we only detect the low mass galaxies in the UV if they are relatively unobscured, and as redshift increases we detect fewer and fewer of the obscured ones. This effect could cancel out any intrinsic increase in obscuration with redshift, and cause the observed $L_{250}/L_{NUV}$ to decrease.

In contrast we detect almost exactly the opposite trends in the red sample, and in the high mass end of the green sample, which suggests that the selection bias could hide similar trends in blue galaxies (which generally have much lower stellar mass for the same redshift). Number statistics are poor in the red bin because the selection is naturally biased against red galaxies, { and errors are compounded by the uncertainty on the lensing contamination}. Nevertheless the observed trends of increasing obscuration with increasing redshift and with decreasing stellar mass cannot be explained by the selection bias. These trends are both perfectly consistent with the trends in $M_\text{dust}/M_\text{star}$ in Fig.~\ref{fig:mdms_bymass}: a higher amount of dust per stellar mass is almost certain to increase the obscuration of UV light. 
The red galaxy sample therefore {appears to contain} a larger fraction of obscured star-forming galaxies at higher redshifts and lower masses. This is consistent with the findings of \citet{Zhu2010a} in an analysis of the mid-IR colours of optically-selected galaxies at redshifts between 0.1 and 0.5.
%
{ It is also in agreement with \citet{Tojeiro2011}, who stacked SDSS spectra of luminous red galaxies (LRGs), and fitted stellar population models to obtain representative star-formation histories, metallicity and dust content as a function of colour, luminosity and redshift. Their results showed strong correlations of dust extinction with optical luminosity and redshift which are consistent with our own findings. Such agreement between independent measures of the dust extinction is encouraging.
}

In any case we must be careful in the interpretation of $L_{250}/L_{NUV}$ as a tracer of obscuration, in particular due to the potential for $L_{250}$ to be uncorrelated with star formation. We showed in Section~\ref{sec:temps} that the conversion from SPIRE luminosities to $L_\text{TIR}$ is extremely model-dependent, and to plot the ratio $L_\text{TIR}/L_\text{UV}$ using our SED fits for $L_\text{TIR}$ would be misleading when $L_\text{TIR}$ is based only on the SPIRE photometry. In addition, it has been shown by \citet{Wijesinghe2011} that the ratio of $L_\text{TIR}$ (from fitting SEDs to GAMA/\hatlas\ data including PACS and SPIRE) to $L_\text{UV}$ is poorly correlated with other measures of obscuration (the Balmer decrement and UV slope), { probably because they trace a different component of the dust in galaxies.} We therefore hesitate to take this particular analysis any further without the addition of shorter wavelength data to better constrain the IR SED.
}

\begin{figure*}
 \includegraphics[width=\textwidth]{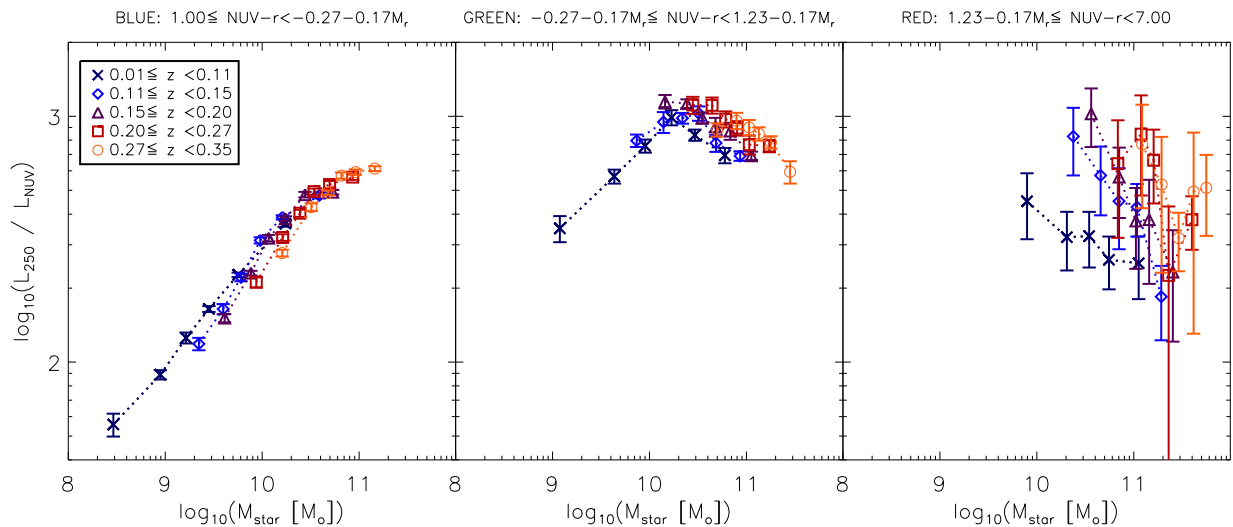}
\caption{Stacked 250\mum/$NUV$ luminosity as a function of $NUV-r$ colour, redshift and stellar mass. This fraction can be used as a proxy for the relative obscuration of star formation, subject to the limitations discussed in the text. Error bars are the statistical $1\sigma$ errors in the bins as described in Section~\ref{sec:errors}. { Data and errors in the red bin incorporate the correction for lensing described in Section~\ref{sec:lens}.}}
\label{fig:stackedIRUV_bymass}
\end{figure*}

\section{Conclusions}
\label{sec:conc}

We have conducted the first submm stacking analysis of a large sample of { about 80,000} galaxies uniformly selected by optical ($r$-band) magnitude. 
We divided the sample by rest-frame colour, absolute magnitude/stellar mass and redshift ($0.01\leq z \leq 0.35$) and stacked into SPIRE maps covering about 126 square degrees at 250, 350 and 500\mum. 
We used a simple (but effective) deblending method to avoid the problem of over-estimating the flux of blended sources when stacking in confused images; this ensures that stacked flux ratios are not biased by the increasing level of confusion at longer wavelengths.
Our main results are summarised below:
\begin{enumerate}
 \item The submm fluxes of all but the most massive optically-selected galaxies are below the $5\sigma$ limits of the \hatlas, yet with the large sample size made possible by the coverage of \hatlas\ and GAMA we are able to probe more than an order of magnitude below these limits using stacking.

 \item We estimate that the total emission from optically-selected galaxies at $r<19.8$ and $z<0.35$ accounts for only $5.0\pm0.4\,\%$ of the cosmic infrared background at 250\mum. At $z<0.28$, where the sample is complete to below $M^\star$, this fraction is $4.2\pm0.3\,\%$. {Of this, roughly 60 per cent originates from blue galaxies, and 20 per cent each from the red and green bins of our sample.} 

{ \item We derive the total $k$- and $e$-corrected luminosity density of the Universe at $z=0$ to be $3.9\pm0.3\times10^{33}$, $1.5\pm0.1\times10^{33}$ and $4.3\pm0.3\times10^{32}$\,W\,Mpc$^{-3}\,h_{70}$ at 250, 350 and 500\mum\ respectively {($h_{70}=H_0$/70\,km\,s$^{-1}$\,Mpc$^{-1}$). }}

{
 \item We show that stacked fluxes of {\it red} galaxies can be significantly contaminated by the lensing of background SMGs. Using models for the lensing amplification distribution and observed lens number counts, we estimate that around 10, 20, and 30 per cent (at 250, 350 and 500\mum\ respectively) of the stacked fluxes is likely to result from lensing. We correct our stacked results for this contamination to red galaxy stacks making reasonable assumptions for the redshift distribution of lensed flux, and include the uncertainty from the lens number counts.
}

 \item We observe a strong dependence of submm luminosity on optical colour ($g-r$) and stellar mass or \Mr, with red galaxies being { up to} an order of magnitude less luminous than blue galaxies of equal stellar mass. {The luminosities of green galaxies are intermediate between the two.} The observed trends of SPIRE luminosities are not strongly dependent on the SED model assumed, { and cannot be explained by lensing, which implies a fundamental difference between the dust emission properties of red and blue galaxies.


 \item We measure cold dust temperatures that vary strongly as a function of stellar mass in blue galaxies, from $\sim11$\,K at $3\times10^8\,\text{M}_\odot$ to $\sim28$\,K at $5\times10^{10}\,\text{M}_\odot$. { Correcting for} the contamination from lensing, red galaxies have dust temperatures $\sim${ 16\,K} at all stellar masses between $3\times10^9-8\times10^{10}\,\text{M}_\odot$ at $z<0.35$. {The dust temperatures of green galaxies appear to have a greater scatter (with mean $T=19.4$\,K) but are not correlated with stellar mass as with the blue; this is indicative of a mixed population.} {Temperature values depend on the assumption of a constant emissivity parameter $\beta=2$. }

 \item { The temperature variation can account for much of the difference in luminosities between red and blue galaxies; however it is not responsible for an increase in luminosity with redshift by a factor around 2 for blue galaxies at a given stellar mass. This appears to be due to an increase in the dust masses of galaxies of all stellar masses by a factor of $3-4$ between $z\sim0$ and $z\sim0.3$. The red sample exhibit a stronger luminosity evolution, for which the likely explanation is also dust mass evolution. Due to the lensing uncertainty we cannot rule out evolution in the temperatures of the red sample, although there is no temperature evolution in the other colour bins.}

 \item { We fit the evolution in $L_{250}$ (which is not dependent on the temperature) with the function $L(z)\propto(1+z)^\alpha$, and obtain indices for the three colour bins at $M_\text{star}=2-7\times10^{10}\,\text{M}_\odot$. We find $\alpha=4.0\pm0.2$ for blue galaxies, $\alpha=1.1\pm0.4$ for green and { $\alpha=$7.7$\pm$1.6} for red {(the larger error on the evolution of red galaxies is due to the uncertainty on the lensing correction)}. Consistent rates of evolution are also derived for the dust masses. { The evolution suggests} a change in the dominant population of red galaxies, from passive systems at low redshift to obscured star-forming systems at higher redshift.}
 }
 
 \item { The redshift evolution of galaxies classified as green seems to indicate a change in the nature of galaxies selected in this way, from a sample dominated by blue-cloud-like galaxies at low redshift and low \Mr\ (or $M_\text{star}$) to a sample more similar to the red bin at the higher redshift and brighter \Mr. 
}

 \item { Deriving TIR luminosities is problematic with only the SPIRE data, and we show that the results obtained depend sensitively on the SED model used (therefore the dust temperature). The low temperatures implied by the SPIRE colours indicate that a cold model such as the \hatlas\ SED fits \citep{Smith2011b} are more appropriate than earlier models based on {\it IRAS} and SCUBA data (CE01).}

 \item The dust-to-stellar mass ratio is strongly anti-correlated with stellar mass, varying by more than an order of magnitude between $M_\text{star} \sim 10^{8} - 10^{11}\,\text{M}_\odot$. This relationship appears to vary little between different optical colours, although it evolves toward higher values with increasing redshift.
{ These results provide a challenge to dust formation models that rely on a purely stellar source of dust, implying a need for dust formation in supernovae and/or the ISM to reach the high dust masses in galaxies at the lower end of the stellar mass function.}

 \item { We attempt to explore the obscuration of galaxies in our sample using the $L_{250}/L_{NUV}$ ratio, and see that red and green galaxies may become more obscured at increasing redshift and decreasing stellar mass (results for the blue galaxies are unclear due to selection bias). This conclusion is dependent on the assumption that this ratio is a good tracer of obscuration, but due to uncertainties in the heating mechanism for cold dust this may not be valid. Nevertheless, such trends in obscuration are consistent with the trends of luminosity and dust/stellar mass.}

\end{enumerate}

{This study is the first of its kind and provides some tantalising glimpses of the characteristics of emission from dust in normal galaxies. Our understanding of the IR SED of optically-selected galaxies and of the obscuration of star formation would be greatly improved by the availability of data covering the peak of the SED. In a future study we hope to stack data from PACS and {\it WISE} in order to make a much more detailed analysis of the full SED.}

\section*{Acknowledgements}
NB wishes to thank Carlos Hoyos for useful discussions regarding the biases discussed in Appendix~\ref{app:sims}, { and Douglas Scott and Dave Clements for providing thoughtful comments on drafts of this paper.}
This publication makes use of Ned Wright's Javascript Cosmology Calculator \citep{Wright2006}, as well as the IDL Astronomy Library \citep{Landsman1993}.}
The \textit{Herschel}-ATLAS is a project with \textit{Herschel}, which is an ESA space observatory with science instruments provided by European-led Principal Investigator consortia and with important participation from NASA. The \hatlas\ website is \url{http://www.h-atlas.org/}.
GAMA is a joint European-Australasian project based around a spectroscopic campaign using the Anglo-Australian Telescope. The GAMA input catalogue is based on data taken from the Sloan Digital Sky Survey and the UKIRT Infrared Deep Sky Survey. Complementary imaging of the GAMA regions is being obtained by a number of independent survey programs including {\it GALEX} MIS, VST KIDS, VISTA VIKING, {\it WISE}, \textit{Herschel}-ATLAS, GMRT and ASKAP providing UV to radio coverage. GAMA is funded by the STFC (UK), the ARC (Australia), the AAO, and the participating institutions. The GAMA website is \url{http://www.gama-survey.org/}.

\bibliographystyle{mn2e}
\bibliography{GH2011b}

\begin{thebibliography}{}

\bibitem[\protect\citeauthoryear{Adelman-McCarthy et~al.}{Adelman-McCarthy
  et~al.}{2008}]{Adelman-McCarthy2008}
Adelman-McCarthy J.~K.,  et~al., 2008, ApJS, 175, 297

\bibitem[\protect\citeauthoryear{{Auger}, {Treu}, {Bolton}, {Gavazzi},
  {Koopmans}, {Marshall}, {Bundy} \& {Moustakas}}{{Auger}
  et~al.}{2009}]{Auger2009}
{Auger} M.~W.,  {Treu} T.,  {Bolton} A.~S.,  {Gavazzi} R.,  {Koopmans}
  L.~V.~E.,  {Marshall} P.~J.,  {Bundy} K.,    {Moustakas} L.~A.,  2009, ApJ,
  705, 1099

\bibitem[\protect\citeauthoryear{Baldry et~al.}{Baldry
  et~al.}{2010}]{Baldry2010}
Baldry I.~K.,  et~al., 2010, MNRAS, 404, 86

\bibitem[\protect\citeauthoryear{Baldry, Glazebrook, Brinkmann, Ivezic, Lupton,
  Nichol \& Szalay}{Baldry et~al.}{2004}]{Baldry2004}
Baldry I.~K.,  Glazebrook K.,  Brinkmann J.,  Ivezic Z.,  Lupton R.~H.,  Nichol
  R.~C.,    Szalay A.~S.,  2004, ApJ, 600, 681

\bibitem[\protect\citeauthoryear{Bell et~al.}{Bell  et~al.}{2004}]{Bell2004}
Bell E.~F.,  et~al., 2004, ApJ, 608, 752

\bibitem[\protect\citeauthoryear{Bendo et~al.}{Bendo
  et~al.}{2010}]{Bendo2010}
Bendo G.~J.,  et~al., 2010, A\&A, 518, L65

\bibitem[\protect\citeauthoryear{{Blain}}{{Blain}}{1996}]{Blain1996}
{Blain} A.~W.,  1996, MNRAS, 283, 1340

\bibitem[\protect\citeauthoryear{Blain, Barnard \& Chapman}{Blain
  et~al.}{2003}]{Blain2003}
Blain A.~W.,  Barnard V.~E.,    Chapman S.~C.,  2003, MNRAS, 338, 733

\bibitem[\protect\citeauthoryear{{Blain}, {Smail}, {Ivison} \& {Kneib}}{{Blain}
  et~al.}{1999}]{Blain1999}
{Blain} A.~W.,  {Smail} I.,  {Ivison} R.~J.,    {Kneib} J.,  1999, MNRAS, 302,
  632

\bibitem[\protect\citeauthoryear{Blanton et~al.}{Blanton
  et~al.}{2003}]{Blanton2003}
Blanton M.~R.,  et~al., 2003, ApJ, 594, 186

\bibitem[\protect\citeauthoryear{Blanton \& Roweis}{Blanton \&
  Roweis}{2007}]{Blanton2007}
Blanton M.~R.,  Roweis S.,  2007, AJ, 133, 734

\bibitem[\protect\citeauthoryear{Boselli et~al.}{Boselli
  et~al.}{2010a}]{Boselli2010}
Boselli A.,  et~al., 2010a, A\&A, 518, L61

\bibitem[\protect\citeauthoryear{Boselli et~al.}{Boselli
  et~al.}{2010b}]{Boselli2010a}
Boselli A.,  et~al., 2010b, PASP, 122, 261

\bibitem[\protect\citeauthoryear{Bourne, Dunne, Ivison, Maddox, Dickinson \&
  Frayer}{Bourne et~al.}{2011}]{Bourne2011}
Bourne N.,  Dunne L.,  Ivison R.~J.,  Maddox S.~J.,  Dickinson M.,    Frayer
  D.~T.,  2011, MNRAS, 410, 1155

\bibitem[\protect\citeauthoryear{{Browne} et~al.}{{Browne}
  et~al.}{2003}]{Browne2003}
{Browne} I.~W.~A.,  et~al., 2003, MNRAS, 341, 13

\bibitem[\protect\citeauthoryear{Bruzual \& Charlot}{Bruzual \&
  Charlot}{2003}]{Bruzual2003}
Bruzual G.,  Charlot S.,  2003, MNRAS, 344, 1000

\bibitem[\protect\citeauthoryear{Buat et~al.}{Buat  et~al.}{2010}]{Buat2010}
Buat V.,  et~al., 2010, MNRAS, 409, L1

\bibitem[\protect\citeauthoryear{Béthermin, Dole, Cousin \&
  Bavouzet}{Béthermin et~al.}{2010}]{Bethermin2010}
Béthermin M.,  Dole H.,  Cousin M.,    Bavouzet N.,  2010, A\&A, 516, A43

\bibitem[\protect\citeauthoryear{{Cameron}}{{Cameron}}{2011}]{Cameron2011}
{Cameron} E.,  2011, PASA, 28, 128

\bibitem[\protect\citeauthoryear{Chabrier}{Chabrier}{2003}]{Chabrier2003}
Chabrier G.,  2003, PASP, 115, 763

\bibitem[\protect\citeauthoryear{Chary \& Elbaz}{Chary \&
  Elbaz}{2001}]{Chary2001}
Chary R.,  Elbaz D.,  2001, ApJ, 556, 562

\bibitem[\protect\citeauthoryear{Chavez \& Bertone}{Chavez \&
  Bertone}{2011}]{Chavez2011}
Chavez M.,  Bertone E.,  2011, Ap\&SS, p.~269

\bibitem[\protect\citeauthoryear{Connolly et~al.}{Connolly
  et~al.}{2002}]{Connolly2002}
Connolly A.~J.,  et~al., 2002, ApJ, 579, 42

\bibitem[\protect\citeauthoryear{Cortese et~al.}{Cortese
  et~al.}{2006}]{Cortese2006}
Cortese L.,  et~al., 2006, ApJ, 637, 242

\bibitem[\protect\citeauthoryear{Cortese \& Hughes}{Cortese \&
  Hughes}{2009}]{Cortese2009}
Cortese L.,  Hughes T.~M.,  2009, MNRAS, 400, 1225

\bibitem[\protect\citeauthoryear{Cowie, Songaila, Hu \& Cohen}{Cowie
  et~al.}{1996}]{Cowie1996}
Cowie L.~L.,  Songaila A.,  Hu E.~M.,    Cohen J.~G.,  1996, AJ, 112, 839

\bibitem[\protect\citeauthoryear{da Cunha, Eminian, Charlot \&
  Blaizot}{da~Cunha et~al.}{2010}]{Cunha2010a}
da Cunha E.,  Eminian C.,  Charlot S.,    Blaizot J.,  2010, MNRAS, 403, 1894

\bibitem[\protect\citeauthoryear{Dale \& Helou}{Dale \& Helou}{2002}]{Dale2002}
Dale D.~A.,  Helou G.,  2002, ApJ, 576, 159

\bibitem[\protect\citeauthoryear{Damen et~al.}{Damen
  et~al.}{2009a}]{Damen2009}
Damen M.,  et~al., 2009a, ApJ, 690, 937

\bibitem[\protect\citeauthoryear{Damen et~al.}{Damen
  et~al.}{2009b}]{Damen2009a}
Damen M.,  et~al., 2009b, ApJ, 705, 617

\bibitem[\protect\citeauthoryear{Dariush et~al.}{Dariush
  et~al.}{2011}]{Dariush2011}
Dariush A.,  et~al., 2011, MNRAS accepted (arXiv:1106.6195)

\bibitem[\protect\citeauthoryear{Devlin et~al.}{Devlin
  et~al.}{2009}]{Devlin2009}
Devlin M.~J.,  et~al., 2009, Nat, 458, 737

\bibitem[\protect\citeauthoryear{Dole et~al.}{Dole  et~al.}{2006}]{Dole2006}
Dole H.,  et~al., 2006, A\&A, 451, 417

\bibitem[\protect\citeauthoryear{{Draine}}{{Draine}}{1990}]{Draine1990}
{Draine} B.~T.,  1990, in {L.~Blitz} ed., The Evolution of the Interstellar
  Medium Vol.~12 of Astronomical Society of the Pacific Conference Series,
  {Evolution of interstellar dust}.
pp 193--205

\bibitem[\protect\citeauthoryear{Draine et~al.}{Draine
  et~al.}{2007}]{Draine2007}
Draine B.~T.,  et~al., 2007, ApJ, 663, 866

\bibitem[\protect\citeauthoryear{{Draine} \& {Salpeter}}{{Draine} \&
  {Salpeter}}{1979}]{Draine1979}
{Draine} B.~T.,  {Salpeter} E.~E.,  1979, ApJ, 231, 438

\bibitem[\protect\citeauthoryear{{Driver} et~al.}{{Driver}
  et~al.}{2006}]{Driver2006}
{Driver} S.~P.,  et~al., 2006, MNRAS, 368, 414

\bibitem[\protect\citeauthoryear{{Driver} et~al.}{{Driver}
  et~al.}{2009}]{Driver2009}
{Driver} S.~P.,  et~al., 2009, Astron. Geophys., 50, 12

\bibitem[\protect\citeauthoryear{Driver et~al.}{Driver
  et~al.}{2011}]{Driver2010}
Driver S.~P.,  et~al., 2011, MNRAS, 413, 971

\bibitem[\protect\citeauthoryear{{Driver}, {Popescu}, {Tuffs}, {Liske},
  {Graham}, {Allen} \& {de Propris}}{{Driver} et~al.}{2007}]{Driver2007}
{Driver} S.~P.,  {Popescu} C.~C.,  {Tuffs} R.~J.,  {Liske} J.,  {Graham} A.~W.,
   {Allen} P.~D.,    {de Propris} R.,  2007, MNRAS, 379, 1022

\bibitem[\protect\citeauthoryear{Dunne \& Eales}{Dunne \&
  Eales}{2001}]{Dunne2001}
Dunne L.,  Eales S.~A.,  2001, MNRAS, 327, 697

\bibitem[\protect\citeauthoryear{{Dunne}, {Eales} \& {Edmunds}}{{Dunne}
  et~al.}{2003}]{Dunne2003}
{Dunne} L.,  {Eales} S.~A.,    {Edmunds} M.~G.,  2003, MNRAS, 341, 589

\bibitem[\protect\citeauthoryear{Dunne et~al.}{Dunne
  et~al.}{2009a}]{Dunne2009}
Dunne L.,  et~al., 2009a, MNRAS, 394, 3

\bibitem[\protect\citeauthoryear{Dunne et~al.}{Dunne
  et~al.}{2009b}]{Dunne2009a}
Dunne L.,  et~al., 2009b, MNRAS, 394, 1307

\bibitem[\protect\citeauthoryear{Dunne et~al.}{Dunne
  et~al.}{2011}]{Dunne2011}
Dunne L.,  et~al., 2011, MNRAS, 417, 1510

\bibitem[\protect\citeauthoryear{{Dwek} \& {Scalo}}{{Dwek} \&
  {Scalo}}{1980}]{Dwek1980}
{Dwek} E.,  {Scalo} J.~M.,  1980, ApJ, 239, 193

\bibitem[\protect\citeauthoryear{Dye et~al.}{Dye  et~al.}{2009}]{Dye2009}
Dye S.,  et~al., 2009, ApJ, 703, 285

\bibitem[\protect\citeauthoryear{Dye et~al.}{Dye  et~al.}{2010}]{Dye2010}
Dye S.,  et~al., 2010, A\&A, 518, L10

\bibitem[\protect\citeauthoryear{Eales et~al.}{Eales
  et~al.}{2009}]{Eales2009}
Eales S.,  et~al., 2009, ApJ, 707, 1779

\bibitem[\protect\citeauthoryear{Eales et~al.}{Eales
  et~al.}{2010a}]{Eales2010}
Eales S.,  et~al., 2010a, A\&A, 518, L23

\bibitem[\protect\citeauthoryear{Eales et~al.}{Eales
  et~al.}{2010b}]{Eales2010a}
Eales S.,  et~al., 2010b, PASP, 122, 499

\bibitem[\protect\citeauthoryear{{Edmunds}}{{Edmunds}}{2001}]{Edmunds2001}
{Edmunds} M.~G.,  2001, MNRAS, 328, 223

\bibitem[\protect\citeauthoryear{Faber et~al.}{Faber
  et~al.}{2007}]{Faber2007}
Faber S.~M.,  et~al., 2007, ApJ, 665, 265

\bibitem[\protect\citeauthoryear{{Faure} et~al.}{{Faure}
  et~al.}{2008}]{Faure2008}
{Faure} C.,  et~al., 2008, ApJS, 176, 19

\bibitem[\protect\citeauthoryear{{Fixsen}, {Dwek}, {Mather}, {Bennett} \&
  {Shafer}}{{Fixsen} et~al.}{1998}]{Fixsen1998}
{Fixsen} D.~J.,  {Dwek} E.,  {Mather} J.~C.,  {Bennett} C.~L.,    {Shafer}
  R.~A.,  1998, ApJ, 508, 123

\bibitem[\protect\citeauthoryear{Franceschini, Aussel, Cesarsky, Elbaz \&
  Fadda}{Franceschini et~al.}{2001}]{Franceschini2001}
Franceschini A.,  Aussel H.,  Cesarsky C.~J.,  Elbaz D.,    Fadda D.,  2001,
  A\&A, 378, 1

\bibitem[\protect\citeauthoryear{Glenn et~al.}{Glenn
  et~al.}{2010}]{Glenn2010}
Glenn J.,  et~al., 2010, MNRAS, 409, 109

\bibitem[\protect\citeauthoryear{Gomez et~al.}{Gomez
  et~al.}{2011}]{Gomez2011a}
Gomez H.~L.,  et~al., 2011, MNRAS accepted (arXiv:1111.6627)

\bibitem[\protect\citeauthoryear{González-Nuevo et~al.}{González-Nuevo
  et~al.}{2012}]{Gonzalez-Nuevo2011}
González-Nuevo J.,  et~al., 2012, ApJ in press (arXiv:1202.0402)

\bibitem[\protect\citeauthoryear{Goto et~al.}{Goto  et~al.}{2011}]{Goto2011}
Goto T.,  et~al., 2011, MNRAS, 414, 1903

\bibitem[\protect\citeauthoryear{Gott, Vogeley, Podariu \& Ratra}{Gott
  et~al.}{2001}]{Gott2001}
Gott J.~R.,  Vogeley M.~S.,  Podariu S.,    Ratra B.,  2001, ApJ, 549, 1

\bibitem[\protect\citeauthoryear{Greve et~al.}{Greve
  et~al.}{2010}]{Greve2009}
Greve T.~R.,  et~al., 2010, ApJ, 719, 483

\bibitem[\protect\citeauthoryear{Griffin et~al.}{Griffin
  et~al.}{2010}]{Griffin2010}
Griffin M.~J.,  et~al., 2010, A\&A, 518, L3

\bibitem[\protect\citeauthoryear{Gruppioni et~al.}{Gruppioni
  et~al.}{2010}]{Gruppioni2010}
Gruppioni C.,  et~al., 2010, A\&A, 518, L27

\bibitem[\protect\citeauthoryear{{Harayama}, {Eisenhauer} \&
  {Martins}}{{Harayama} et~al.}{2008}]{Harayama2008}
{Harayama} Y.,  {Eisenhauer} F.,    {Martins} F.,  2008, ApJ, 675, 1319

\bibitem[\protect\citeauthoryear{Helou}{Helou}{1986}]{Helou1986}
Helou G.,  1986, ApJ, 311, L33

\bibitem[\protect\citeauthoryear{{Hill} et~al.}{{Hill}
  et~al.}{2011}]{Hill2011}
{Hill} D.~T.,  et~al., 2011, MNRAS, 412, 765

\bibitem[\protect\citeauthoryear{{Hill}, {Thompson}, {Burton}, {Walsh},
  {Minier}, {Cunningham} \& {Pierce-Price}}{{Hill} et~al.}{2006}]{Hill2006}
{Hill} T.,  {Thompson} M.~A.,  {Burton} M.~G.,  {Walsh} A.~J.,  {Minier} V.,
  {Cunningham} M.~R.,    {Pierce-Price} D.,  2006, MNRAS, 368, 1223

\bibitem[\protect\citeauthoryear{{Holwerda}, {Keel}, {Williams}, {Dalcanton} \&
  {de Jong}}{{Holwerda} et~al.}{2009}]{Holwerda2009}
{Holwerda} B.~W.,  {Keel} W.~C.,  {Williams} B.,  {Dalcanton} J.~J.,    {de
  Jong} R.~S.,  2009, AJ, 137, 3000

\bibitem[\protect\citeauthoryear{Hopkins, Younger, Hayward, Narayanan \&
  Hernquist}{Hopkins et~al.}{2010}]{Hopkins2010}
Hopkins P.~F.,  Younger J.~D.,  Hayward C.~C.,  Narayanan D.,    Hernquist L.,
  2010, MNRAS, 402, 1693

\bibitem[\protect\citeauthoryear{{Hu}}{{Hu}}{1999}]{Hu1999}
{Hu} W.,  1999, ApJ, 522, L21

\bibitem[\protect\citeauthoryear{Hwang et~al.}{Hwang
  et~al.}{2010}]{Hwang2010}
Hwang H.~S.,  et~al., 2010, MNRAS, 409, 75

\bibitem[\protect\citeauthoryear{Irwin}{Irwin}{2010}]{Irwin2010a}
Irwin M.,  2010, UKIRT Newsletter, 26, 14

\bibitem[\protect\citeauthoryear{James, Dunne, Eales \& Edmunds}{James
  et~al.}{2002}]{James2002}
James A.,  Dunne L.,  Eales S.,    Edmunds M.~G.,  2002, MNRAS, 335, 753

\bibitem[\protect\citeauthoryear{Karim et~al.}{Karim
  et~al.}{2011}]{Karim2010}
Karim A.,  et~al., 2011, ApJ, 730, 61

\bibitem[\protect\citeauthoryear{{Kauffmann} et~al.}{{Kauffmann}
  et~al.}{2003}]{Kauffman2003}
{Kauffmann} G.,  et~al., 2003, MNRAS, 341, 33

\bibitem[\protect\citeauthoryear{Kennicutt}{Kennicutt}{1998}]{Kennicutt1998}
Kennicutt R.~C.,  1998, ARA\&A, 36, 189

\bibitem[\protect\citeauthoryear{Kurczynski \& Gawiser}{Kurczynski \&
  Gawiser}{2010}]{Kurczynski2010}
Kurczynski P.,  Gawiser E.,  2010, AJ, 139, 1592

\bibitem[\protect\citeauthoryear{{Landsman}}{{Landsman}}{1993}]{Landsman1993}
{Landsman} W.~B.,  1993, in {R.~J.~Hanisch, R.~J.~V.~Brissenden, \& J.~Barnes}
  ed., Astronomical Data Analysis Software and Systems II Vol.~52 of
  Astronomical Society of the Pacific Conference Series, {The IDL Astronomy
  User's Library}.
p.~246

\bibitem[\protect\citeauthoryear{Landy \& Szalay}{Landy \&
  Szalay}{1993}]{Landy1993}
Landy S.~D.,  Szalay A.~S.,  1993, ApJ, 412, 64

\bibitem[\protect\citeauthoryear{{Lapi} et~al.}{{Lapi}
  et~al.}{2011}]{Lapi2011}
{Lapi} A.,  et~al., 2011, ApJ, 742, 24

\bibitem[\protect\citeauthoryear{{Law}, {Gordon} \& {Misselt}}{{Law}
  et~al.}{2011}]{Law2011}
{Law} K.-H.,  {Gordon} K.~D.,    {Misselt} K.~A.,  2011, ApJ accepted

\bibitem[\protect\citeauthoryear{Lawrence et~al.}{Lawrence
  et~al.}{2007}]{Lawrence2007}
Lawrence A.,  et~al., 2007, MNRAS, 379, 1599

\bibitem[\protect\citeauthoryear{Le~Floc'h et~al.}{Le~Floc'h
  et~al.}{2005}]{LeFloc'h2005}
Le~Floc'h E.,  et~al., 2005, ApJ, 632, 169

\bibitem[\protect\citeauthoryear{{Leeuw}, {Sansom}, {Robson}, {Haas} \&
  {Kuno}}{{Leeuw} et~al.}{2004}]{Leeuw2004}
{Leeuw} L.~L.,  {Sansom} A.~E.,  {Robson} E.~I.,  {Haas} M.,    {Kuno} N.,
  2004, ApJ, 612, 837

\bibitem[\protect\citeauthoryear{Lilly, Le~Fevre, Hammer \& Crampton}{Lilly
  et~al.}{1996}]{Lilly1996}
Lilly S.~J.,  Le~Fevre O.,  Hammer F.,    Crampton D.,  1996, ApJ, 460, L1

\bibitem[\protect\citeauthoryear{Madau, Ferguson, Dickinson, Giavalisco,
  Steidel \& Fruchter}{Madau et~al.}{1996}]{Madau1996}
Madau P.,  Ferguson H.~C.,  Dickinson M.~E.,  Giavalisco M.,  Steidel C.~C.,
  Fruchter A.,  1996, MNRAS, 283, 1388

\bibitem[\protect\citeauthoryear{Magnelli, Elbaz, Chary, Dickinson, Le~Borgne,
  Frayer \& Willmer}{Magnelli et~al.}{2009}]{Magnelli2009}
Magnelli B.,  Elbaz D.,  Chary R.~R.,  Dickinson M.,  Le~Borgne D.,  Frayer
  D.~T.,    Willmer C. N.~A.,  2009, A\&A, 496, 57

\bibitem[\protect\citeauthoryear{Marsden et~al.}{Marsden
  et~al.}{2009}]{Marsden2009}
Marsden G.,  et~al., 2009, ApJ, 707, 1729

\bibitem[\protect\citeauthoryear{Martin et~al.}{Martin
  et~al.}{2007}]{Martin2007}
Martin D.~C.,  et~al., 2007, ApJS, 173, 342

\bibitem[\protect\citeauthoryear{{Matsuura} et~al.}{{Matsuura}
  et~al.}{2011}]{Matsuura2011}
{Matsuura} M.,  et~al., 2011, Science, 333, 1258

\bibitem[\protect\citeauthoryear{{Morgan} \& {Edmunds}}{{Morgan} \&
  {Edmunds}}{2003}]{Morgan2003}
{Morgan} H.~L.,  {Edmunds} M.~G.,  2003, MNRAS, 343, 427

\bibitem[\protect\citeauthoryear{{Morrissey} et~al.}{{Morrissey}
  et~al.}{2005}]{Morrissey2005}
{Morrissey} P.,  et~al., 2005, ApJ, 619, L7

\bibitem[\protect\citeauthoryear{Negrello et~al.}{Negrello
  et~al.}{2010}]{Negrello2010}
Negrello M.,  et~al., 2010, Science, 330, 800

\bibitem[\protect\citeauthoryear{{Negrello}, {Gonz{\'a}lez-Nuevo},
  {Magliocchetti}, {Moscardini}, {De Zotti}, {Toffolatti} \&
  {Danese}}{{Negrello} et~al.}{2005}]{Negrello2005}
{Negrello} M.,  {Gonz{\'a}lez-Nuevo} J.,  {Magliocchetti} M.,  {Moscardini} L.,
   {De Zotti} G.,  {Toffolatti} L.,    {Danese} L.,  2005, MNRAS, 358, 869

\bibitem[\protect\citeauthoryear{{Negrello}, {Perrotta}, {Gonz{\'a}lez-Nuevo},
  {Silva}, {de Zotti}, {Granato}, {Baccigalupi} \& {Danese}}{{Negrello}
  et~al.}{2007}]{Negrello2007}
{Negrello} M.,  {Perrotta} F.,  {Gonz{\'a}lez-Nuevo} J.,  {Silva} L.,  {de
  Zotti} G.,  {Granato} G.~L.,  {Baccigalupi} C.,    {Danese} L.,  2007, MNRAS,
  377, 1557

\bibitem[\protect\citeauthoryear{{Oguri} et~al.}{{Oguri}
  et~al.}{2006}]{Oguri2006}
{Oguri} M.,  et~al., 2006, AJ, 132, 999

\bibitem[\protect\citeauthoryear{Oliver et~al.}{Oliver
  et~al.}{2010a}]{Oliver2010}
Oliver S.~J.,  et~al., 2010a, A\&A, 518, L210

\bibitem[\protect\citeauthoryear{Oliver et~al.}{Oliver
  et~al.}{2010b}]{Oliver2010a}
Oliver S.~J.,  et~al., 2010b, MNRAS, 405, 2279

\bibitem[\protect\citeauthoryear{Pannella et~al.}{Pannella
  et~al.}{2009}]{Pannella2009}
Pannella M.,  et~al., 2009, ApJ, 698, L116

\bibitem[\protect\citeauthoryear{Paradis, Bernard \& Mény}{Paradis
  et~al.}{2009}]{Paradis2009}
Paradis D.,  Bernard J.-P.,    Mény C.,  2009, A\&A, 506, 745

\bibitem[\protect\citeauthoryear{Pascale et~al.}{Pascale
  et~al.}{2011}]{Pascale2010}
Pascale E.,  et~al., 2011, MNRAS, 415, 911

\bibitem[\protect\citeauthoryear{Pilbratt et~al.}{Pilbratt
  et~al.}{2010}]{Pilbratt2010}
Pilbratt G.~L.,  et~al., 2010, A\&A, 518, L1

\bibitem[\protect\citeauthoryear{{Planck Collaboration} et~al.}{{Planck
  Collaboration}  et~al.}{2011a}]{PlanckCollaboration2011b}
{Planck Collaboration} et~al., 2011a, arXiv:1101.2032

\bibitem[\protect\citeauthoryear{{Planck Collaboration} et~al.}{{Planck
  Collaboration}  et~al.}{2011b}]{PlanckCollaboration2011a}
{Planck Collaboration} et~al., 2011b, arXiv:1101.2045

\bibitem[\protect\citeauthoryear{Poglitsch et~al.}{Poglitsch
  et~al.}{2010}]{Poglitsch2010}
Poglitsch A.,  et~al., 2010, A\&A, 518, L2

\bibitem[\protect\citeauthoryear{Popescu, Tuffs, Dopita, Fischera, Kylafis \&
  Madore}{Popescu et~al.}{2011}]{Popescu2011}
Popescu C.~C.,  Tuffs R.~J.,  Dopita M.~A.,  Fischera J.,  Kylafis N.~D.,
  Madore B.~F.,  2011, A\&A, 527, 109

\bibitem[\protect\citeauthoryear{Popescu, Tuffs, Völk, Pierini \&
  Madore}{Popescu et~al.}{2002}]{Popescu2002}
Popescu C.~C.,  Tuffs R.~J.,  Völk H.~J.,  Pierini D.,    Madore B.~F.,  2002,
  ApJ, 567, 221

\bibitem[\protect\citeauthoryear{{Pozzi} et~al.}{{Pozzi}
  et~al.}{2004}]{Pozzi2004}
{Pozzi} F.,  et~al., 2004, ApJ, 609, 122

\bibitem[\protect\citeauthoryear{{Puget}, {Abergel}, {Bernard}, {Boulanger},
  {Burton}, {Desert} \& {Hartmann}}{{Puget} et~al.}{1996}]{Puget1996}
{Puget} J.-L.,  {Abergel} A.,  {Bernard} J.-P.,  {Boulanger} F.,  {Burton}
  W.~B.,  {Desert} F.-X.,    {Hartmann} D.,  1996, A\&A, 308, L5

\bibitem[\protect\citeauthoryear{{Rho} et~al.}{{Rho}  et~al.}{2008}]{Rho2008}
{Rho} J.,  et~al., 2008, ApJ, 673, 271

\bibitem[\protect\citeauthoryear{Rigby et~al.}{Rigby
  et~al.}{2011}]{Rigby2010}
Rigby E.~E.,  et~al., 2011, MNRAS, 415, 2336

\bibitem[\protect\citeauthoryear{Rodighiero et~al.}{Rodighiero
  et~al.}{2010}]{Rodighiero2009}
Rodighiero G.,  et~al., 2010, A\&A, 515, 8

\bibitem[\protect\citeauthoryear{Rowan-Robinson et~al.}{Rowan-Robinson
  et~al.}{2008}]{Rowan-Robinson2008}
Rowan-Robinson M.,  et~al., 2008, MNRAS, 386, 697

\bibitem[\protect\citeauthoryear{Rowlands et~al.}{Rowlands
  et~al.}{2012}]{Rowlands2011}
Rowlands K.,  et~al., 2012, MNRAS, 419, 2545

\bibitem[\protect\citeauthoryear{Santini et~al.}{Santini
  et~al.}{2010}]{Santini2010}
Santini P.,  et~al., 2010, A\&A, 518, L154+

\bibitem[\protect\citeauthoryear{{Saunders}, {Rowan-Robinson}, {Lawrence},
  {Efstathiou}, {Kaiser}, {Ellis} \& {Frenk}}{{Saunders}
  et~al.}{1990}]{Saunders1990}
{Saunders} W.,  {Rowan-Robinson} M.,  {Lawrence} A.,  {Efstathiou} G.,
  {Kaiser} N.,  {Ellis} R.~S.,    {Frenk} C.~S.,  1990, MNRAS, 242, 318

\bibitem[\protect\citeauthoryear{{Sauvage}, {Tuffs} \& {Popescu}}{{Sauvage}
  et~al.}{2005}]{Sauvage2005}
{Sauvage} M.,  {Tuffs} R.~J.,    {Popescu} C.~C.,  2005, Space Science Reviews,
  119, 313

\bibitem[\protect\citeauthoryear{Savoy, Welch \& Fich}{Savoy
  et~al.}{2009}]{Savoy2009}
Savoy J.,  Welch G.~A.,    Fich M.,  2009, ApJ, 706, 21

\bibitem[\protect\citeauthoryear{Schiminovich et~al.}{Schiminovich
  et~al.}{2007}]{Schiminovich2007}
Schiminovich D.,  et~al., 2007, ApJS, 173, 315

\bibitem[\protect\citeauthoryear{Schlegel, Finkbeiner \& Davis}{Schlegel
  et~al.}{1998}]{Schlegel1998}
Schlegel D.~J.,  Finkbeiner D.~P.,    Davis M.,  1998, ApJ, 500, 525

\bibitem[\protect\citeauthoryear{{Schmidt}}{{Schmidt}}{1968}]{Schmidt1968}
{Schmidt} M.,  1968, ApJ, 151, 393

\bibitem[\protect\citeauthoryear{{Sedgwick} et~al.}{{Sedgwick}
  et~al.}{2011}]{Sedgwick2011}
{Sedgwick} C.,  et~al., 2011, MNRAS, 416, 1862

\bibitem[\protect\citeauthoryear{{Serjeant}}{{Serjeant}}{2010}]{Serjeant2010a}
{Serjeant} S.,  2010, in {V.~P.~Debattista \& C.~C.~Popescu} ed., American
  Institute of Physics Conference Series Vol.~1240 of American Institute of
  Physics Conference Series, {The Dark and Dusty Side of Galaxy Evolution}.
pp 29--32

\bibitem[\protect\citeauthoryear{Serjeant et~al.}{Serjeant
  et~al.}{2008}]{Serjeant2008}
Serjeant S.,  et~al., 2008, MNRAS, 386, 1907

\bibitem[\protect\citeauthoryear{{Serjeant} \& {Harrison}}{{Serjeant} \&
  {Harrison}}{2005}]{Serjeant2005}
{Serjeant} S.,  {Harrison} D.,  2005, MNRAS, 356, 192

\bibitem[\protect\citeauthoryear{Silva, Granato, Bressan \& Danese}{Silva
  et~al.}{1998}]{Silva1998}
Silva L.,  Granato G.~L.,  Bressan A.,    Danese L.,  1998, ApJ, 509, 103

\bibitem[\protect\citeauthoryear{Smith et~al.}{Smith
  et~al.}{2011a}]{Smith2011a}
Smith D. J.~B.,  et~al., 2011a, MNRAS, 416, 857

\bibitem[\protect\citeauthoryear{Smith et~al.}{Smith
  et~al.}{2011b}]{Smith2011b}
Smith D. J.~B.,  et~al., 2011b, MNRAS submitted

\bibitem[\protect\citeauthoryear{Smith et~al.}{Smith
  et~al.}{2011c}]{Smith2012}
Smith M. W.~L.,  et~al., 2011c, ApJ submitted

\bibitem[\protect\citeauthoryear{{Stetson}}{{Stetson}}{1987}]{Stetson1987}
{Stetson} P.~B.,  1987, PASP, 99, 191

\bibitem[\protect\citeauthoryear{Stevens, Amure \& Gear}{Stevens
  et~al.}{2005}]{Stevens2005}
Stevens J.~A.,  Amure M.,    Gear W.~K.,  2005, MNRAS, 357, 361

\bibitem[\protect\citeauthoryear{Stickel, Klaas \& Lemke}{Stickel
  et~al.}{2007}]{Stickel2007}
Stickel M.,  Klaas U.,    Lemke D.,  2007, A\&A, 466, 831

\bibitem[\protect\citeauthoryear{{Strateva} et~al.}{{Strateva}
  et~al.}{2001}]{Strateva2001}
{Strateva} I.,  et~al., 2001, AJ, 122, 1861

\bibitem[\protect\citeauthoryear{Takagi et~al.}{Takagi
  et~al.}{2007}]{Takagi2007}
Takagi T.,  et~al., 2007, MNRAS, 381, 1154

\bibitem[\protect\citeauthoryear{Taylor et~al.}{Taylor
  et~al.}{2011}]{Taylor2011}
Taylor E.~N.,  et~al., 2011, MNRAS, p.~1907

\bibitem[\protect\citeauthoryear{Temi, Brighenti \& Mathews}{Temi
  et~al.}{2009}]{Temi2009}
Temi P.,  Brighenti F.,    Mathews W.~G.,  2009, ApJ, 707, 890

\bibitem[\protect\citeauthoryear{{Temi}, {Brighenti}, {Mathews} \&
  {Bregman}}{{Temi} et~al.}{2004}]{Temi2004}
{Temi} P.,  {Brighenti} F.,  {Mathews} W.~G.,    {Bregman} J.~D.,  2004, ApJS,
  151, 237

\bibitem[\protect\citeauthoryear{Tojeiro, Percival, Heavens \& Jimenez}{Tojeiro
  et~al.}{2011}]{Tojeiro2011}
Tojeiro R.,  Percival W.~J.,  Heavens A.~F.,    Jimenez R.,  2011, MNRAS, 413,
  434

\bibitem[\protect\citeauthoryear{{Tresse}, {Maddox}, {Loveday} \&
  {Singleton}}{{Tresse} et~al.}{1999}]{Tresse1999}
{Tresse} L.,  {Maddox} S.,  {Loveday} J.,    {Singleton} C.,  1999, MNRAS, 310,
  262

\bibitem[\protect\citeauthoryear{{Tuffs} et~al.}{{Tuffs}
  et~al.}{2002}]{Tuffs2002}
{Tuffs} R.~J.,  et~al., 2002, ApJS, 139, 37

\bibitem[\protect\citeauthoryear{Viero et~al.}{Viero
  et~al.}{2010}]{Viero2010}
Viero M.~P.,  et~al., 2010, MNRAS in press (arXiv:1008.4359)

\bibitem[\protect\citeauthoryear{Vlahakis, Dunne \& Eales}{Vlahakis
  et~al.}{2005}]{Vlahakis2005}
Vlahakis C.,  Dunne L.,    Eales S.,  2005, MNRAS, 364, 1253

\bibitem[\protect\citeauthoryear{White, Helfand, Becker, Glikman \& de
  Vries}{White et~al.}{2007}]{White2007}
White R.~L.,  Helfand D.~J.,  Becker R.~H.,  Glikman E.,    de Vries W.,  2007,
  ApJ, 654, 99

\bibitem[\protect\citeauthoryear{{Wijesinghe} et~al.}{{Wijesinghe}
  et~al.}{2011}]{Wijesinghe2011}
{Wijesinghe} D.~B.,  et~al., 2011, MNRAS, 415, 1002

\bibitem[\protect\citeauthoryear{{Wright}}{{Wright}}{2006}]{Wright2006}
{Wright} E.~L.,  2006, PASP, 118, 1711

\bibitem[\protect\citeauthoryear{Wyder et~al.}{Wyder
  et~al.}{2007}]{Wyder2007}
Wyder T.~K.,  et~al., 2007, ApJS, 173, 293

\bibitem[\protect\citeauthoryear{Yi et~al.}{Yi  et~al.}{2005}]{Yi2005}
Yi S.~K.,  et~al., 2005, ApJ, 619, L111

\bibitem[\protect\citeauthoryear{{Zehavi} et~al.}{{Zehavi}
  et~al.}{2011}]{Zehavi2010}
{Zehavi} I.,  et~al., 2011, ApJ, 736, 59

\bibitem[\protect\citeauthoryear{Zheng, Bell, Rix, Papovich, Le~Floc'h, Rieke
  \& Pérez-González}{Zheng et~al.}{2006}]{Zheng2006}
Zheng X.~Z.,  Bell E.~F.,  Rix H.-W.,  Papovich C.,  Le~Floc'h E.,  Rieke
  G.~H.,    Pérez-González P.~G.,  2006, ApJ, 640, 784

\bibitem[\protect\citeauthoryear{Zhu et~al.}{Zhu  et~al.}{2011}]{Zhu2010a}
Zhu G.,  et~al., 2011, ApJ, 726, 110

\end{thebibliography}

\appendix
\section{Estimating Fluxes of Blended Sources Below the Noise Level}
\label{app:flux}
\subsection{Deblending individual sources}
When stacking sources we must be careful not to over-count the flux in blended sources, which would lead to over-estimation of stacked fluxes { especially} in bins whose galaxies are more clustered, and in the longer wavelength images which have lower resolution. Since many sources are close to or below the noise level in the images it is impossible to model them from the images themselves, and since submm flux is poorly correlated with optical flux, we have no other prior information to base models on. We must therefore make some simplifying assumptions in order to avoid over-counting.

Consider that two or more sources may be blended, but we do not know the true flux of either or their brightness ratio. How can we decide how much of the blended flux to attribute to each source?  We first make the assumption that all sources are unresolved and therefore have a shape given by the PSF. We treat the image pixel-by-pixel and assume that the fractional contribution to a pixel from each of the nearby blended sources is dependent only on the distance of that pixel from the source in question. The concept is visualised in one dimension in Fig.~\ref{fig:dfa-eg}. Panel (a) shows two sources $A$ and $B$ at positions $x_A$ and $x_B$, whose fluxes are distributed across the image as $f_A(x)$ and $f_B(x)$ (Jy\,pixel$^{-1}$). The sources are blended in the image and the measured data (solid line) is given by 
\begin{equation}
 f_\text{tot}(x)=f_A(x)+f_B(x)
\label{eqn:f_tot}
\end{equation}
In Fig.~\ref{fig:dfa-eg}{(b)} the sources are modeled by PSFs of equal height ($p_A$, $p_B$; thin black lines). We can convolve the image data with the PSF of $A$ to give the function
\begin{equation}
 F_{p,A}(x')=p_A*f_\text{tot}.
\label{eqn:fun_p}
\end{equation}
To compute the convolution the PSF is shifted so that the peak is at the origin. The convolution is a function of the offset $x'$, and the source flux is strictly given by the value at $x'=x_A$,
\begin{equation}
f_{p,A}=F_{p,A}(x_A)/\Sigma p_A^2
\label{eqn:flux_p}
\end{equation}
(and similarly for $B$). The division by the sum of the PSF squared normalises the flux. However, both $f_{p,A}$ and $f_{p,B}$ now contain too much flux because both include all of the flux that is blended. This blended flux would therefore be counted twice if the sources were stacked.
Instead of doing this, the PSF $p_A$ can be weighted by the function 
\begin{equation}
 g_A(x)=p_A(x)/[p_A(x) + p_B(x)]
\label{eqn:g}
\end{equation}
which is simply the fractional contribution from the PSF $p_A$ at position $x$ to the total $p_A+p_B$.
Thus we can replace $p_A$ and $p_B$ with `deblended' PSFs
\begin{equation}
 q_A(x)=g_A(x)\,p_A(x)
\label{eqn:q}
\end{equation}
(and similarly $q_B(x)$) which are given by the thick grey lines in Fig.~\ref{fig:dfa-eg}{(b)}. Convolving the image data with each of these deblended PSFs gives {a more conservative estimate of the total flux}:
\begin{equation}
 F_{q,A}(x')=q_A*f_\text{tot}
\label{eqn:fun_q}
\end{equation}
Again the flux is given by the value of the convolution at $x'=x_A$:
\begin{equation}
f_{q,A}=F_{q,A}(x_A)/\Sigma p_A^2
\label{eqn:flux_q}
\end{equation}
The total (deblended) flux $f_{q,A}+f_{q,B}$ is the same as the total input flux under the functions in Fig.~\ref{fig:dfa-eg}{(a)}, whereas the total of $f_{p,A}+f_{p,B}$ is greater because blended flux has been double-counted. This deblending method always conserves total flux, whatever the ratio of the fluxes.

On the other hand, the individual fluxes measured using equation~(\ref{eqn:flux_q}) are not exactly correct because blended flux is shared evenly between the two sources, whereas ideally it should be distributed according to the flux ratio of the sources. Hence in this example the recovered flux of $A$ is too low and that of $B$ too high (this effect is worsened by closer proximity of the sources). However with no prior information on the true flux ratio this is the best estimate that can be made. 

\begin{figure}
 \includegraphics[width=0.45\textwidth]{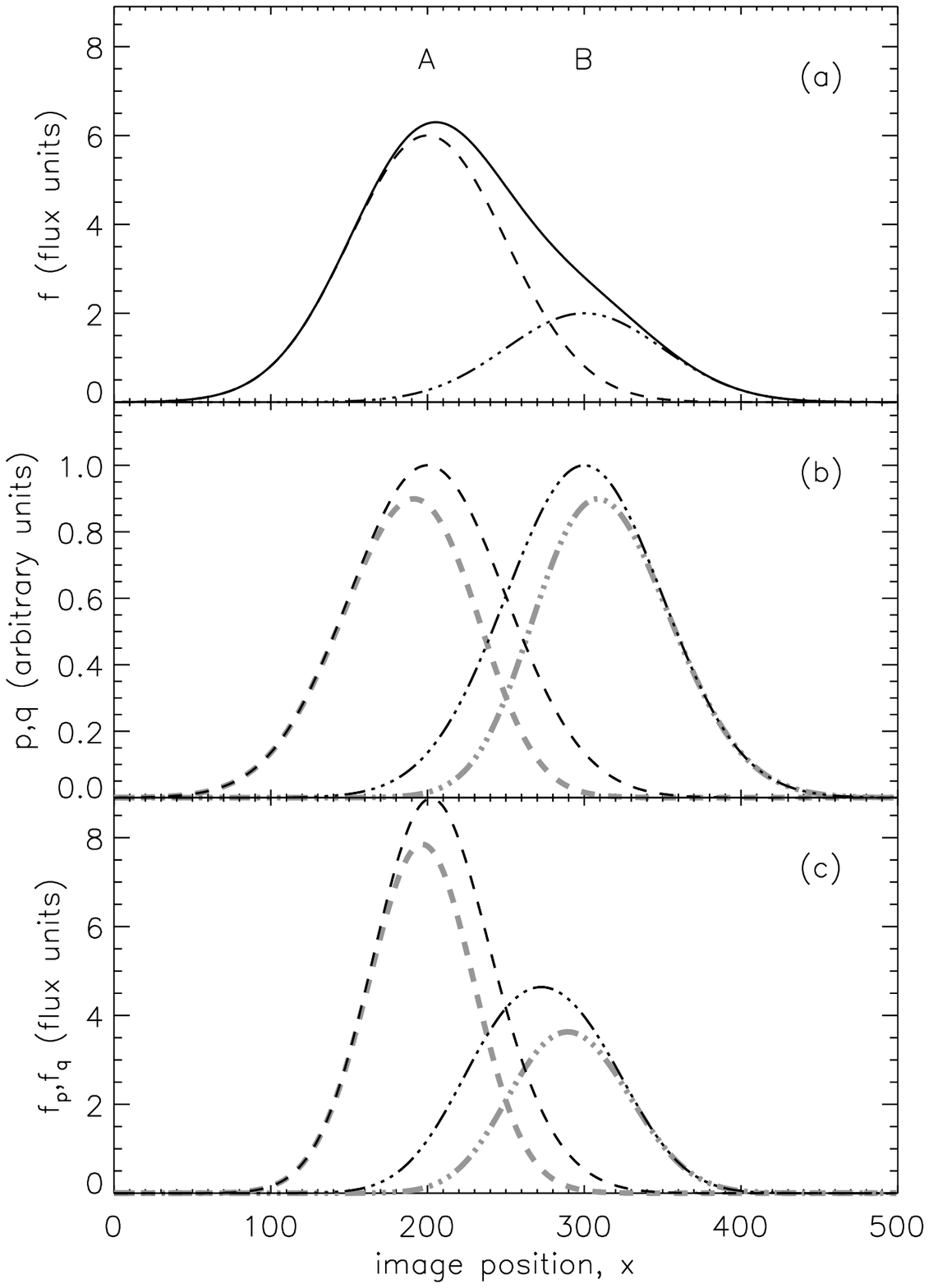}
\caption{\textbf{(a)} Two simulated point sources $A$ and $B$ in a one-dimensional image, {modeled with the same PSF but different normalisations,} represented by the dashed and dash-dotted curves respectively. The total flux in the image as a function of position $x$ is given by equation~(\ref{eqn:f_tot}) and is represented by the solid line.
\textbf{(b)} The thin black lines are the PSFs, $p_A$ and $p_B$, centred at $x=200$ and $x=300$ respectively. Both PSFs have width $\sigma=50$. The thick grey lines are the PSFs weighted for deblending, $q_A$ and $q_B$, given by equation~(\ref{eqn:q}).
\textbf{(c)} The reconstructed sources {given by the image data (solid line in panel a) weighted by the PSFs in the middle panel.} The thin black lines are obtained using the unweighted PSFs $p$ in equation~(\ref{eqn:fun_p}), and the thick grey lines using the weighted PSFs $q$ in equation~(\ref{eqn:fun_q}).}
\label{fig:dfa-eg}
\end{figure}

To generalise this method to a two-dimensional image with multiply-blended sources we start with an image array of the same dimensions as the data image, filled with values of zero. To this we add a PRF for every source in the input catalogue, centred on the pixel where the source is located and interpolated from the PSF with a small offset to correctly account for sub-pixel-scale positioning. In the region of an isolated source this image will be identical to the individual PRF, but where sources are blended it is equal to the sum of the PRFs (analogous to the sum of the thin black lines in Fig.~\ref{fig:dfa-eg}{(b)}). Thus all multiple blends are automatically accounted for, and the image we have constructed is analogous to the denominator in equation~(\ref{eqn:g}) (i.e. $p_A + p_B + ...$). For each source $i$ we derive the weighting function $g_i(x,y)$ as the ratio of the PRF to a cutout region of our all-PRFs image, as in equation~(\ref{eqn:g}). In other words, the weight given to the flux in a pixel $(x,y)$ is the value of the PRF of the target at $(x,y)$ divided by the sum total of the contributions of all PRFs in that pixel. We measure the flux of each source by convolving the {\it data} image (Jy\,pixel$^{-1}$) with the weighted PRF, as in equation~(\ref{eqn:fun_q}). 
{We tested the method in simulated maps with realistic source densities and using the PRFs of the three SPIRE bands. We found that the correct mean and median fluxes were always recovered when stacking, and that convolving with the PRF without any deblending always led to an overestimate of the median and mean fluxes.}

{The deblending technique is carried out before any binning, so all catalogue sources in the field are automatically deblended, not just those in the same bin as the target in question. We note that the method is essentially very similar to the `global deblending' technique described by \citet{Kurczynski2010}, which was demonstrated in that paper to minimise bias and variance in the stacks. 

\subsection{Comparison to a statistical approach}
The problem of stacking into confused maps is not a new one, and other methods for removing the excess flux due to blending have been used in the literature. The advantages of the method described above are that it automatically takes into account blending between objects in different bins, and also allows for the possibility of different bins having different amounts of blending (e.g. due to the stronger clustering of more massive and red galaxies -- \citealp{Zehavi2010}). To check that the results of our method are reasonable, we compare them to a simple statistical approach to calculate the average fraction of the stacked flux which results from multiple counting of blended sources \citep{Serjeant2008,Bourne2011}. Even a randomly distributed sample would lead to some excess signal from the superposition of multiple targets, but the average excess can be measured by simply stacking random positions in the map and subtracting this signal from the target stack. We do this already when we perform background subtraction (see Section~\ref{sec:stacking}). If however there is any clustering in the sample, then the probability of a target being superimposed on another target is greater than that of a random position being superimposed on a target, so the excess signal from blending in the target stack is not fully cancelled out by the background subtraction. { Following \Citet{Serjeant2008}}, the fractional flux contribution due to clustering is given by 
\begin{equation}
 F=n\int_0^\infty w(\theta) \text{e}^{-\theta^2/2\sigma^2} 2\pi \theta \text{d}\theta
\label{eqn:clustercorr}
\end{equation}
where { $n$ is the background source density,} $w(\theta)$ is the two-point angular correlation function of the sample and $\sigma$ is the width of the Gaussian beam profile in the map that we stack into. The function $w(\theta)$ describes the excess probability of a background source appearing at an angular distance $\theta$ from a target position, compared with a random distribution. We measured this using the \citet{Landy1993} estimator which counts pairs between the data ($D$) and random ($R$) positions as a function of separation $\theta$:
\begin{equation}
 w(\theta)=\dfrac{DD-2DR-RR}{RR}
\label{eqn:landyszalay}
\end{equation}
We counted pairs in 40 radial bins logarithmically spaced between 4 and 180 arcsec. The results, averaged over the three fields, are shown in Fig.~\ref{fig:correlation} together with a power-law fit by linear regression, described by $w(\theta)=(0.012 \pm 0.001) \theta^{(-0.76 \pm 0.03)}$. This fit is in good agreement with previous results from SDSS $r$-limited data \citep{Connolly2002}.

\begin{figure}
\includegraphics[width=0.45\textwidth]{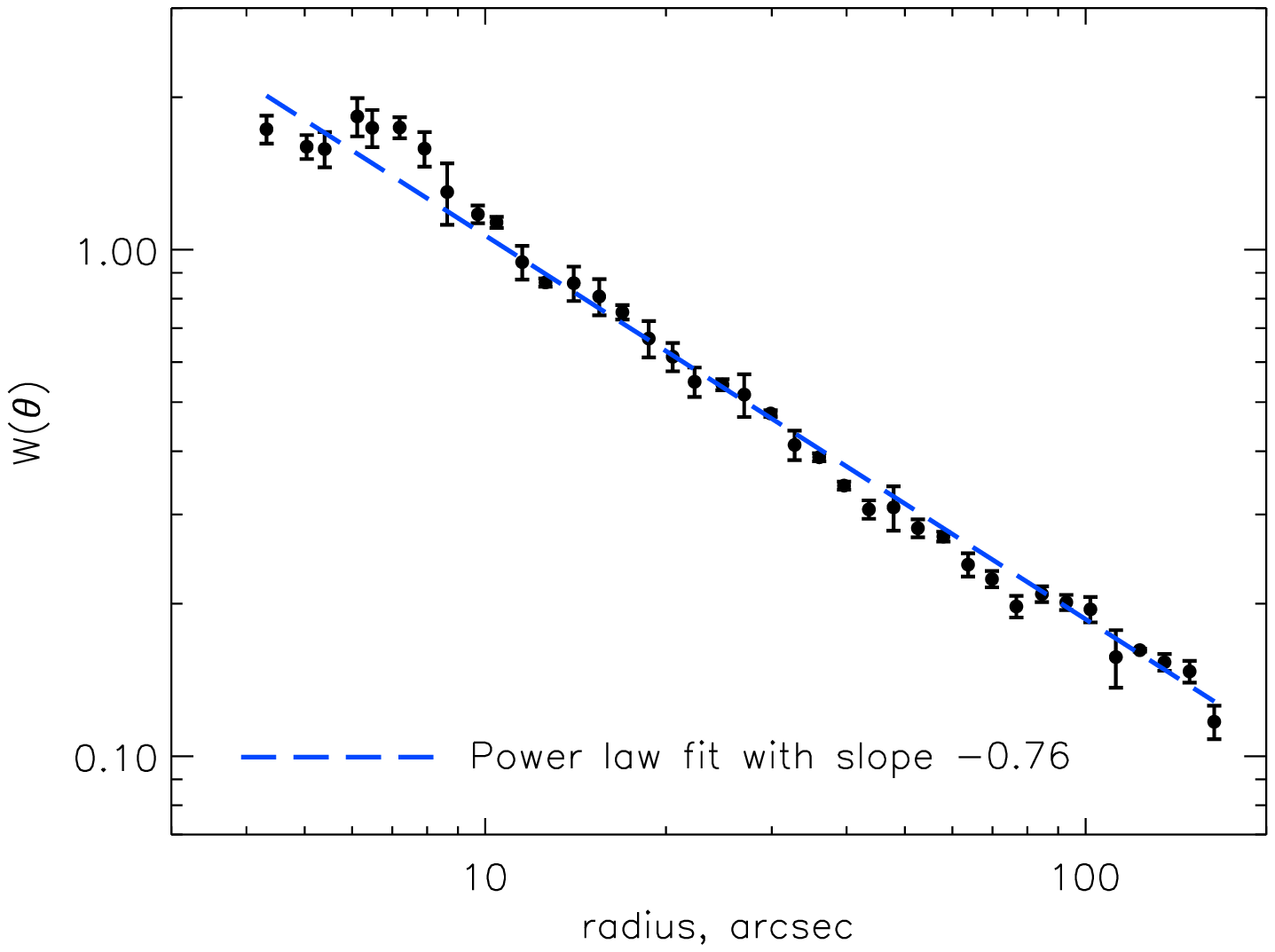}
\caption{The two-point angular correlation function of the GAMA catalogue used in this work, averaged over the G09, G12 and G15 fields. Error bars are the standard errors between the values obtained in the three fields. A power-law fit by linear regression (with free slope and normalisation) gives $w(\theta)=(0.012 \pm 0.001) \theta^{(-0.76 \pm 0.03)}$.}
\label{fig:correlation}
\end{figure}

By substituting into equation~(\ref{eqn:clustercorr}) the fit to $w(\theta)$ and the beam sizes at 250, 350 and 500\mum\ (18, 25, and 35 arcsec FWHM), we obtained the fractional contribution to stacked fluxes due to blending. The corrected flux is then obtained by multiplying the stacked flux by $C=1/(1+F)$. Results are summarised in Table~\ref{tab:clustercorr}. We see that this average statistical correction factor is broadly similar to the typical correction to individual fluxes using the deblending technique (equation~(\ref{eqn:flux_q})). The deblending technique has the advantage of correcting the flux of each target individually, so however the sample is binned the appropriate average correction is made. This is not the case using the correlation function of the entire catalogue, since that only gives a single correction factor. It would be possible to calculate separate correction factors for each bin, using the cross-correlation between the bin and the full sample (as in \citealp{Bourne2011}), but the uncertainties would be significantly increased since each bin contains a relatively small number of objects.

\begin{table}
\caption{Flux correction factors $C=1/(1+F)$ based on equation~(\ref{eqn:clustercorr}), compared with the typical/average ratios of deblended to non-deblended flux using equations (\ref{eqn:flux_q}) and (\ref{eqn:flux_p}).}

\begin{tabular}{ c c c}
 \hline
 & Statistical & Mean (median) \\
 & correction ($C$) & deblending correction \\
 \hline
 250\mum & $0.9702 \pm 0.0001$ & 0.937 (1.000) \\
 350\mum & $0.9550 \pm 0.0002$ & 0.932 (0.987) \\
 500\mum & $0.9326 \pm 0.0003$ & 0.989 (0.902) \\
 \hline

\end{tabular}

\label{tab:clustercorr}
\end{table}

}

\section{Simulations of Bias in the Median}
\label{app:sims}
{
We simulated power-law distributions of fluxes with Gaussian noise added, and found that in certain cases the median measured flux (true flux plus noise) was biased high with respect to the median true flux, as a result of noise in the measured values. 
The amount of bias depends on (i) the flux limits of the distribution; (ii) the $1\sigma$ noise level in relation to the flux limits; and (iii) the slope of the power law describing the underlying flux distribution.
The bias only becomes apparent when considering distributions with a median signal-to-noise less than $5\sigma$. 

In order to ascertain the level of bias that could be present in our stacks we must consider the shape of the underlying (true) flux distribution of sources in the stacks. In order to do so we must look at the distribution of fluxes in the much brighter \hatlas\ detected sample, which are not dominated by noise. 
The situation is helped considerably by the fact that we bin the sample according to \Mr\ and redshift, meaning that each bin is likely to have a limited range of fluxes with a distribution determined by the LF.
To estimate the flux limits in a given bin we can look at the submm fluxes of the galaxies with the highest optical fluxes. \citetalias{Dunne2011} show the distribution of $r$ magnitude and $S_{250}$ in the SDP ID catalogue. At $r<16$ the catalogue contains the full range of submm fluxes, which at a given $r$ spans 1.3\,dex in $S_{250}$.
We inspected the Phase 1 reliable IDs with good spectroscopic redshifts (250\mum\ sources matched to SDSS data using the same method as \citealt{Smith2011a}; these will be described by Hoyos,~{ C.} et al. in preparation). We found the same range { of 1.3\,dex} in $r<16$ sources in Phase 1 as in SDP. 

{ 
We therefore assume that any bin of \Mr\ and redshift will have fluxes within a limited range. The actual limits of this range, $S_\text{max}$ and $S_\text{min}$, will depend on the range of \Mr\ and $z$ sampled by the bin (although the majority of fluxes at all \Mr\ and $z<0.35$ will lie within the range 0.1-100\,mJy). However, since the range in a bin is determined by the LF, we can safely assume that the logarithmic range $R=\log_{10}(S_\text{max}/S_\text{min})=1.3$ will be consistent in all bins.
%
}
Similar ranges of 1.3\,dex are expected at all three SPIRE wavelengths (although of course fluxes at different wavelengths will be offset with respect to each other due to the shape of the SED). Within these limits the fluxes are assumed to follow a power-law distribution: in the Phase 1 IDs this is approximately described by differential number counts $dN/dS\propto S^{-2.5}$, although this is unreliable due to the incompleteness of the ID catalogue. 
A similar slope is apparent in the 250\mum\ number counts from $P(D)$ analysis of the HerMES maps \citep{Glenn2010}, although there is some evidence in that data for a shallower slope at $S_{250}\lesssim 10$\,mJy.
{
We must however remember that the number counts in our bins will not follow the overall submm number counts, since our bins contain only a narrow distribution of \Mr\ and particularly of redshift. 
In sufficiently narrow redshift bins the distribution of the counts will approach the submm LF: this has a slope of $-1.01$ at the faint end ($L< L^\star$) of the \hatlas\ LFs derived by \citetalias{Dunne2011}. 
We can therefore be confident that in finite redshift bins the slope will be intermediate between $-1$ and $-2.5$. 
\citet{Lapi2011} have modelled the number counts to fit the data from \hatlas, HerMES and BLAST. We split these into redshift bins ($\delta z =0.05$) and found that the faint-end differential counts at $z<0.35$ follow a slope of approximately $-2$. Our redshift bins have similar widths ($0.05 \lesssim \delta z \lesssim 0.11$) hence it is reasonable to assume that the flux distributions in the bins will have a similar slope.
}

For these reasons we conclude that a simulated flux distribution that spans a range $R=1.3$\,dex with a power law $dN/dS\propto S^{-2}$ is representative of our bins. We therefore created 16 simulated distributions with these parameters, with a range of lower limits between $0.1-30$\,mJy: the corresponding upper limits are 20 times larger, given by $R=1.3$. The resulting distributions have medians in the range $0.3-70$\,mJy, which is sufficient to cover all the bins in our real data. We added random noise to these true flux distributions, where the noise values were drawn from a Gaussian distribution with zero mean, and $\sigma$ given by the average total noise level (instrumental plus confusion) in the Phase 1 maps: $\sigma=6.7, 7.9, 8.8$\,mJy\,beam$^{-1}$ at 250, 350, 500\mum\ respectively. We then compared median measured flux (true flux plus noise) to the median true flux, and quantified the bias factor as the ratio of measured to true median flux.

We corrected the measured median fluxes in our stacked data by interpolating the relationship between true median flux and measured median flux from the simulations. This was done separately for each of the three bands (since the bias behaves differently as a result of the different noise levels). All correction factors are in the range $0.6-1.0$, which is generally small in comparison to the range of stacked fluxes resulting from true differences between the bins.

{
For these corrections we have assumed that the slope of differential number counts is $-2$ and that fluxes in each bin lie in a range given by $R=1.3$\,dex. However, the bias factor depends on both of these variables, as we show in Fig.~\ref{fig:sims}. To test the sensitivity of our results to these parameters, we tried correcting our results by the bias factors obtained using different values. We tried slopes of $dN/dS$ ranging from $-0.5$ to $-4$ and $R$-values between 1 and 2. The level of bias varies, indicating that there is some uncertainty in the correction, but we note that the corrections for slopes between $-1.5$ and $-2$ give almost identical corrections (for $R=1.3$). We are sure that the slope is between $-1$ and $-2.5$, and over this range the bias corrections do not vary by more than 15\%. Increasing the range $R$ has a greater impact on the correction; however the value of $1.3$ is well motivated and in narrow bins of \Mr\ and redshift there is good reason to expect the flux range to be limited.

Crucially, all of our conclusions remain valid, and all trends remain significant, for any combination of slope ($-0.5$ to $-4$) and range (1 to 2). This is equally true if we make no corrections. 

An alternative to these corrections would be to use the mean instead of the median when stacking. The mean is not altered by the effects of noise as we found the median to be; however the mean will be highly biased simply by the skewed shape of the distribution. 
In fact the bias in the mean at {\it all} flux levels is equal to the maximum bias seen in the median at the lowest flux levels. The reason for this is simple: at the lowest flux levels, where noise dominates over the true flux, the distribution of measured fluxes closely resembles the noise distribution -- a symmetrical Gaussian -- but instead of being centred on zero as the noise is, it has the same mean as the true flux distribution. The mean is not altered by the addition of noise if the mean noise is zero. The measured flux distribution is therefore symmetrical in this case, hence the median is equal to the mean. Thus the maximum amount of bias in the median occurs when the shape of the distribution becomes dominated by the noise rather than by the true fluxes.

In conclusion, we choose to use the corrected median estimator rather than the mean, because at all but the lowest fluxes the median is a better descriptor of the underlying (true) flux distribution. At the lowest fluxes the median becomes biased and approaches the mean of the distribution. However as a result of our binning scheme we can make a good estimate of the shape of the underlying flux distribution and can be reasonably certain of the correction factors for the bias. The fact that all of our ultimate conclusions remain valid for any reasonable choice of the distribution indicates the robustness of our results to these corrections.
}

\begin{figure*}
 \includegraphics[width=0.33\textwidth]{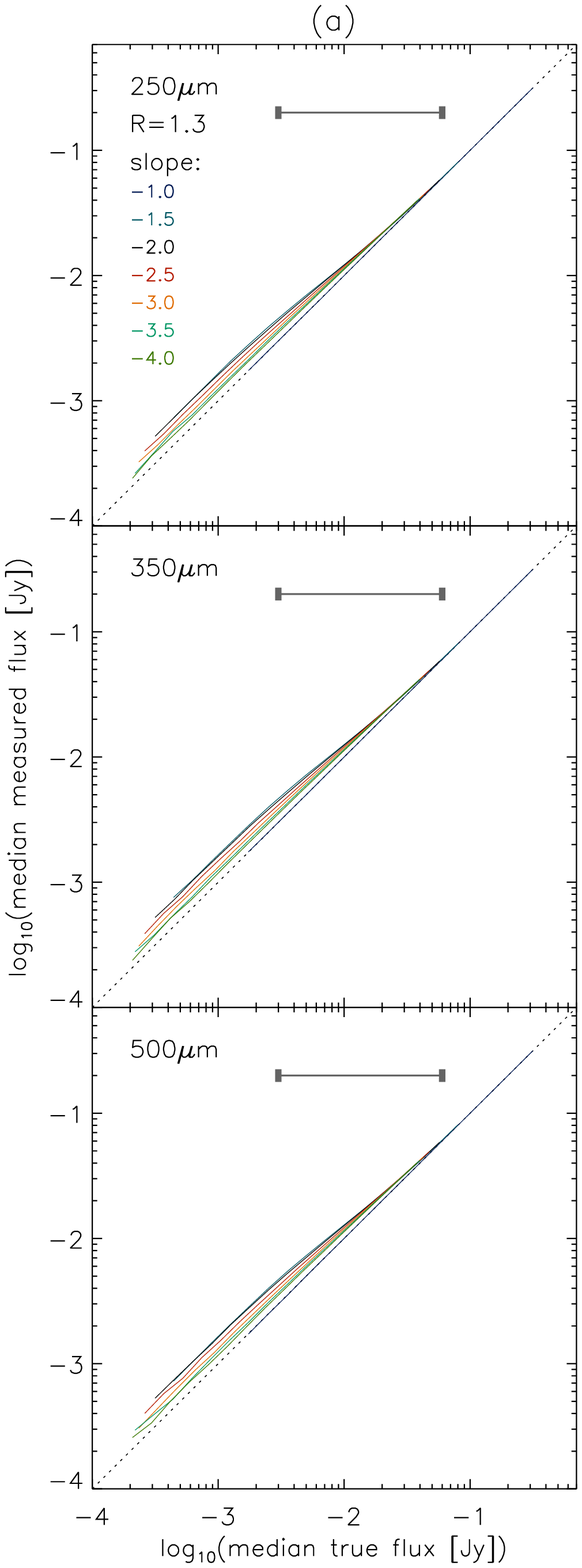}
 \includegraphics[width=0.33\textwidth]{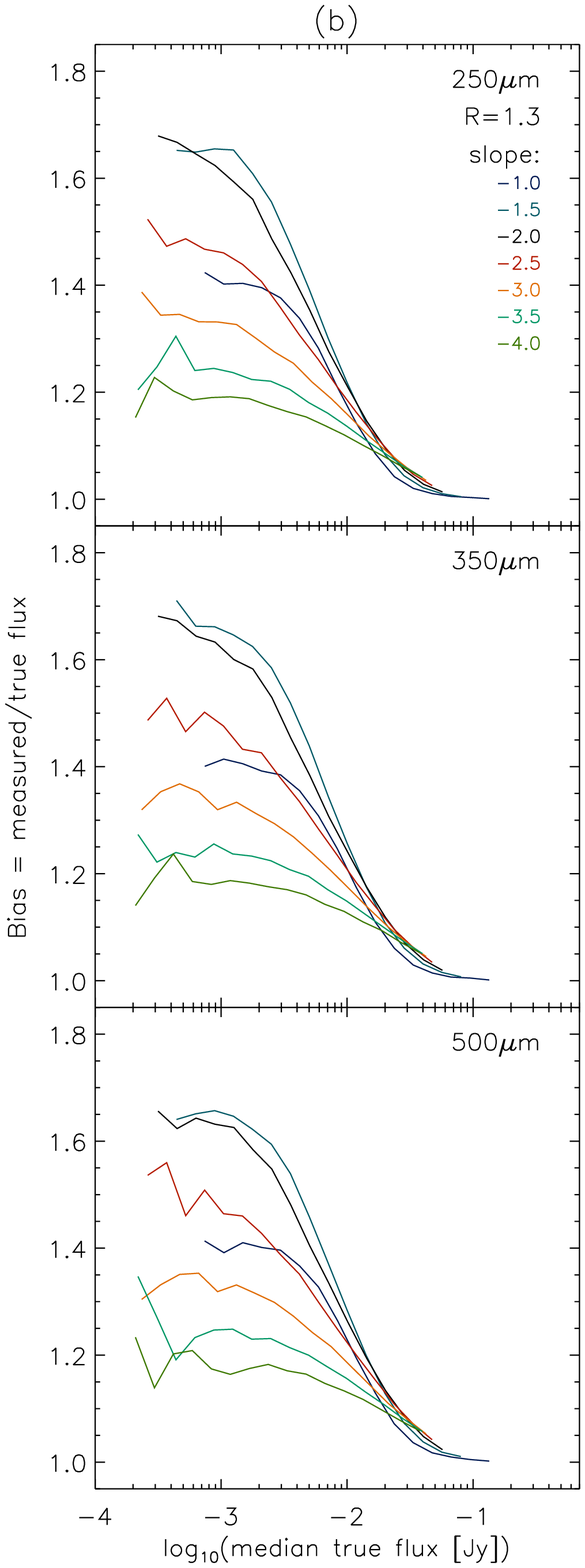}
 \includegraphics[width=0.33\textwidth]{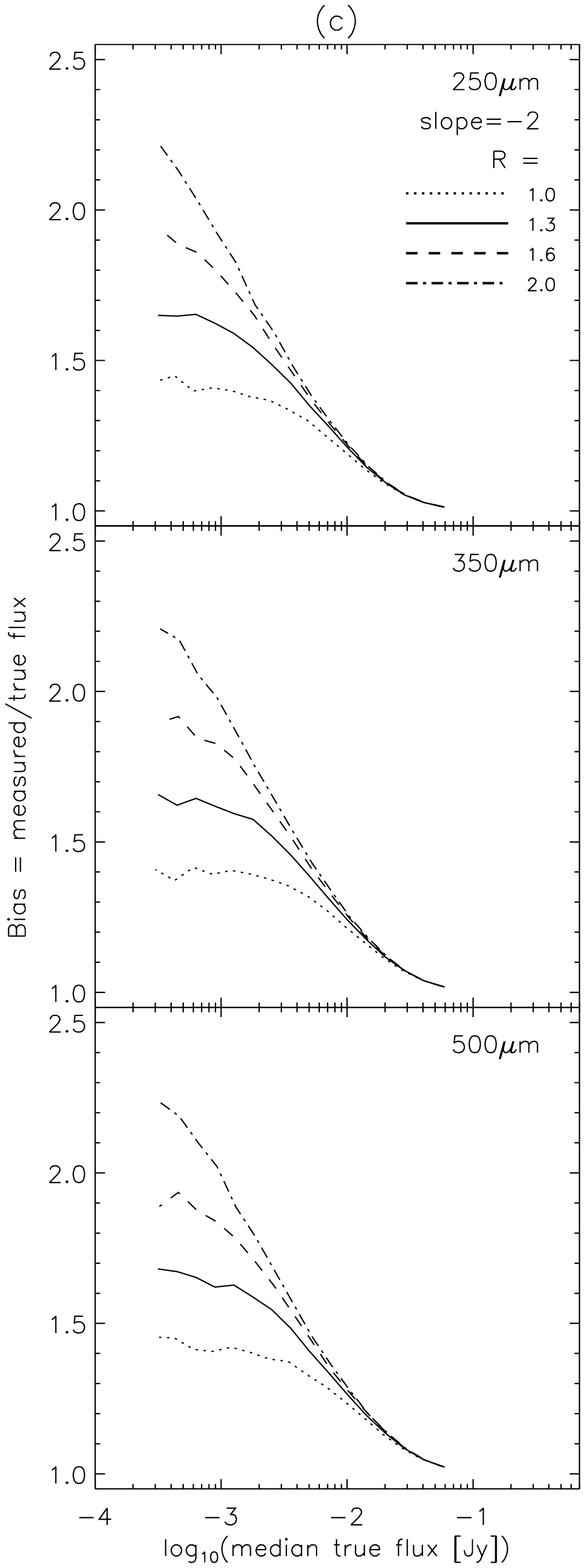}
\caption{{Results of a set of stacking simulations: {\bf (a)} Comparing median measured flux and median true flux, for distributions in various flux ranges. The range in each simulation is given by $R=\log_{10}(S_\text{max}/S_\text{min})=1.3$ (the length of the grey horizontal bar), and the simulations were run 16 times with different $S_\text{min}, S_\text{max}$ values to produce distributions with a range of median (true) fluxes. Each coloured line in the Figure connects 16 data points, one point from each simulation, showing how the median measured flux (after adding noise) varies as the median true flux is varied. Noise is drawn from a Gaussian distribution centred on zero, with $\sigma=6.7, 7.9, 8.8$\,mJy\,beam$^{-1}$ at 250, 350, 500\mum\ respectively. Different coloured lines correspond to sets of simulations with different slopes of $dN/dS$ between $-1$ and $-4$. {\bf (b)} Bias factor (ratio of median measured to median true flux) as a function of median true flux, for the same set of simulations, showing how the bias depends on the slope of $dN/dS$ for $R=1.3$. Note that bias does not vary monotonically with slope for fixed $R$, but is greatest for a slope of $-1.5$ in this case. {\bf (c)} Bias factor as a function of median true flux, showing the effect of increasing the range of true fluxes. Different lines correspond to sets of simulations with different values of $R$ between 1 and 2; in each of these the slope of $dN/dS$ is $-2$.}
}
\label{fig:sims}
\end{figure*}

}

\section{Postage Stamps, Histograms and Spectral Energy Distributions of Stacks}
\label{app:figs}
{
In Figures~\ref{fig:appc1} -- \ref{fig:appc4} we choose some example bins to show the stacked postage-stamp images, the distribution of measured fluxes, and the SED data and model. The Figures below contain the following information:

Postage stamps of the stack in the three SPIRE bands are shown with contours at signal-to-noise levels of 5, 10, 15, 20, 25, 50, 100, 150, 200 \& 250. The images are each 41 pixels square, corresponding to $3^{\prime}25^{\prime\prime}$ at 250\mum\ and $6^{\prime}50^{\prime\prime}$ in the other two bands. 
These images are illustrative only, and were made by stacking in the PSF-filtered, background-subtracted SPIRE maps (we do not measure fluxes in these maps but using the deblend filter method described in Appendix~\ref{app:flux}). 
The flux and signal-to-noise reached in the central pixel of each postage stamp agree with the stacked values in Table~\ref{tab:flux}, although the postage stamps show slightly boosted fluxes due to blending not being accounted for. Similar agreement was found in all stacks, although only a selection are shown here for brevity.

We also show histograms of the measured fluxes in the stack (red) and of a set of fluxes measured at random positions in the background (blue), as described in Section~\ref{sec:results} . The number shown is the KS probability that these two samples were drawn from the same distribution. Beneath these is plotted the single-component SED fitted to the three stacked fluxes (plotted in the rest frame), assuming $\beta=2$. The SED is fitted to obtain the temperature which is printed over the SED, as described in Section~\ref{sec:disc}.
}
\clearpage
\begin{figure}
\includegraphics[width=0.45\textwidth]{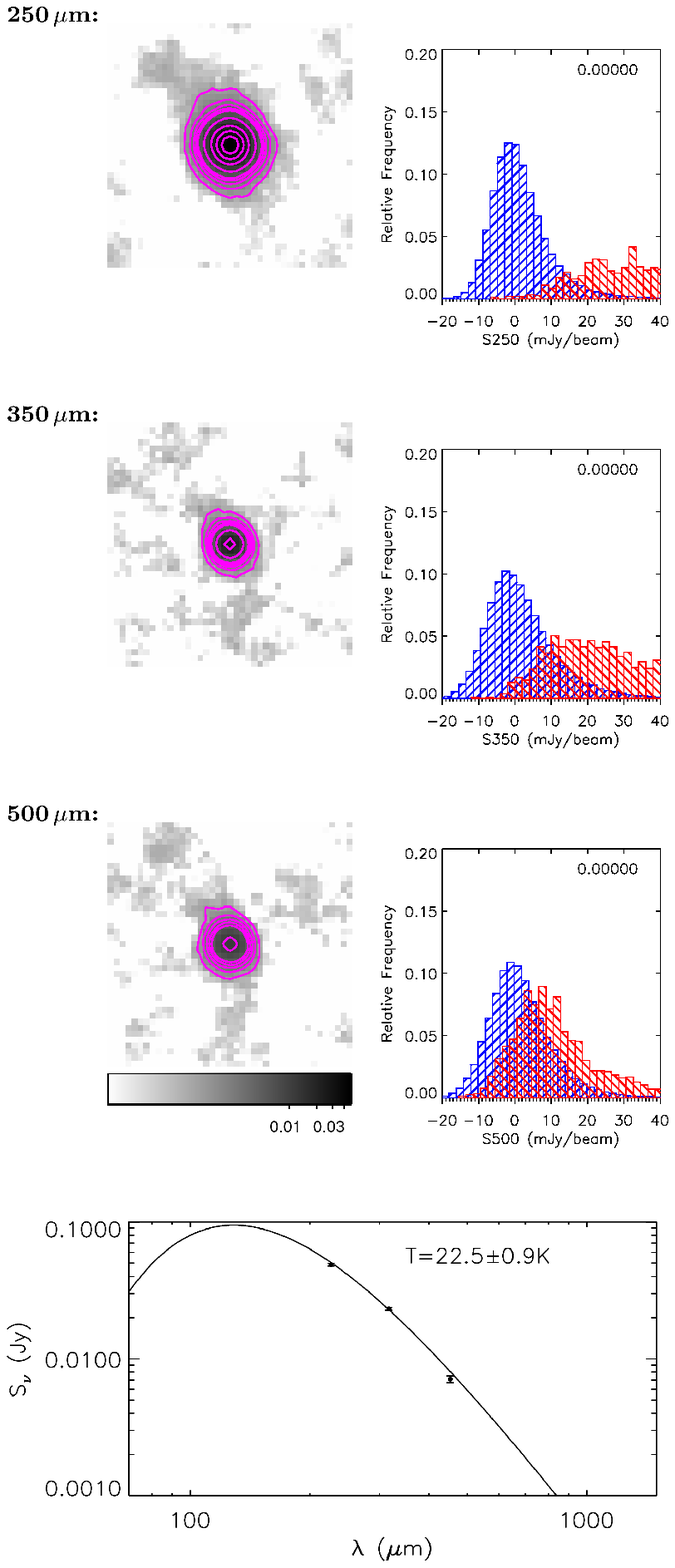}
\caption{A stack of galaxies with blue $g-r$ colours, median $\Mr=-21.1$, median $z=0.11$. This stack is one of the brightest in submm flux, and contains 1567 objects. Stacked, PSF-filtered postage stamps are shown with a logarithmic greyscale between $0.0001-0.05$\,Jy\,beam$^{-1}$ and signal-to-noise contours at 5, 10, 15, 20, 25, 50, 100, 150, 200, 250. Histograms of the measured fluxes in the stack (red) and in a random background stack (blue) are shown with the KS probability that they were drawn from the same distribution. Also plotted is the rest-frame SED fit with the temperature indicated, assuming $\beta=2$. More details are given in the text { of Appendix \ref{app:figs}}.}
\label{fig:appc1}
\end{figure}

\begin{figure}
\includegraphics[width=0.45\textwidth]{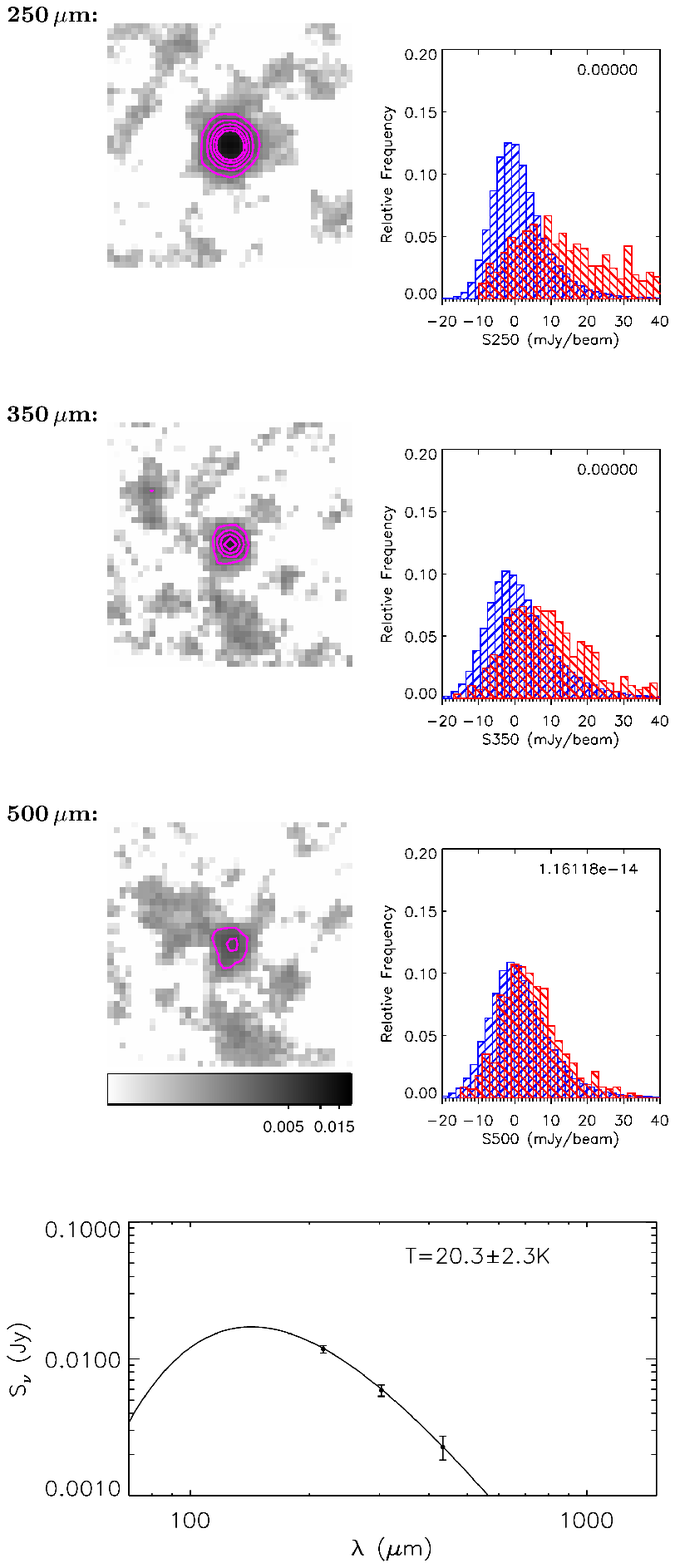}
\caption{A stack of galaxies with green $g-r$ colours, median $\Mr=-21.1$, median $z=0.15$. This stack has moderate submm flux, and contains 570 objects. Plots are as in Fig.~\ref{fig:appc1} except that the postage stamps are plotted on a logarithmic greyscale between $0.0001-0.02$\,Jy\,beam$^{-1}$.}
\end{figure}

\begin{figure}
\includegraphics[width=0.45\textwidth]{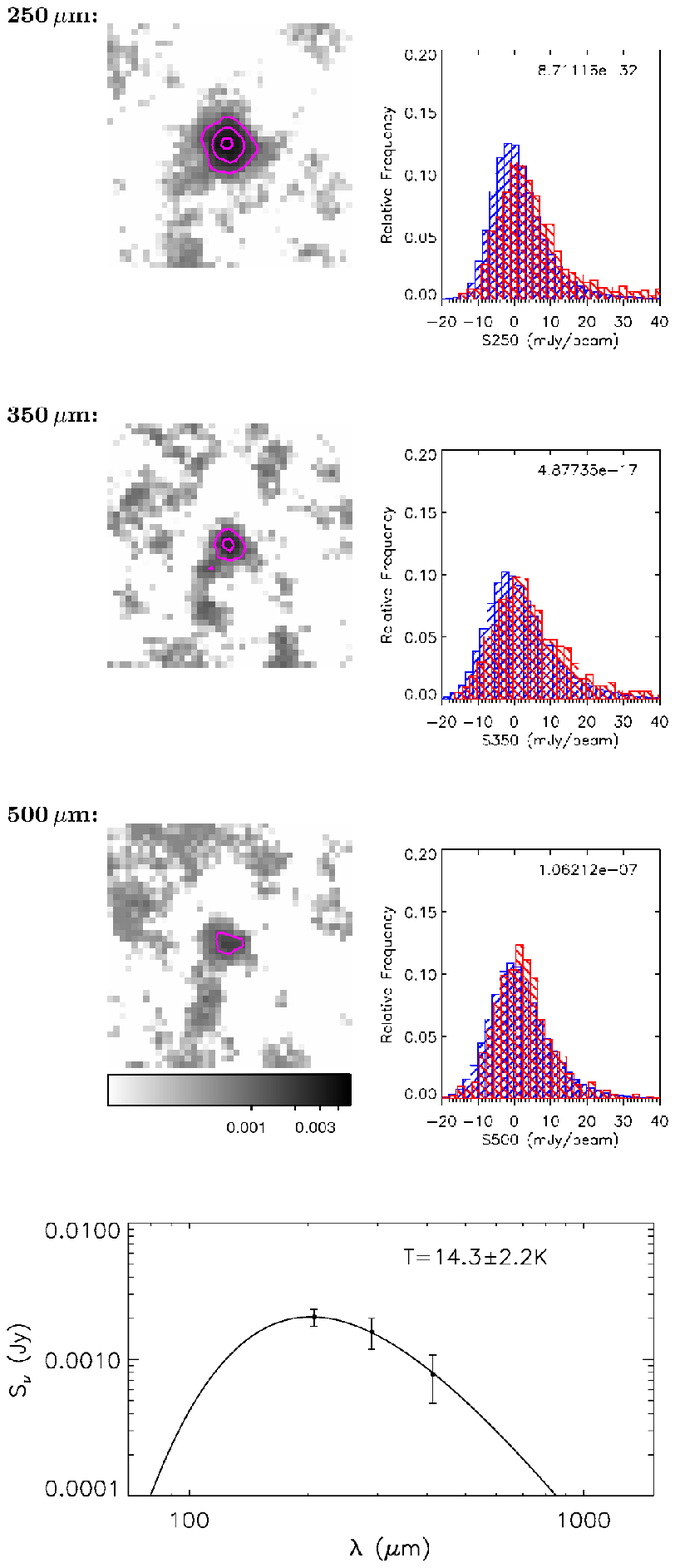}
\caption{A stack of galaxies with red $g-r$ colours, median $\Mr=-21.7$, median $z=0.21$. This stack has among the faintest submm fluxes, and contains 1111 objects. Plots are as in Fig.~\ref{fig:appc1} except that the postage stamps are plotted on a logarithmic greyscale between $0.0001-0.005$\,Jy\,beam$^{-1}$.}
\end{figure}

\begin{figure}
\includegraphics[width=0.45\textwidth]{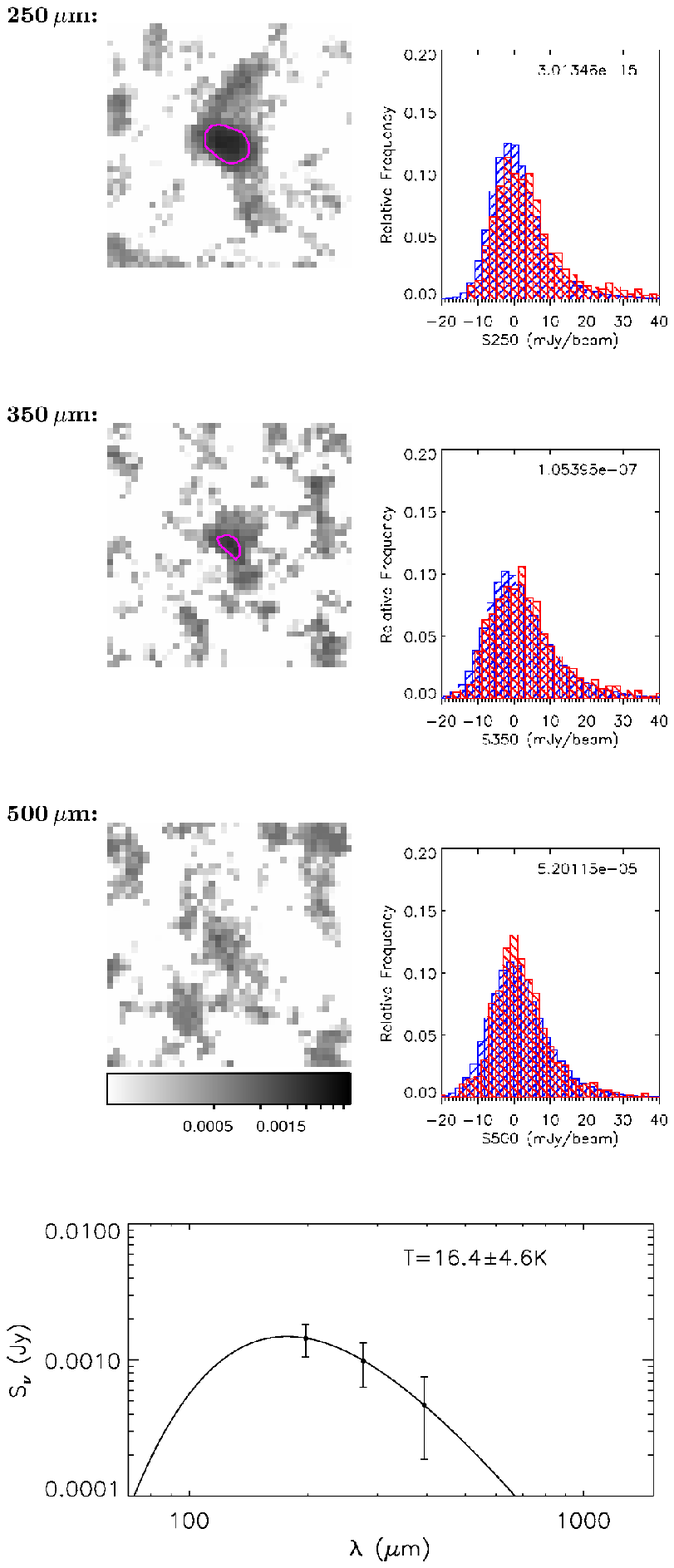}
\caption{A stack of galaxies with red $g-r$ colours, median $\Mr=-22.5$, median $z=0.27$. This stack has among the faintest submm fluxes, and contains 1002 objects. Plots are as in Fig.~\ref{fig:appc1} except that the postage stamps are plotted on a logarithmic greyscale between $0.0001-0.004$\,Jy\,beam$^{-1}$.}
\label{fig:appc4}
\end{figure}

\clearpage

\end{document}